\documentclass[12pt,a4paper]{iopart}

\newcommand{\vect}[1]{\boldsymbol{\mathbf{#1}}}

\expandafter\let\csname equation*\endcsname=\relax
\expandafter\let\csname endequation*\endcsname=\relax

\usepackage{amsmath,amsfonts,amsthm}
\usepackage{color}
\usepackage{graphicx}
\usepackage{hyperref}
\usepackage{subcaption}
\usepackage{tikz}
\usetikzlibrary{arrows}
\usetikzlibrary{calc}
\usetikzlibrary{patterns}
\usetikzlibrary{scopes}

\newcommand{\<}{\langle}
\renewcommand{\>}{\rangle}

\numberwithin{equation}{section}

\begin{document}
\title{Self-attracting polymers in two dimensions with three low-temperature phases}
\author{A Bedini$^1$, A L Owczarek$^1$ and T Prellberg$^2$}
\address{$^1$ Department of Mathematics and Statistics,
  The University of Melbourne, Vic 3010, Australia.}
\address{$^2$ School of Mathematical Sciences, Queen Mary University
  of London, Mile End Road, London E1 4NS, UK.}
\ead{abedini@ms.unimelb.edu.au, owczarek@unimelb.edu.au, t.prellberg@qmul.ac.uk}

\begin{abstract}
We study  via Monte Carlo simulation a generalisation of the so-called vertex interacting self-avoiding walk (VISAW) model on the square lattice. The configurations are actually not self-avoiding walks  but rather  restricted self-avoiding trails (bond avoiding paths)  which may visit a site of the lattice twice provided the path does not cross itself: to distinguish this subset of trails we shall call these configurations \emph{grooves}. Three distinct interactions are added to the configurations: firstly the VISAW interaction, which is associated with doubly visited sites, secondly a nearest neighbour interaction in the same fashion as the canonical interacting self-avoiding walk (ISAW) and thirdly, a stiffness energy to enhance or decrease the probability of bends in the configuration.

In addition to the normal high temperature phase we find three low temperature phases: (i) the usual amorphous liquid drop-like ``globular" phase, (ii) an anisotropic ``$\beta$-sheet'' phase with dominant configurations consisting of aligned long straight segments, which has been found in semi-flexible nearest neighbour ISAW models, and (iii) a maximally dense phase, where the all sites of the path are associated with doubly visited sites (except those of the boundary of the configuration), previously observed in interacting self-avoiding trails. 

We construct a phase diagram using the fluctuations of the energy parameters and three order parameters. The $\beta$-sheet and maximally dense phases do not seem to meet in the phase space and are always separated by either the extended or globular phases. We focus attention on the transition between the extended and maximally dense phases, as that is the transition in the original VISAW model. We find that for the path lengths considered there is a range of parameters where the transition is first order and it is otherwise continuous. 
\end{abstract}

\maketitle

\section{Introduction}
\label{sec:introduction}

There are many lattice models of a single polymer in solution that take account of different physical scenarios and capture different aspects of such a system, for example by modelling the inclusion of  hydrogen bonding or stiffness. Importantly, a closer analysis of the behaviour of different lattice models has shown that details in the modelling of polymers, such as excluded volume and interactions, affect the phase structure and the universality class of the transitions between different phases. In this work we consider a general model which contains many of these aspects within one single model.
 
The first ingredient in all these models is the type of lattice path that is considered. The simplest example is the unrestricted random walk, which is a lattice path that may visit sites and bonds of the lattice multiple times. On the other hand the classical model of polymers is based on the \emph{self-avoiding walk}, which is a lattice path that cannot visit either bonds or sites of the lattice more than once. There are however several configuration systems used that lie between these two extremes. One commonly used for modelling polymers is the self-avoiding trail, which is a lattice path that is bond avoiding but may visit sites multiple times. Importantly, this path may both touch at a site, or may cross. 

Consider the subset, $\mathcal G_n\subseteq \mathcal T_n$, of $n$-step trails $\mathcal T_n$ on the square lattice with the added restriction that no crossings are allowed. This is an important type of configuration since they appear in the high temperature expansion of the $O(n)$ model on the square lattice (see \cite{Fu2013} for a recent example). Let us call these configurations \emph{grooves}. Given that the type of underlying configurations seems to be related to different phase behaviour, we have used this new terminology to clearly distinguish this configuration type, which is intermediate between trails and walks. We note that grooves are the configurations found for a special multi-critical point of the $O(n)$ model, known as the Bl\"{o}te-Nienhuis point or BN-point \cite{blote1989a-a}.

\begin{figure}[ht!]
  \centering
  \includegraphics{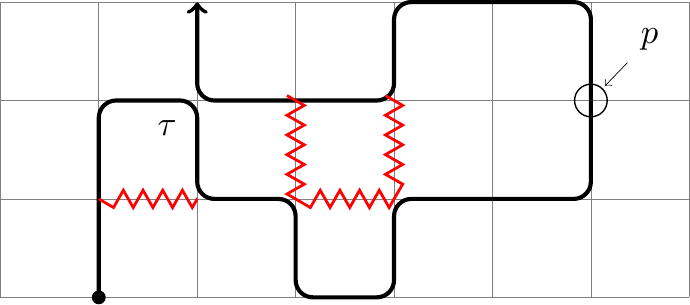}
  \caption{An example of a \emph{groove} of $n=18$ steps with one
    ($c = 1$) collision, five ($s = 5$) straight segments, and four ($m=4$) nearest neighbours. There are $v_1=17$ singly visited sites and $v_2=c=1$ doubly visited sites. The total weight of this configuration is $\tau p^5 \omega^4$.}
  \label{fig:configuration1}
\end{figure}

The second ingredient in a lattice model of polymers in solution is the set of interactions that are associated with various combinatorial features of the configurations. Given a groove $\psi_n \in \mathcal G_n$, we associate the following set of Boltzmann weights: a weight $\tau$ for \emph{site interaction}, i.e.\ for a site that the groove visits more than once, such as occuring in the interacting self-avoiding trail model (ISAT), a weight $p$ for every straight segment, which moderates stiffness in the polymer, and finally a weight $\omega$ for each \emph{nearest-neighbour interaction}, i.e.\ for a pair of visited sites that are nearest-neighbours on the lattice but non sequential in the groove. This nearest-neighbour interaction occurs in the canonical interacting self-avoiding walk (ISAW) model of polymer collapse. Figure~\ref{fig:configuration1} shows an example of a configuration belonging to $\mathcal G_n$ and its associated weights.

There has been long term interest in special cases of this model over and above the canonical ISAW model, which has a weak second order phase transition between a high temperature expanded phase and a low-temperature phase. One can distinguish these phases for example by the different scaling of the size of the polymer as given by its radius of gyration. The associated exponent takes the value $\nu=3/4$ in the high-temperature phase and $\nu=1/2$ in the low-temperature phase. In the low temperature phase the polymer is disordered and dense, but not maximally dense: it has been described as a liquid-like disordered drop. The phase transition known as the $\theta$-point is conjectured in two dimensions to have a cusp singularity in a convergent specific heat with the associate finite length exponent $\alpha\phi=-1/7$. The size exponent at the transition is $\nu=4/7$. The ISAW model is realised by letting $\tau=0$ and $p=1$.

On the other hand setting $\omega = 1$ and $p=1$ gives the so-called ``vertex interacting self-avoiding walk" (VISAW), which is a misnomer since the configurations are actually grooves (as described above) and not self-avoiding walks. This model first arose as a specialisation of an integrable lattice model introduced by Bl\"ote and Nienhuis in 1989 \cite{blote1989a-a} related to the Izergin-Korepin vertex model. In fact, that model included more generally a stiffness parameter $p$. The model was studied numerically \cite{foster2003a-a,foster2003b-a} for various values of $p$ including $p=1$. It has also been studied as part of a generalisation of the interacting self-avoiding trail model (asymmetric ISAT) that distinguishes collisions from crossings, using transfer matrix calculations \cite{foster2011} and Monte Carlo simulations \cite{bedini2013}. The Monte Carlo work \cite{bedini2013} concluded that the transition in the VISAW $(\tau, \omega, p)=(\tau_c,1,1)$ would apparently seem to be a strong second order transition. The exponent describing the finite length singularity in the specific heat has been estimated \cite{bedini2013} as $\alpha\phi\approx 0.69$, which is in accord with the divergence seen in the ISAT model of $0.68(5)$ \cite{owczarek2006c-:a}, estimated from much longer configurations. The size exponent of VISAW has not been measured using Monte Carlo, though the ISAT model has $\nu=1/2$ with a scaling form containing multiplicative logarithmic corrections: this clearly differs from the $\theta$-point ISAW value of $4/7$. 
In work on the asymmetric ISAT  \cite{bedini2013} the low temperature phase of VISAW (with $p=1$) was considered; it was found to have a similar structure to the low temperature phase in the ISAT model. This low temperature phase is not the same as the globular, liquid-like phase in the low temperature region of ISAW: for VISAW and ISAT the low temperature phase is maximally dense, i.e.\ every site of the path, apart from those on the outside of a dense ball, is associated with a doubly visited site of the lattice.

In contrast, setting $p = 0$ excludes the possibility of straight segments and by also setting $\omega=1$ the model is known as interacting self-avoiding trails on the L-lattice (\emph{LSAT}). Note that by setting $p=0$ the trails do not cross themselves and so are actually \emph{grooves}. The model can also be mapped onto ISAW on the Manhattan lattice \cite{bradley1989a-a}. A kinetic growth algorithm that produces long L-lattice trails maps to configurations with Boltzmann weights $(\tau, \omega, p)=(2,1,0)$: an analysis of simulations of very long configurations \cite{prellberg1994a-:a} has shown that the transition is $\theta$-like with a weak second order transition with an exponent conjectured to be the same as the $\theta$-point value of  $\alpha\phi=-1/7$. The scaling of the size of the polymer given by the radius of gyration or similar is described by the exponent $\nu=4/7$. Interestingly, the low temperature phase of the LSAT model has not been previously considered. Of course it has been implicitly assumed that it would be like the ISAW model, i.e.\ disordered and dense but not maximally dense.

The parameter space of the semi-flexible VISAW model ($\omega=1$) also includes the special multi-critical point of the $O(n)$ model found in \cite{blote1989a-a}, known as the Bl\"{o}te-Nienhuis point or BN-point. The location of this point is given by $(\tau, \omega, p)=(\tau_{BN}, 1, p_{BN})$ with  $p_{BN} = 0.275899\ldots$, and $\tau_{BN} = 2.630986\ldots$. Recently there have been various investigations of the BN-point itself \cite{foster2003a-a,bedini2013d-:a}. In the Monte Carlo work \cite{bedini2013d-:a} the size exponent was estimated as close to $\nu=4/7$ in agreement with earlier work \cite{foster2003a-a}. The specific heat exponent was not estimated and since the study looked at just that point the low temperature phase was not considered. Very recently, a theoretical framework for the BN-point has been elucidated where it has been proposed that this point is related to a particularly unusual conformal field theory with continuous exponents \cite{vernier2015a-a}.

So while the semi-flexible VISAW model has been studied previously by Monte Carlo at several values of $p$, there are clearly some outstanding gaps in our knowledge. To gain a wider perspective, the generalised model we study here allows us to interpolate between the semiflexible VISAW model and the canonical interacting self-avoiding walk (ISAW) model directly. The set of $n$-step self-avoiding walks $\mathcal S_n$ is a subset of grooves, $\mathcal S_n \subseteq \mathcal G_n$. Hence, if we set $\tau = 0$ we suppress all multiple visits to sites and recover self-avoiding walks as configurations. We are now left with two Boltzmann weights $\omega$ and $p$ which gives us the semi-flexible interacting self-avoiding walk model or semi-flexible ISAW model. This model was studied on the square lattice in \cite{Krawczyk2010}.

\begin{figure}[ht!]
\centering
\includegraphics[width=0.8\textwidth]{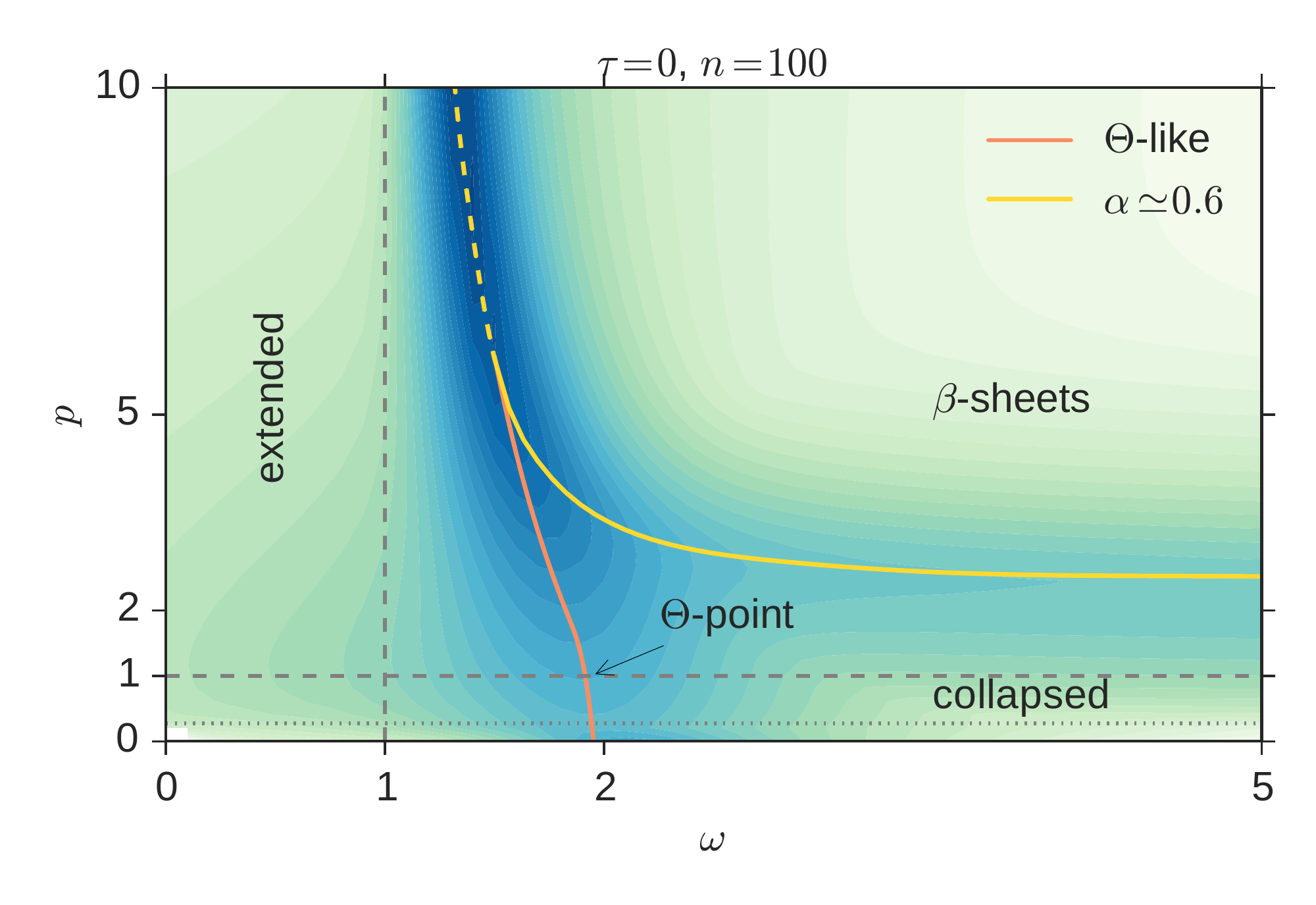}
\caption{A density plot of the largest eigenvalue of fluctuations of the semi-flexible ISAW model ($\tau=0$) and a schematic phase diagram inferred from this  that is in accord with previous work. Darker shades indicate larger values of the fluctuations. Later we shall see that appropriate  order parameters/indicator variables  reinforce these conclusions. There are three phases: extended, collapsed (globular) and $\beta$-sheet-like. Solid lines represent the approximate location of a second-order phase transitions. The dashed line indicates first order behaviour.}
\label{fig:sfisawpd}
\end{figure}

To present a phase diagram that can be readily compared to the data we describe below, we present the schematic phase diagram for semi-flexible ISAW. This has been obtained from our own data and is in accord with the phase diagram elucidated in \cite{krawczyk2009semi} previously. In Figure~\ref{fig:sfisawpd} we see a density plot of a generalised specific heat and a schematic phase diagram as per \cite{krawczyk2009semi}. 

The data in Figure~\ref{fig:sfisawpd} is a slice obtained from the dataset 3P by fixing $\tau = 0$ (see table \ref{appendix:table}). There are three phases: two that occur in the canonical ISAW model, that is the extended and globular phases, and a third which is an anisotropic collapsed phase that can be described as $\beta$-sheets. These two-dimensional $\beta$-sheets are actually parallel lines of polymer in the $x$ or $y$ directions. The extended and globular collapsed phases are separated by a line of weak second-order transitions. The $\theta$-point lies on this line and the conjecture is that  the entire line belongs to the $\theta$-universality class. On the other hand the extended phase and $\beta$-sheet-like phase are separated by a line of first order transition points. According to \cite{krawczyk2009semi}, the $\beta$-sheet-like phase and the collapsed phase are separated by a line of second-order phase transitions associated with a positive $\alpha\phi$-exponent (around $0.4$) and therefore a diverging specific-heat.

As can be seen from this summary of the models that are contained within our model there are very different transitions and different collapsed low temperature phases. This mirrors experimental evidence that physical polymers can have different low temperature phases \cite{experiments,experiments2,experiments3}. Here we investigate our generalised model filling in the gaps in our knowledge and providing a coherent picture of how the transitions and where the different low temperature phases exist.

\section{The Model}
\label{sec:model}

\subsection{Quantities}

Given an $n$-step groove $\psi_n\in\mathcal G_n$ with $c$ collisions, $s$ straight segments and $m$ nearest neighbours, we assign to this groove a Boltzmann weight $ \tau^c \omega^m p^s$. Summing over all $n$-step grooves, we arrive at the partition function
\begin{equation}
  \label{eq:canonical-pf}
  Z_n(\tau, \omega, p) = \sum_{\psi_n\in\mathcal G_n}\ \tau^c \omega^m p^s
  .
\end{equation} 
If we denote the number of singly visited sites of the path as $v_1$ and the number of doubly visited sites by $v_2$ we can infer their values from $n$ and $c$ by observing that $n+1 = v_1 +2 v_2$. We find that $v_2 = c$ and $v_1 = n +1 - 2 c$.

The average of any quantity $Q$ over the ensemble set of grooves $\mathcal G_n$ is given generically by
\begin{equation}
  \langle Q \rangle_n(\tau, \omega, p) = \frac{1}{Z_n(\tau, \omega,p)}\sum_{\psi_n\in\mathcal
    G_n} Q(\psi_n) \, \tau^c \omega^m p^s 
  .
\end{equation}
In this paper we are interested in the following quantities. We calculate a measure of the size of the polymer, $\langle R^2 \rangle_n $, by using the mean end-to-end distance $\langle R_e^2 \rangle_n $ defined as follows: Letting any $n$-step path $\psi_n$ on a lattice by a
sequence ${{\bf r}_0, {\bf r}_1, \ldots, {\bf r}_n}$ of vector positions
of the vertices of that path the average-square end-to-end distance is given by
\begin{equation}
\langle R_e^2 \rangle_n = \langle {\bf r}_n \cdot {\bf r}_n \rangle\; ,
\end{equation}
In the above formulae we use ${\bf r}_0 \equiv {\bf0}$.

We are actually more interested in analysing the polymer density which is given by
\begin{equation}
\rho_n = \frac{n}{R^2}\,.
\end{equation}
As noted in the introduction one of the phases we have is anisotropic so to assist in identifying such a phase we define an  anisotropy by
\begin{equation}
\zeta = \left< \left(\frac{n_x - n_y}{n_x + n_y}\right)^2 \right>\,,
\end{equation}
where $n_x$ and $n_y$ are the number of bonds parallel to the $x$-axis and $y$-axis, respectively. The number of singly visited sites per unit length is given by
\begin{equation}
\langle v_1 \rangle(n) = 1 - 2\frac{\langle c \rangle}{(n+1)}\,.
\end{equation}

\subsubsection{Scaling}
\label{subsec:scaling}

The size of the polymer defined above is expected to scale as
\begin{equation}
  R^2_n \sim C_R n^{2\nu}\; ,
\end{equation}
where the amplitude $C_R$ is non-universal and temperature dependent, while $\nu$ is expected to be universal, depending only on the phase.

For high temperature it is always expected that $\nu=3/4$ which is the value for non-interacting self-avoiding walks and self-avoiding trails so
\begin{equation}
\rho_n \sim 1/n^{1/2}\,,
\end{equation}
which goes to zero in the extended phase. For each of the low temperature phases we expect $\nu=1/2$ so that $\rho$ will attain a non-zero positive value in the thermodynamic limit.

The scaling of the peak value of the specific heat $c_n$ or indeed any similar measure such as the largest eigenvalue of the fluctuations matrix at the collapse transition is expected to behave as
\begin{equation}
  c_n^{peak} \sim A n^{\alpha \phi} + C_{x},
\end{equation}
assuming the transition is second order. The need to consider the background constant $C_{x}$ depends on whether $\alpha\phi$ is positive or negative.

For any isotropic phase one expects that
\begin{equation}
n_x  \sim  C_a n_y \sim \frac{C_a}{1+C_a}  n \,,
\end{equation}
so that $\zeta$ will converge to a constant less than one. If it is fully isotropic $C_a=1$ and, moreover,
\begin{equation}
n_x  - n_y \sim o(n) \,.
\end{equation}
so that $\zeta$ converges to zero in the thermodynamic limit.

On the other hand for a fully anisotropic phase one expects that
\begin{equation}
{\min(n_x,n_y)} = o(n)\,,
\end{equation}
so that 
\begin{equation}
\| n_x  - n_y \|  \sim n
\end{equation}
and $\zeta$ will converge to one in such a phase.

For a maximally dense phase one may expect that 
\begin{equation}
\left< v_1 \right>(n) = o(n)\,.
\end{equation}

\section{Simulation and methodology}

We studied our generalised model  for various slices of the three-dimensional parameter space $(\tau, p, \omega)$ using Monte Carlo simulations.

Our simulations were based on the flatPERM algorithm \cite{prellberg2004flat} which is a flat histogram version of the Pruned and Enriched Rosenbluth Method (PERM) developed in \cite{grassberger1997pruned}. The PERM algorithm generates a polymer configuration kinetically, which is to say that each growth step is selected at random from all possible growth steps. During this process the algorithm keeps track of a weight factor to correct the sample bias. At each growth step, configurations with very high weight relative to other configurations of the same size are duplicated (or \emph{enriched}) while configurations with low weight or that cannot be grown any further (i.e. because they are ``trapped" or because they have reached a limit length) are discarded (or \emph{pruned}). A single iteration is then concluded when all configurations have been pruned. The total number of samples generated during each iteration depends on the specifics of the problem at hand and on the details of the enriching/pruning strategy. This simple mechanism produces valid configurations with many steps very efficiently but introduces correlation between samples which are grown from a same smaller configuration. We keep account of these correlations in two ways: first, we count ``effective'' samples by the fraction of independent steps, and second, we run around 10 completely independent runs for each simulation to obtain an a posteriori measure of the statistical error.

FlatPERM extends this method by cleverly choosing the enrichment and pruning steps to generate for each polymer size $n$ a quasi-flat historgram in some choosen micro-canonical quantities $\vect{k}=(k_1,k_2,\dotsc,k_{\ell})$ and producing an estimate $W_{n,\vect{k}}$ of the total weight of the walks of length $n$ at fixed values of $\vect{k}$. From the total weight one can access physical quantities over a broad range of temperatures through a simple weighted average, e.g.
\begin{align}
  \< \mathcal O \>_n(\vect{\rho}) = \frac{\sum_{\mathbf{k}} \mathcal O_{n,\mathbf{k}}\,
   \left(\prod_j \rho_j^{k_j}\right) \, W_{n,\mathbf{k}}}{\sum_\mathbf{k} \left(\prod_j \rho_j^{k_j}\right) \, W_{n,\mathbf{k}}}.
\end{align}
The quantities $k_j$ may be any subset of the physical parameters of the model. For the results presented in this paper, we have run different simulations with different choices for the micro-canonical quantites $\vect{k}$ depending on region of the parameter space of interest. We point out that the number of micro-canonical quantities has a dramatic impact on the memory requirement of algorithm which might need to keep in memory many multi-dimensional histograms like $W_{n,\vect{k}}$ and $\mathcal O_{n,\mathbf{k}}$. The complete list of our simulation runs is provided in Table \ref{appendix:table}.

\begin{table}
\centering
\footnotesize
\begin{tabular}{|l|l|l|l|l|l|l|}
\hline
Name       & Weight          & Parameters   & Iterations       & Max& Samples at& Eff. samples \\
           &                 &              &                  &length & max length & at max length \\
\hline
3P         &                  & $c, m, s$    & $9.8 \cdot 10^5$ & 100 &
      & $1.5 \cdot 10^{10}$ \\
\hline
SF         & $\omega=1$               & $c, s$       & $2.5 \cdot 10^6$ & 256 &
$1.3 \cdot 10^{11}$ & $2.9 \cdot 10^9$ \\
\hline
SF-NN      & $\tau=5$         & $m, s$       & $3.0 \cdot 10^5$ & 256 &
$8.6 \cdot 10^{10}$ & $1.2 \cdot 10^9$ \\
\hline
VI & $\omega=0.5$, $p = p_{BN}$ & $c$        & $1.3 \cdot 10^6$ & 1024 &
$5.7 \cdot 10^9$    & $2.8 \cdot 10^7$ \\
\hline
VI-2       & $\omega=1$, $p=1$                 & $c$          & $1.8 \cdot 10^7$ & 1024 &
$4.6 \cdot 10^{10}$ & $5.2 \cdot 10^8$ \\
\hline
VI-3       & $\omega=1$, $p = p_{BN}$     & $c$          & $5.6 \cdot 10^6$ & 1024 &
$2.5 \cdot 10^{10}$ & $1.3 \cdot 10^8$ \\
\hline
VI-4       & $\omega=0.5$, $p=1$     & $c$          & $4.3 \cdot 10^6$ & 1024 &
$1.5 \cdot 10^{10}$ & $1.2 \cdot 10^8$ \\
\hline
VI L-lattice  & $\omega=1$, $p=0$         & $c$          & $5.3 \cdot 10^3$ & 1000 &
$1.3 \cdot 10^{10}$ & $4.5 \cdot 10^8$ \\
\hline
VI-2 L-lattice & $\omega=0.5$, $p=0$ & $c$     & $2.3 \cdot 10^6$ & 1000 &
$7.6 \cdot 10^9$ & $8.3 \cdot 10^7$ \\
\hline
VI-NN L-lattice  & $p=0$      & $c$, $m$     & $2.0 \cdot 10^6$ & 256 &
$1.1 \cdot 10^{11}$ & $3.0 \cdot 10^9$ \\
\hline
VI-NN       & $p = p_{BN}$    & $c$, $m$     & $2.2 \cdot 10^6$ & 256 &
$1.5 \cdot 10^{11}$ & $2.3 \cdot 10^9$ \\
\hline
VI-NN-2     &$p=1$                 & $c$, $m$     & $6.3 \cdot 10^6$ & 256 &
$3.0 \cdot 10^{11}$ & $7.9 \cdot 10^9$ \\
\hline
NN         & $\tau=1$, $p=7$  & $m$          & $10^3$           & 512 &
$6.0 \cdot 10^6$    & $2.7 \cdot 10^4$ \\
\hline
NN-2       & $\tau=0.5$, $p=1$       & $m$          & $1.6 \cdot 10^6$ & 1024 &
$1.6 \cdot 10^{10}$ & $8.6 \cdot 10^7$ \\
\hline
\end{tabular}
\caption{Summary of flatPERM simulations. The simulations range from uniform sampling across all four parameters $n,c,m,s$ with the length of configurations restricted by a maximal length of $n=100$ (data set 3P) to uniform sampling across only two parameters, with Boltzmann weights for the other parameters at fixed values. For example in the data set NN-2, two Boltzmann weights are fixed at $\tau=0.5$ and $p=1$, and uniform sampling is performed across $n$ and $m$. This enables simulations for longer lengths, allowing to sample up to $n=1024$. 
\label{appendix:table}
}
\end{table}

From a simulation with micro-canonical quantities, it is possible to obtain a mixed distribution that is canonical in one of the quantities by fixing its associated weight. For example: from $W_{n,k_1,k_2,k_3}$ we can obtain ``a slice'' at fixed value of $\rho_3$ by computing
\[
  {\tilde W}_{n,k_1,k_2} = \sum_{k_3}\ \rho_3^{k_3}\, W_{n,k_1,k_2,k_3} \,.
\]
The normalisation of the distribution is irrelevant. 

To obtain a landscape of possible phase transitions, we compute the largest eigenvalue of the matrix of second derivatives of the free energy with respect to $\vect{\rho}$ effectively measuring the strength of the fluctuations and covariance in $\vect{k}$. E.g. given the flatPERM weights $W_{n,\vect{k}}$, we first compute the partition function:
\begin{equation}
  Z_n(\vect{\rho}) = \sum_\mathbf{k} \left(\prod_j \rho_j^{k_j}\right) \, W_{n,\mathbf{k}}
  ,
\end{equation}
and then we find the maximum eigenvalue of the Hessian matrix
\begin{equation}
  H_{ij} = \frac{\partial^2 \log f_n(\vect{\rho})}{\partial \rho_i\, \partial \rho_j}
  ,
\end{equation}
where $f_n(\vect{\rho}) = \log\, Z_n(\vect{\rho})$ is the free energy.

More explicitedly, in the diagrams where $p$ is fixed the remaining parameters are $\vect{\rho} = (\tau, \omega)$ and we plot the eigenvalue of the matrix defined by
\begin{equation}
  H = \begin{pmatrix}
   \frac{\partial^2 f}{\partial \tau^2} & \frac{\partial^2 f}{\partial \tau\, \partial \omega} \\[0.3em]
   \frac{\partial^2 f}{\partial \tau\, \partial \omega} & \frac{\partial^2 f}{\partial \omega^2} 
 \end{pmatrix}.
\end{equation}
In the diagrams where $\omega$ is fixed the remaining parameters are $\vect{\rho} = (\tau, p)$ and we plot the eigenvalue of the matrix defined by
\begin{equation}
  H = \begin{pmatrix}
   \frac{\partial^2 f}{\partial \tau^2} & \frac{\partial^2 f}{\partial \tau\, \partial p} \\[0.3em]
   \frac{\partial^2 f}{\partial \tau\, \partial p} & \frac{\partial^2 f}{\partial p^2} 
 \end{pmatrix}.
\end{equation}
Finally, when $\tau$ is fixed, the remaining parameters are $\vect{\rho} = (p, \omega)$ and we plot the eigenvalue of the matrix defined by
\begin{equation}
  H = \begin{pmatrix}
   \frac{\partial^2 f}{\partial p^2} & \frac{\partial^2 f}{\partial p\, \partial \omega} \\[0.3em]
   \frac{\partial^2 f}{\partial p\, \partial \omega} & \frac{\partial^2 f}{\partial \omega^2} 
 \end{pmatrix}.
\end{equation}

We caution that the landscape of possible phase transitions obtained in such a way is a pseudo-phase diagram for finite-size systems, so some care needs to be taken when extrapolating to the thermodynamic limit, in particular to distinguish crossover regions from actual phase transitions. Also, phase boundaries are expected to shift when increasing the system size. To support our conclusions,
we therefore investigate supplemental information, such as the finite-size scaling of the specific heat in boundary regions. Where boundary regions meet, we expect to find higher-order critical points. The precise location of these points, let alone their nature, is difficult to determine from our simulations.

\clearpage

\section{Semi-flexible VISAW: ($\omega=1$)}
\label{sec:sfvisaw}

We started our new investigation by examining the semi-flexible VISAW model for all values of stiffness $p$. This would allow us to see how the BN point belongs to a critical line. To do so, we simulated a VISAW model with additional stiffness. 

\begin{figure}[ht!]
\centering
\includegraphics[width=0.7\textwidth]{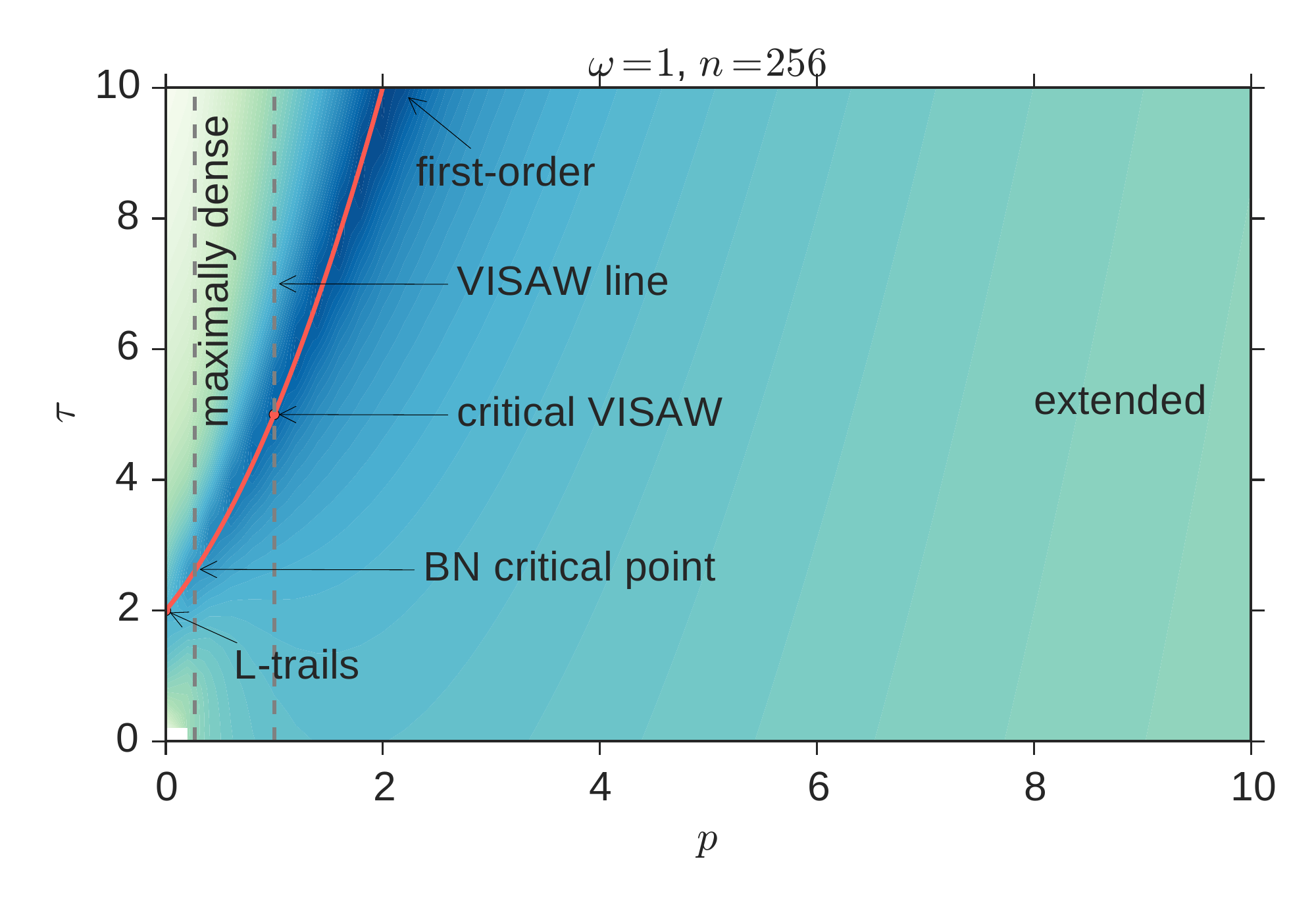}
\caption{A density plot of the largest eigenvalue of fluctuations of the semi-flexible VISAW model ($\omega=1$) and a schematic phase diagram inferred from this and the order parameters/indicator variables of  polymer density  and singly visited sites. Darker shades indicate larger values of the fluctuations. There seems to be only two phases: extended and maximally dense. The solid line (red) represents the estimated location of second order phase transitions. Data is obtained from the dataset SF.}
\label{fig:SFVISAW_phase_diagram}
\end{figure}

A density plot of the largest eigenvalue of fluctuations of a semi-flexible VISAW model is shown in Figure~\ref{fig:SFVISAW_phase_diagram}. We found that the BN point seems to lie on a line of second-order phase transitions that extends from the L-lattice trails point to high value of $\tau$ and $p$, connecting the BN point and the VISAW critical point. For some high value of $\tau$ the transition seems to turn first order. The L-lattice trails point is known to be in the $\theta$-point universality class, while the critical VISAW has a strongly divergent specific-heat, as it was reported in \cite{bedini2013}. Therefore along the line the transition starts weak and becomes stronger and stronger until it turns first order.

\begin{figure}[ht!]
\centering
\includegraphics[width=0.7\textwidth]{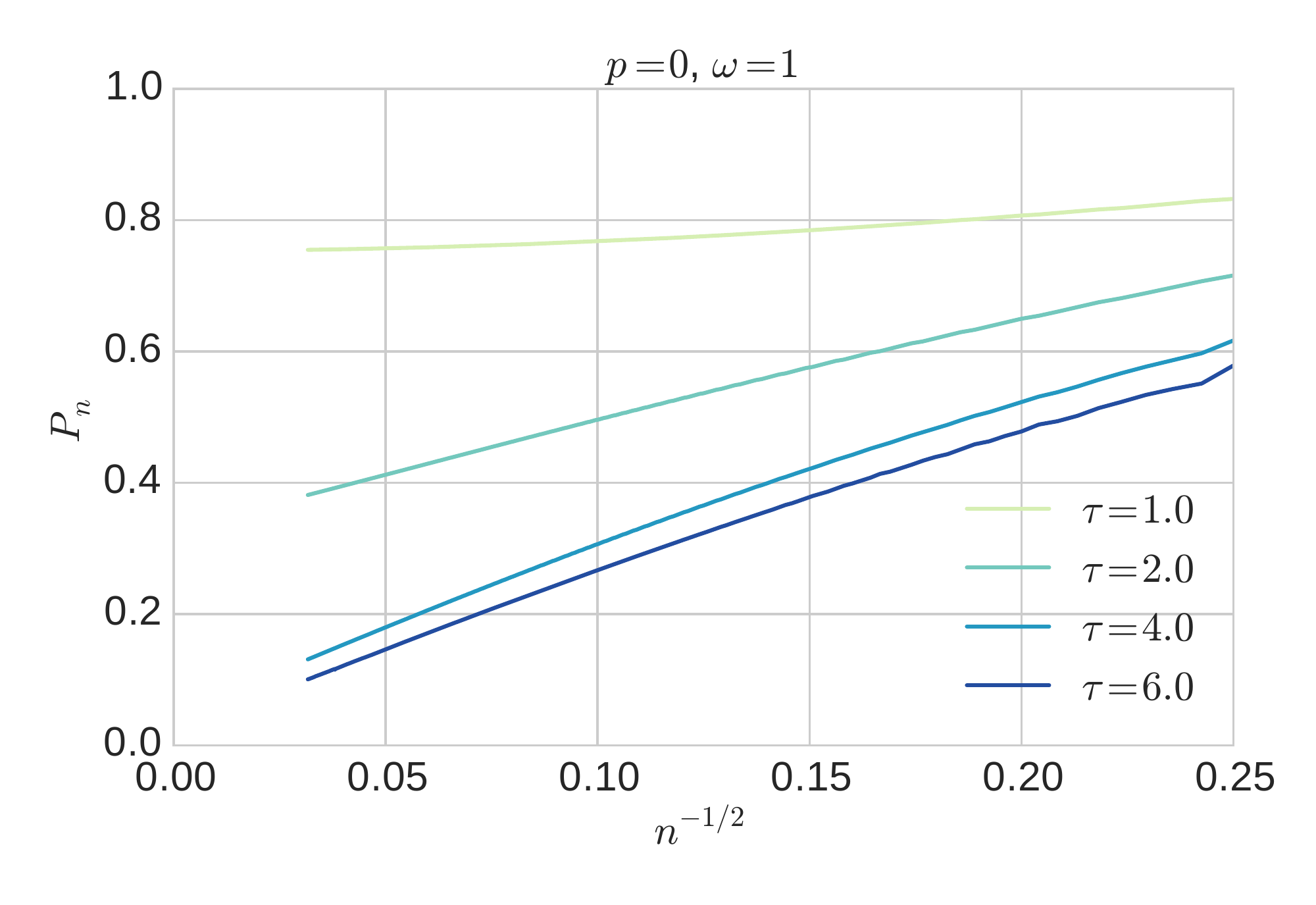}
\caption{A plot of the number of singly visited sites per unit length versus $n^{-1/2}$ for $p=0$ and $\omega=1$ and for several values of $\tau$. 
For $\tau=4$ and $\tau=6$, this quantity goes to zero as $n$ diverges. This indicates the existence of the maximally dense phase where all but surface sites are involved in collisions and so are doubly visited by the polymer. Data is obtained from the dataset VI L-lattice.}
\label{fig:VISAW_low_temperature_omega_eq_1_p_eq_0}
\end{figure}

As done in \cite{bedini2013} we considered the fraction of sites visited only once as an order parameter. In Figure~\ref{fig:VISAW_low_temperature_omega_eq_1_p_eq_0} we show that number of singly visited sites per unit length in the maximally dense phase goes to zero as length diverges. This indicates the existence of the maximally dense phase where all but surface sites are involved in collisions and so are doubly visited by the polymer.

\begin{figure}[ht!]
\centering
\includegraphics[width=0.7\textwidth]{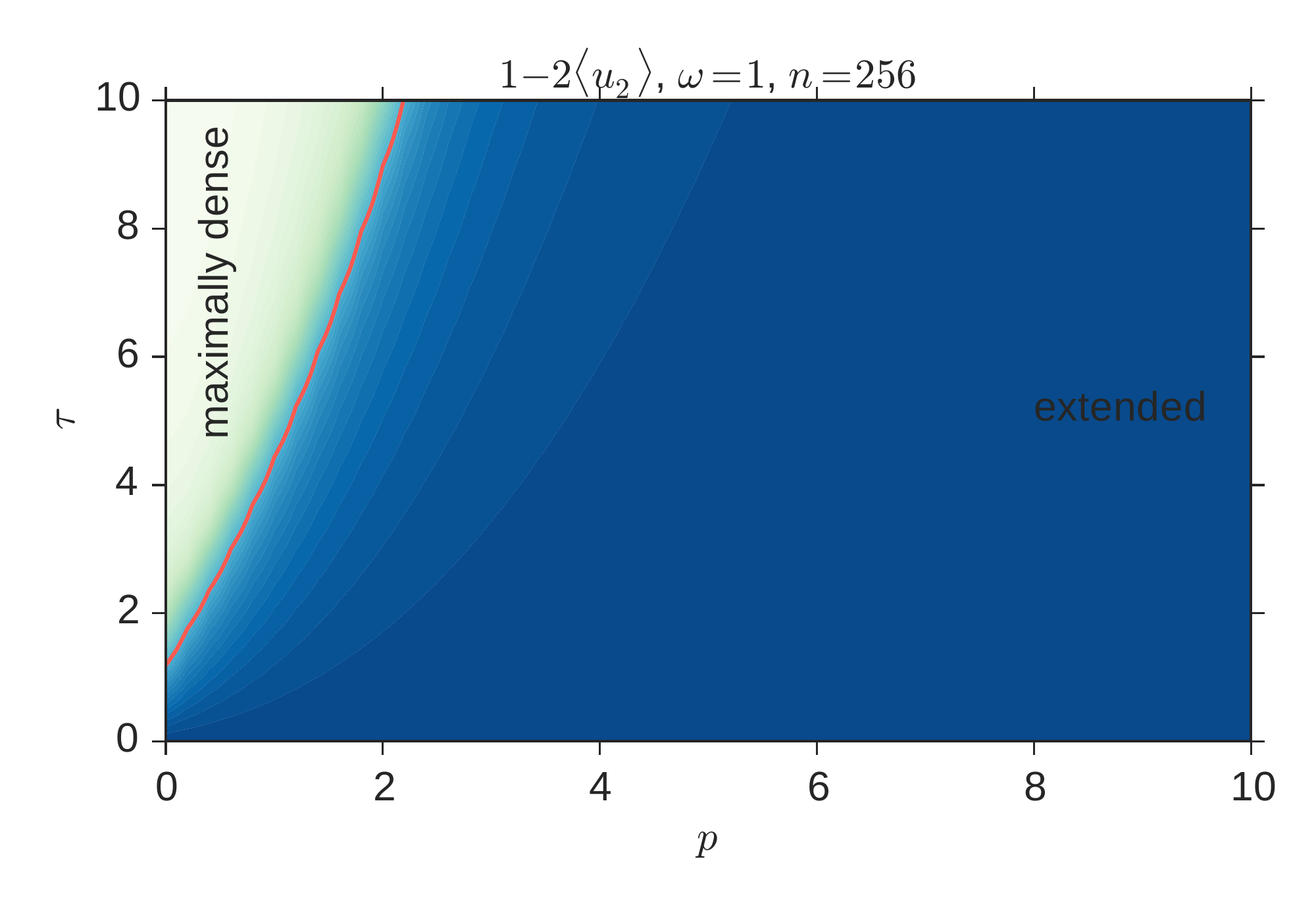}
\caption{A density plot of the number of singly visited sites per unit length indicating the region that is maximally dense for $\omega=1$. The solid line (red) shows the approximate location of second order phase transitions. Data is obtained from the dataset SF.}
\label{fig:VISAW_density}
\end{figure}

In Figure \ref{fig:VISAW_density} we show a density plot of the fraction of singly visited sites indicating that the maximally dense region extends all the way to the line of second order phase transitions.

\begin{figure}
\centering
\includegraphics[width=\textwidth]{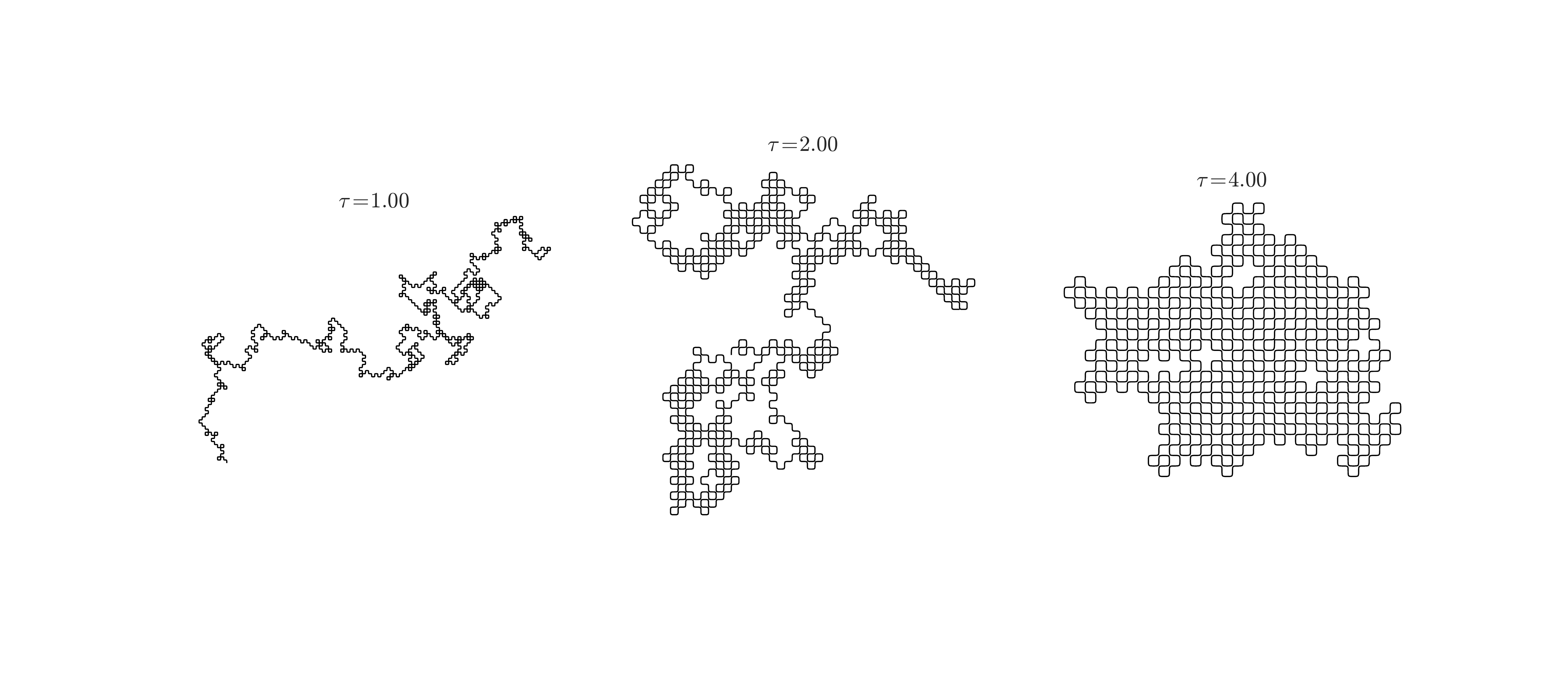}
\caption{Some typical configurations for $\omega = 1$ and $p=0$ at $\tau=1$, $\tau=2$ (at the location of the transition) and $\tau=4$. We see that the low temperature ($\tau=4$) configuration appears to be almost maximally dense. Data is obtained from the dataset VI L-lattice.}
\label{fig:VISAW_configs}
\end{figure}

Inspecting typical configurations at various parameter values further confirms this scenario. In Figure \ref{fig:VISAW_configs} we show three configurations for $\omega=1$ and $p=0$ at $\tau=1$ (swollen), $\tau=2$ (critical) and $\tau=4$. (Note that the choice of $p=0$ forces a right-angle turn after every single step, so these are configurations on the L-lattice.) One clearly sees the typical characteristics of a swollen chain for $\tau=1$, and observes a clear difference between this configuration and the configuration at $\tau=2$, which is critical. In contrast, the configuration for $\tau=4$ is close to maximally dense with only a few small internal defects.

\section{The full model}
\label{sec:full-study}

Our model has three Boltzmann weights and so three parameters that can be varied. The models of previous focus have been two parameter slices of the space, namely the semi-flexible ISAW that is defined by $\tau=1$ and the semi-flexible VISAW that is defined by $\omega=1$. Because simulations of longer lengths can be made when we fix certain Boltzmann weights we have analysed our model in slices. To analyse the phase diagram in a slice we have used three main pieces of information. Firstly we have used the appropriate Hessian and considered the landscape of the largest eigenvalue of that matrix: this gives us an indication of divergences in appropriate fluctuations, equivalently specific heats. We have looked at scaling of those possible divergences. When we suspect a first order transition inferred from linear or super linear scaling of the divergence we consider the distribution of the appropriate micro-canonical quantity so look for emerging bimodality. We have also considered the order parameters associated with the known phases in the semi-flexible ISAW and VISAW models to indicate which phases occur in which regions of the parameter space. Finally we have looked at ``typical" (random) configurations to re-enforce our conjectures.

In the following sections we consider two-dimensional slices of the parameter space, first by fixing $\tau$, then by fixing $\omega$, and finally by fixing the stiffness $p$. 

The first key observation to make is that the order parameters clearly indicate that the three low temperature phases from the semi-flexible ISAW and VISAW models occur in different parts of the three dimensional phase space and so that there are at least three low temperature phases.

\subsection{$\tau$ slices}
Before looking at the range of $\tau$ slices we consider $\tau=5$, first to see how the phase diagram changes from $\tau=1$ when we fix at a value of $\tau$ that should induce the maximally dense phase in some part of the phase diagram but also to demonstrate the method we are using to conjecture phase diagrams.

\subsubsection{$\tau=5$}

\begin{figure}[ht!]
\centering
\includegraphics[width=0.7\textwidth]{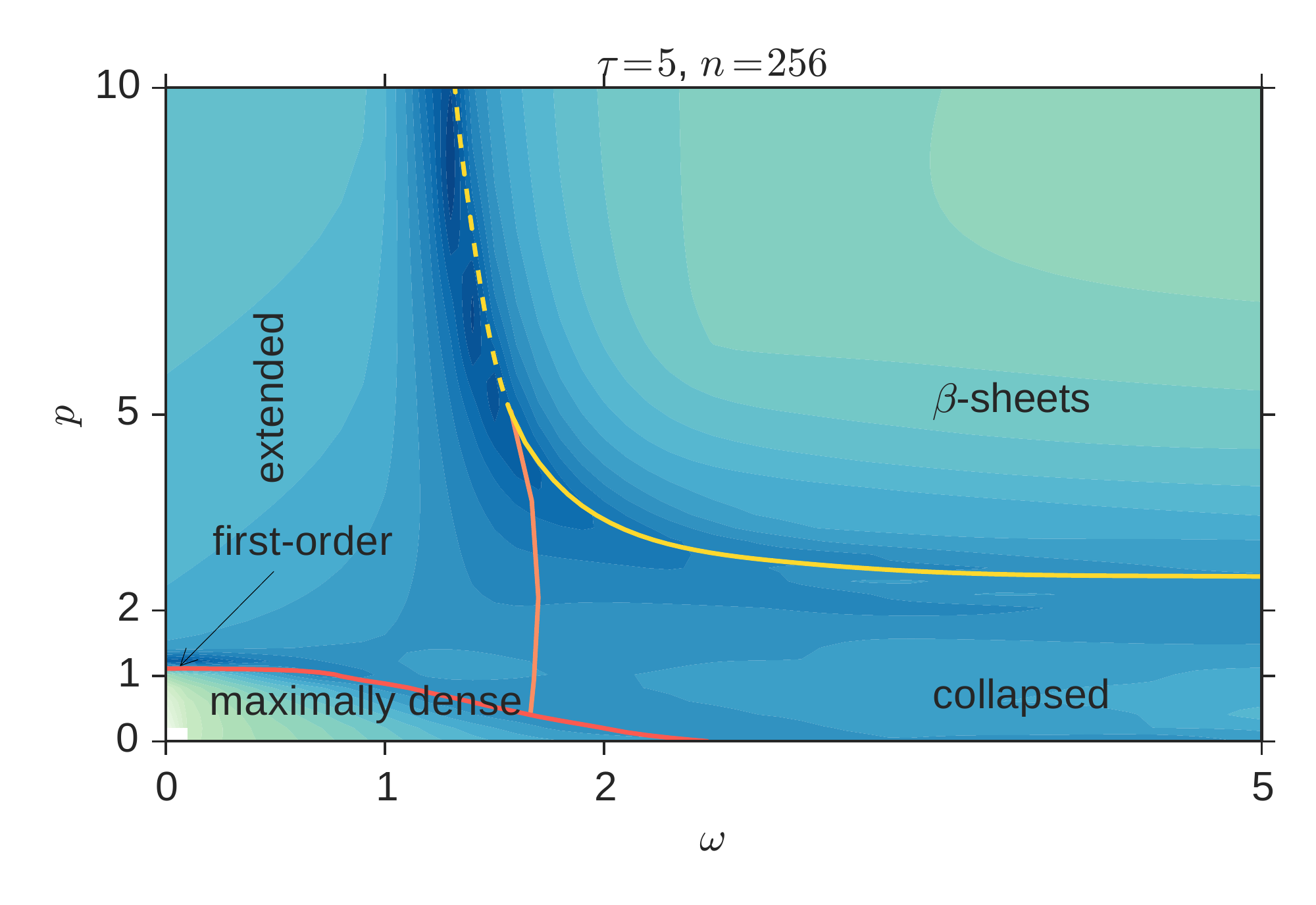}
\caption{Plot of the maximum eigenvalue of the matrix of fluctuations for $\tau=5$. Peaks in this plot can be used to infer the phase diagram. Darker shades indicate larger values of the fluctuations. Solid lines show the approximate location of second order phase transitions and dashed lines indicate first order behaviour. As indicated by the arrow, the phase transition turns first-order as it approaches the vertical axis. The data for this plot is obtained from the dataset SF-NN.}
\label{fig:VISAW_phase_diagram_tau_eq_5}
\end{figure}

\begin{figure}[ht!]
\centering
\includegraphics[width=0.55\textwidth]{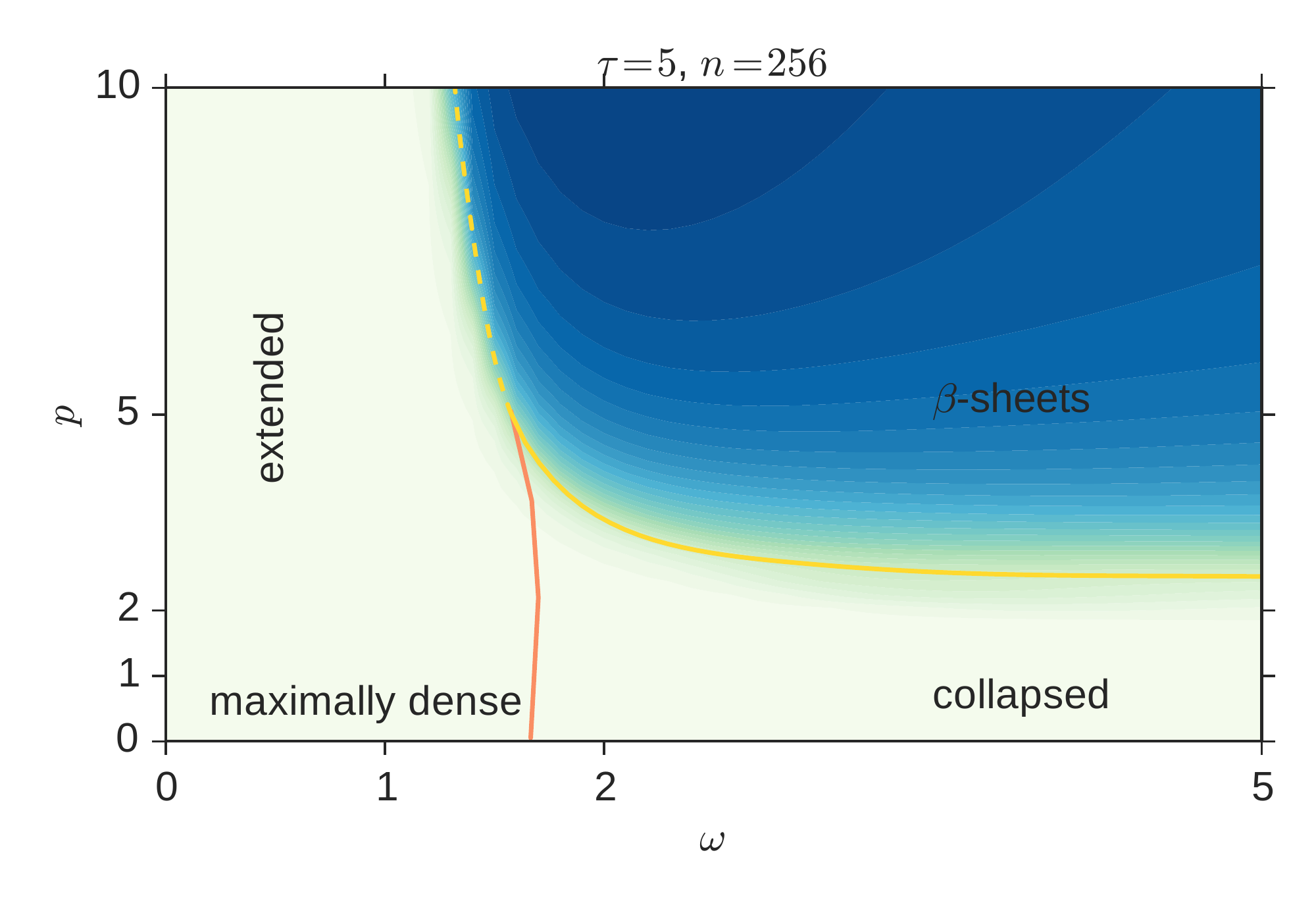}
\includegraphics[width=0.55\textwidth]{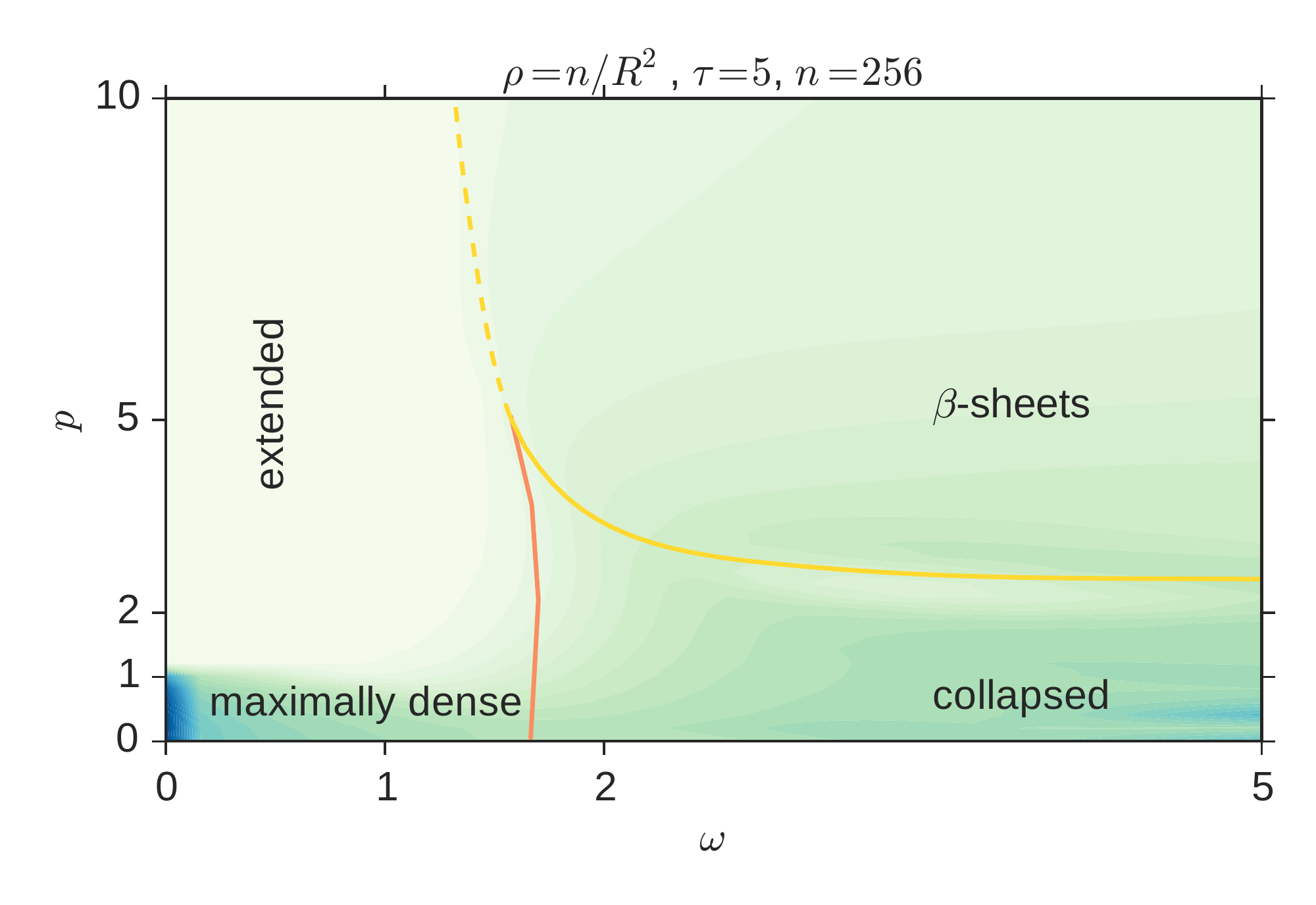}
\includegraphics[width=0.55\textwidth]{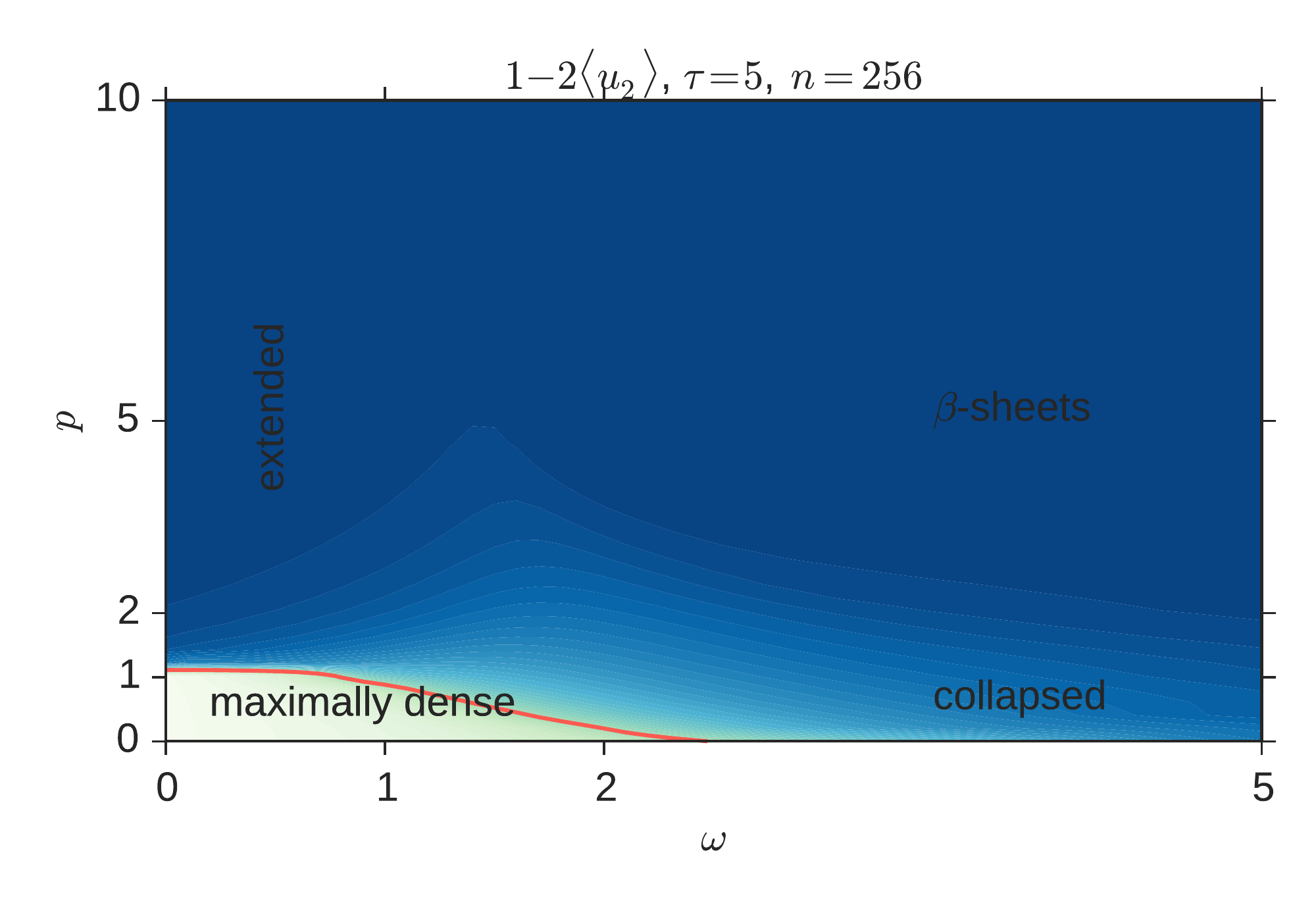}
\caption{For $\tau=5$ we plot three order parameter/indicator variables that can be used to re-enforce the conclusions regarding the phase diagram made from the peaks in the fluctuations. In the top diagram the anisotropy parameter is plotted: it is clearly large for large $p$ and $\omega$ --- this is a good indicator for the $\beta$-sheet phase. In the middle diagram the density is plotted: it is clearly small (tending to zero) in the extended phase for small $\omega$ excluding very small $p$. In the bottom diagram the fraction of singly visited sites is plotted: in the maximally dense phase is elected to go to zero. These plots are all obtained from the same dataset SF-NN.}
\label{fig:order_parameters_tau_eq_5}
\end{figure}

In Figure~\ref{fig:VISAW_phase_diagram_tau_eq_5} a plot of the maximum eigenvalue of the matrix of fluctuations is given for the slice $\tau=5$. Darker shades indicate larger values of the fluctuations. There is clearly a strong peak in these fluctuations for large $p$ separating regions of small and large $\omega$. We then identify this transition to be between an extended phase and a $\beta$-sheet phase given schematically by the dashed phase boundary drawn for $p >5$: the two phases are identified using the order parameter plots seen in Figure~\ref{fig:order_parameters_tau_eq_5}. 
We have then verified that the divergence of the fluctuations is at least linear in length and that at least at some point along that line the distribution of nearest neighbour contacts $m$ is bimodel. The other lines are all seen to be second order transitions. We find that a maximally dense phase does exist for very small $p$ and moderately small $\omega$ in the bottom left corner of the diagram. 
Not all the conjectured phase boundaries are equally well delineated, with the extended - collapsed boundary the least defined: this is not to be unexpected as the transition should be in the $\theta$-universality class which has a convergent specific heat singularity.  In general this phase boundary is the most difficult to estimate. 
\begin{figure}[ht!]
\centering
\includegraphics[width=\textwidth]{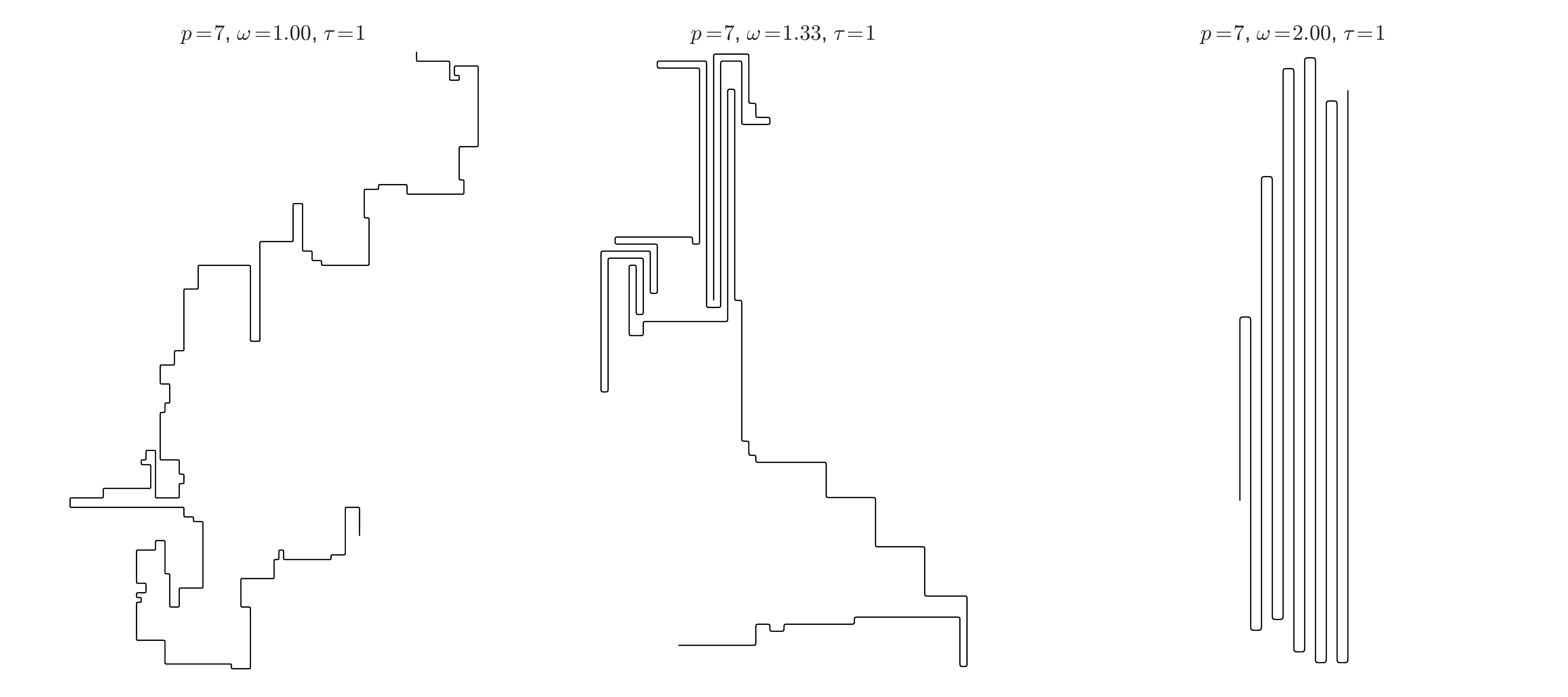}
\caption{Here are some typical configurations in the plane $\tau = 1$ at $p=7$. Of particular interest is the $\beta$-sheet type configuration. These configurations come from the dataset NN.}
\label{fig:tau-eq-5}
\end{figure}
Finally to re-enforce our identification of phases we have considered typical configurations. In Figure~\ref{fig:tau-eq-5} we show a typical configuration in the part of the phase diagram we label as $\beta$-sheets (actually for $p=7$ in this instance): it clearly shows the anisotropic structure we expect.

\begin{figure}[ht!]
\centering
\includegraphics[width=0.45\textwidth]{VISAW_phase_diagram_tau_eq_0}
\includegraphics[width=0.45\textwidth]{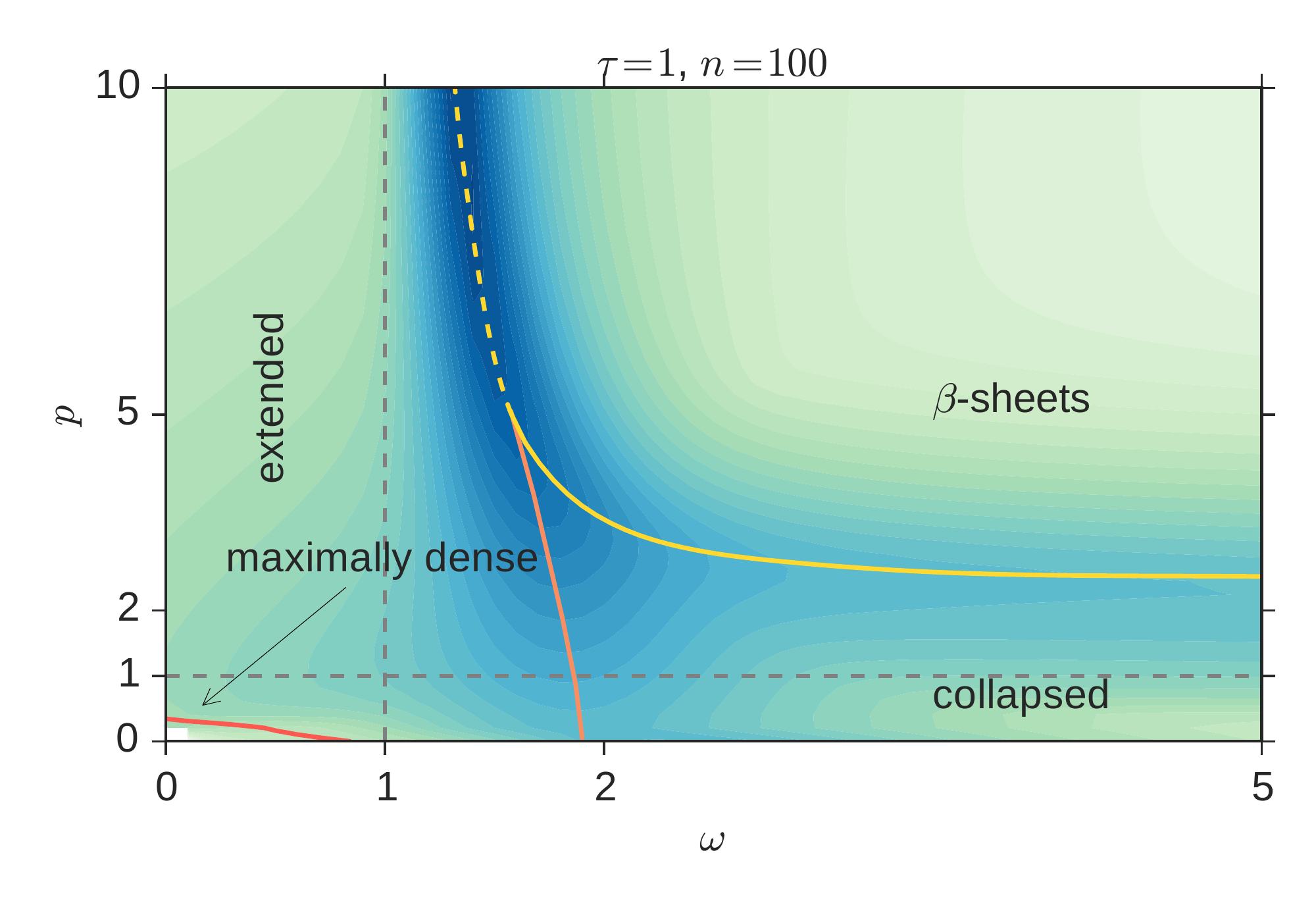}

\includegraphics[width=0.45\textwidth]{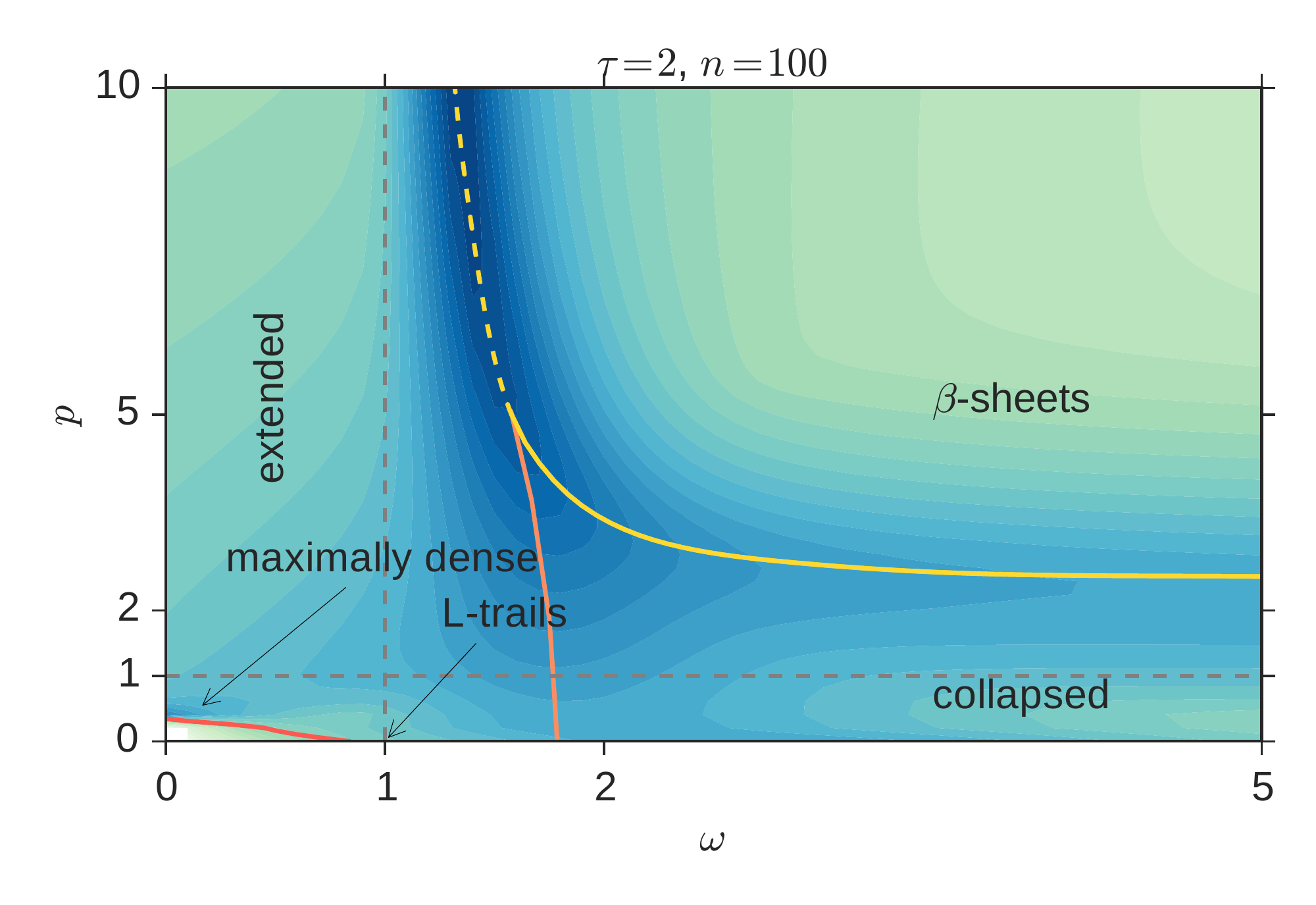}
\includegraphics[width=0.45\textwidth]{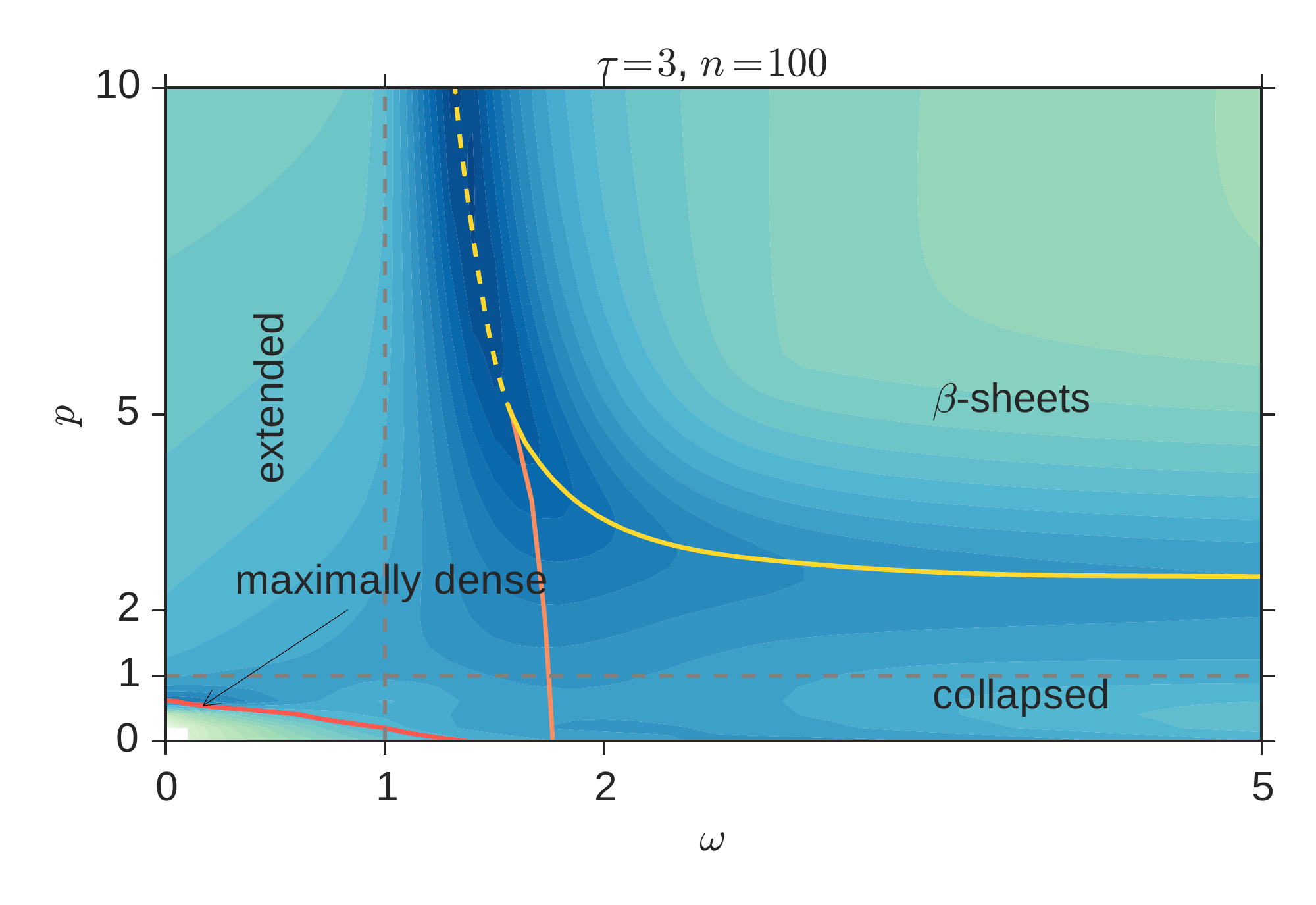}

\includegraphics[width=0.45\textwidth]{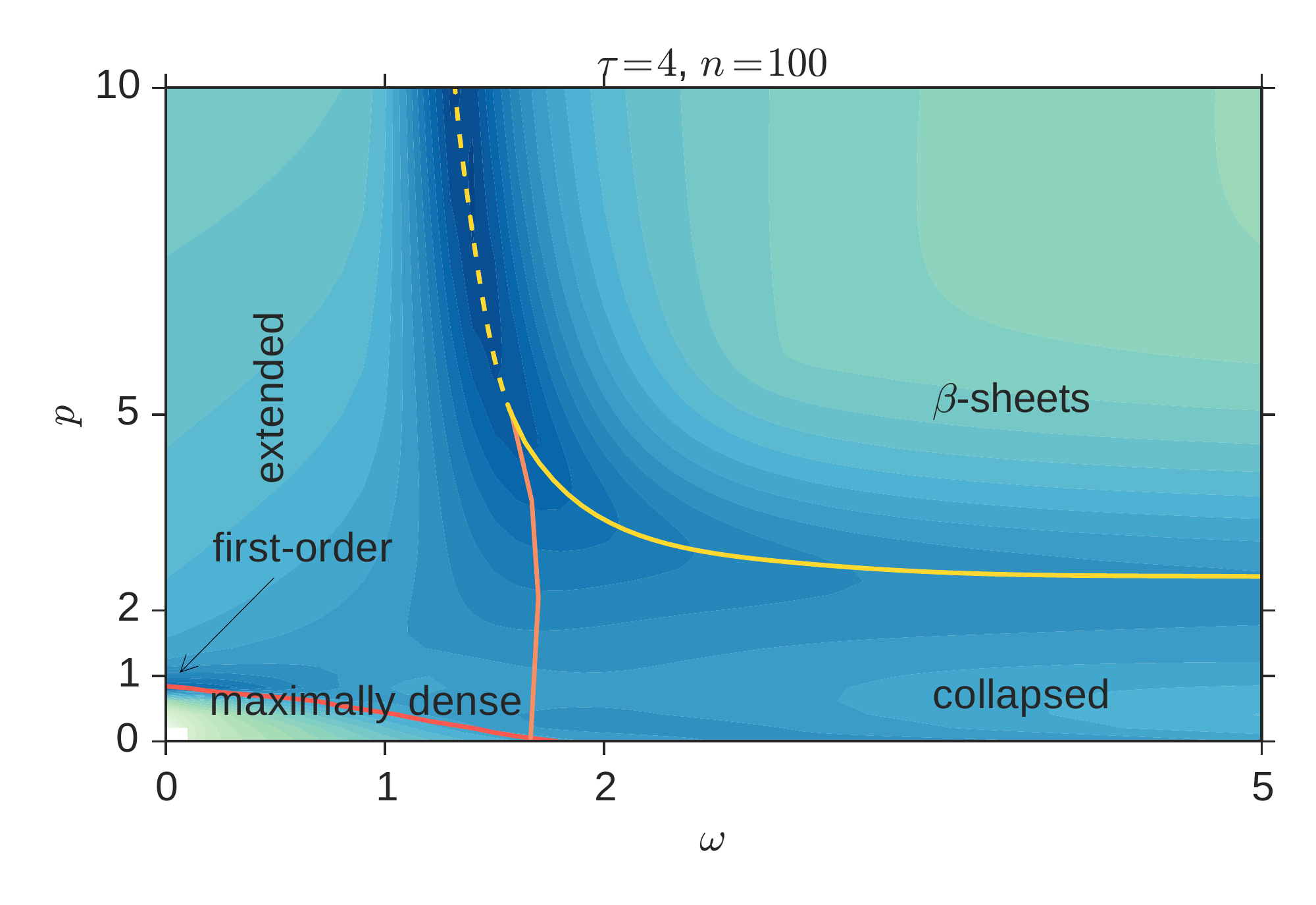}
\includegraphics[width=0.45\textwidth]{VISAW_phase_diagram_tau_eq_5}

\includegraphics[width=0.45\textwidth]{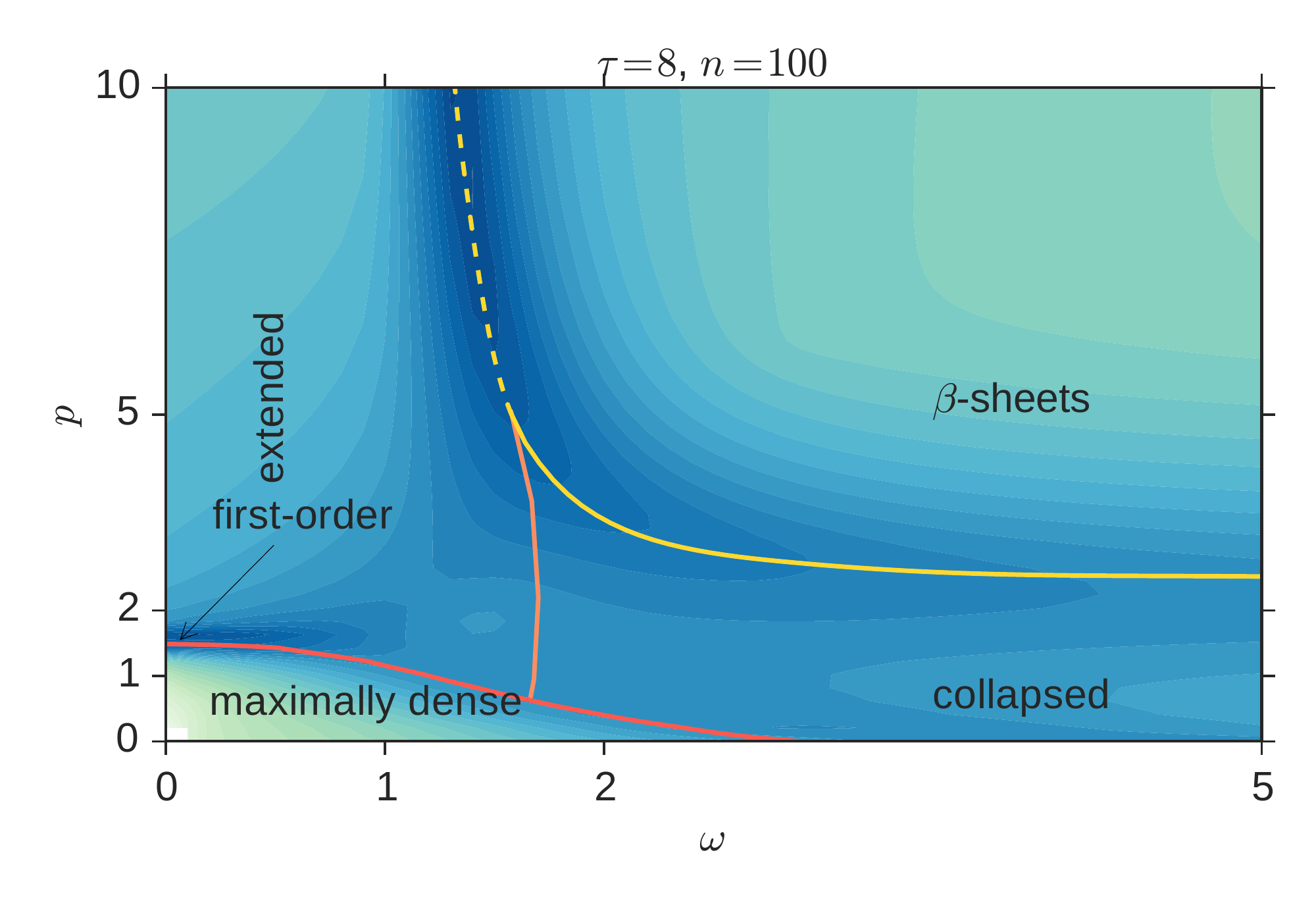}

\caption{For values of $\tau$ ranging from $\tau=0$ (top left) to $\tau=8$ (bottom) a density plot of the larger eigenvalue is displayed. Information from this has been combined with the three order parameter indicator variables mentioned above to infer a phase diagram schematically drawn in each diagram. We could not determine the phase boundary between the extended and collapsed phases for $\tau >2 $ with much accuracy: it could move to smaller values of $\omega$ especially for small $p$. As in Figure~\ref{fig:VISAW_phase_diagram_tau_eq_5} the phase transition turns first-order as it approaches the vertical axis. These plots come from an appropriate slice of the dataset 3P, except the phase diagram at $\tau = 5$ which comes from dataset SF-NN.}
\label{fig:blah}
\end{figure}

One key observation to make is the maximally dense phase and the $\beta$-sheet phase never meet and that there are only ever three phases meeting at a point in the phase diagram.
\subsubsection{All $\tau$}

Starting with the semi-flexible ISAW model again at $\tau=0$ we now compare how the phase diagram changes with changing $\tau$.  We specifically consider $\tau=0, 1, 2, 3,4, 5$ and $8$. The associated density plots of the largest eigenvalue are shown in Figure \ref{fig:blah}. The extended, $\beta$-sheet and collapse (globular) phases exist at each value of $\tau$ and exist for roughy in the same regions of parameter space: the $\beta$-sheet phase always exists when $p$ and $\omega$ are both large enough. For large enough $\tau$ a region of the maximally dense phase appears for small $p$ and small $\omega$. This region increases in size with increasing $\tau$.

\clearpage

\subsection{$\omega$ slices}
\begin{figure}[ht!]
\centering
\includegraphics[width=0.45\textwidth]{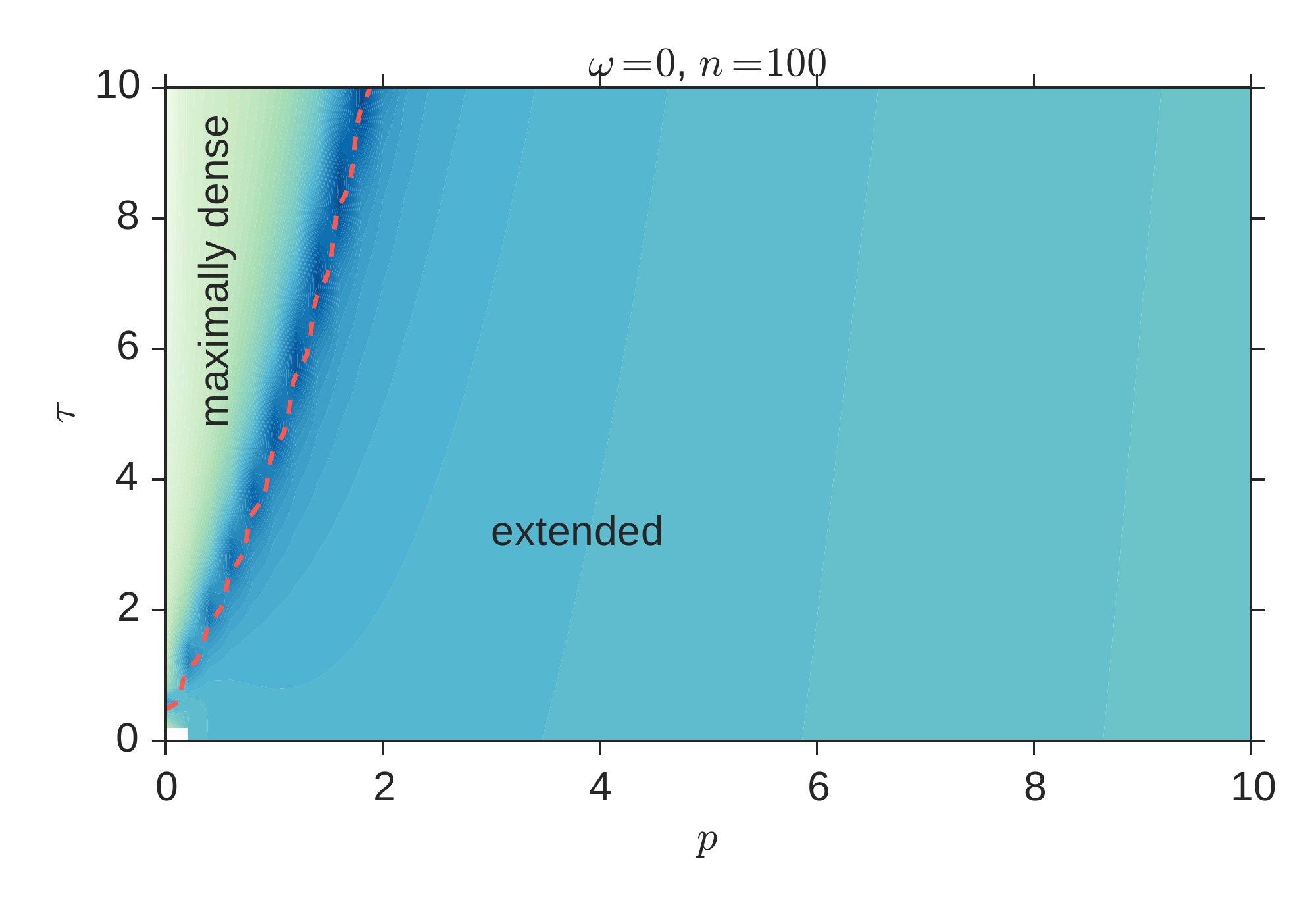}
\includegraphics[width=0.45\textwidth]{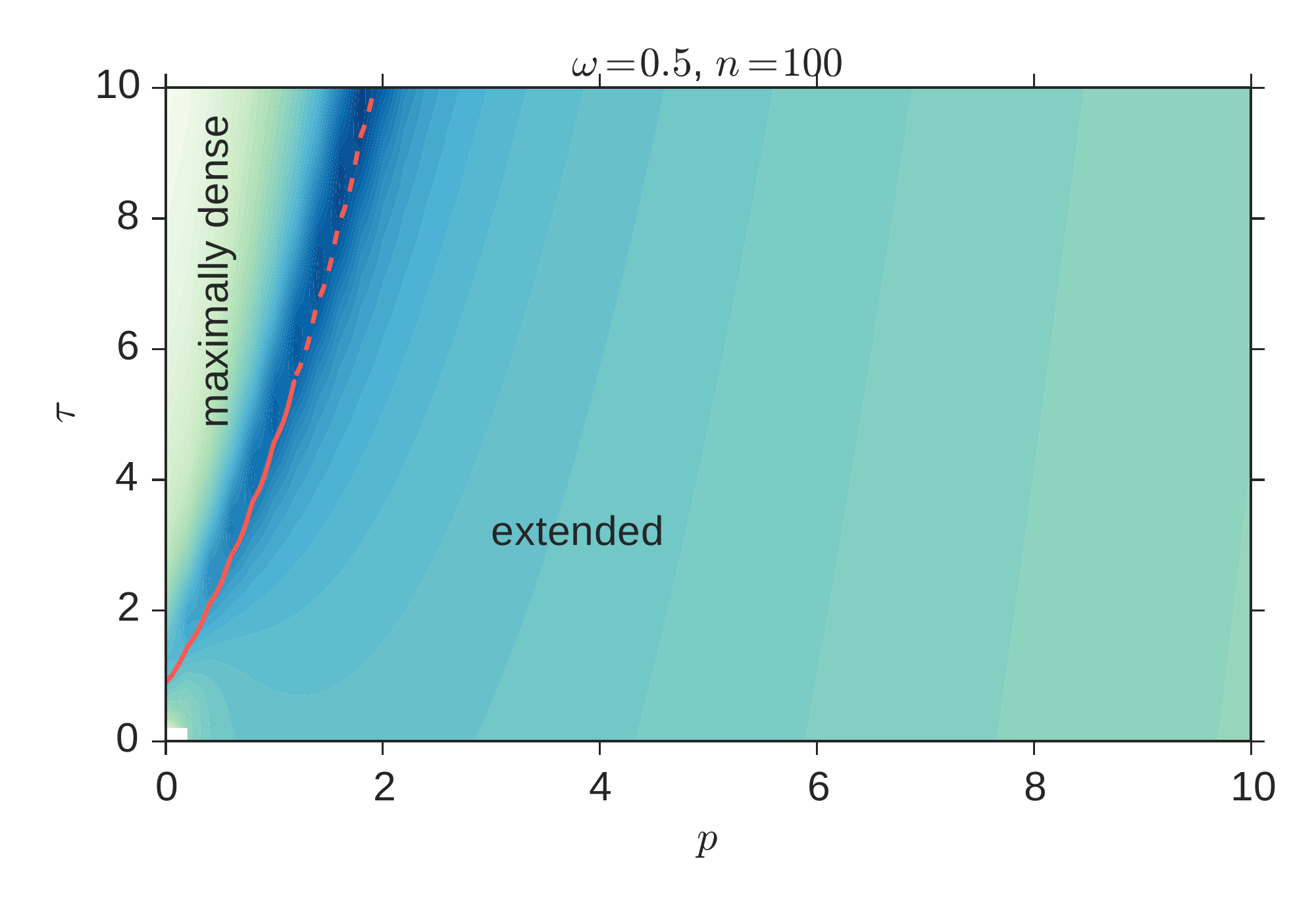}
\includegraphics[width=0.45\textwidth]{VISAW_phase_diagram_omega_eq_1.pdf}
\includegraphics[width=0.45\textwidth]{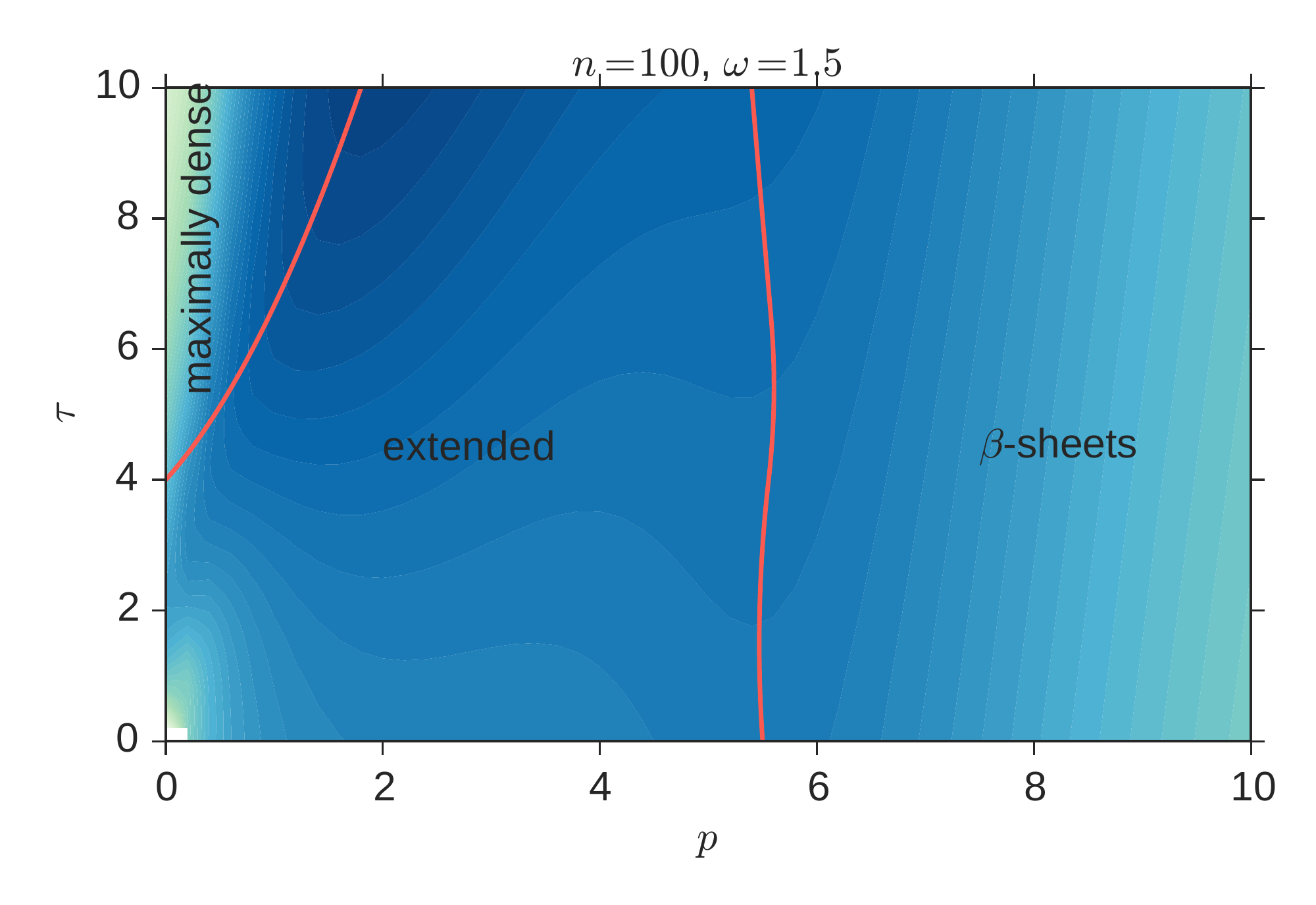}
\includegraphics[width=0.45\textwidth]{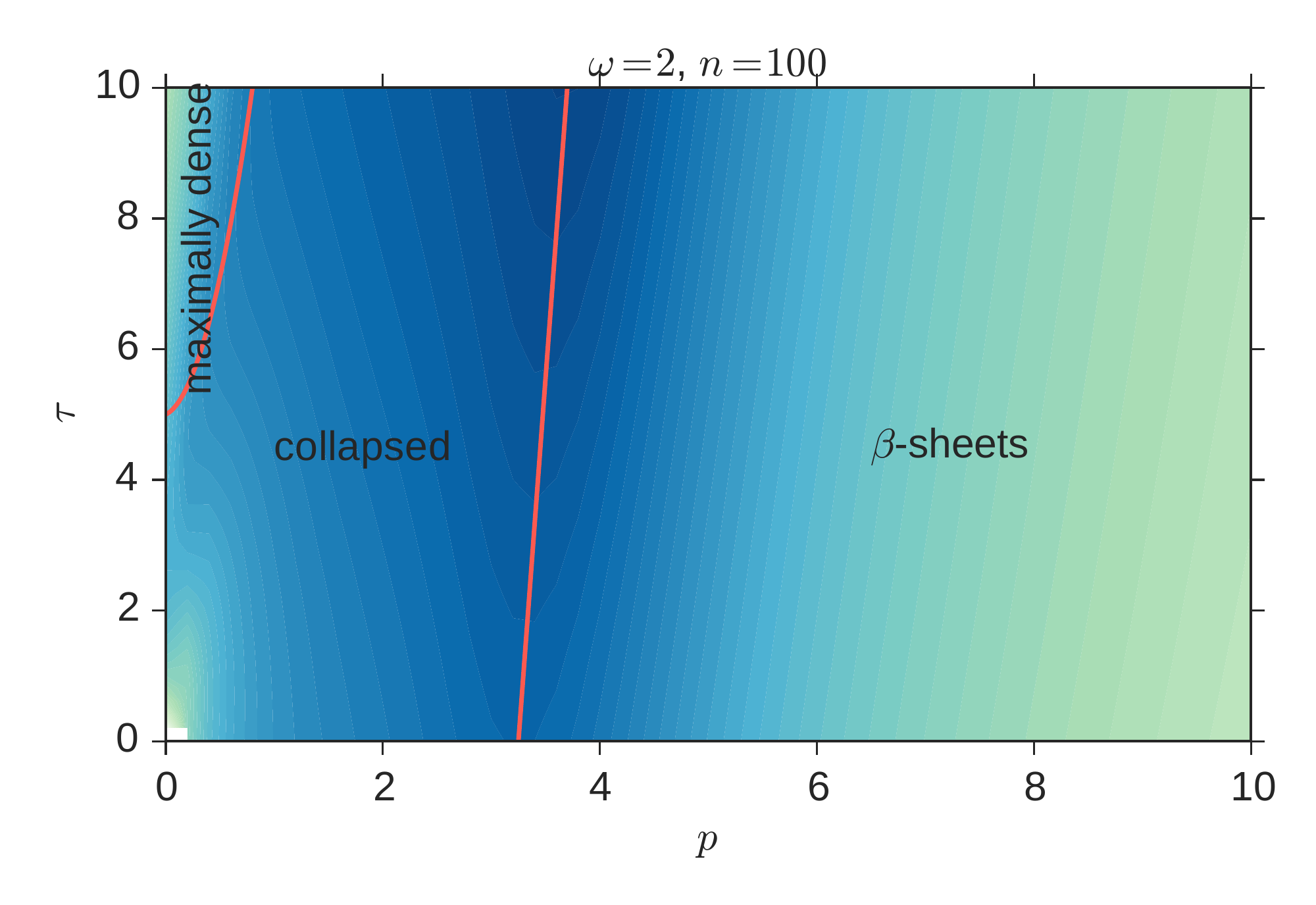}
\includegraphics[width=0.45\textwidth]{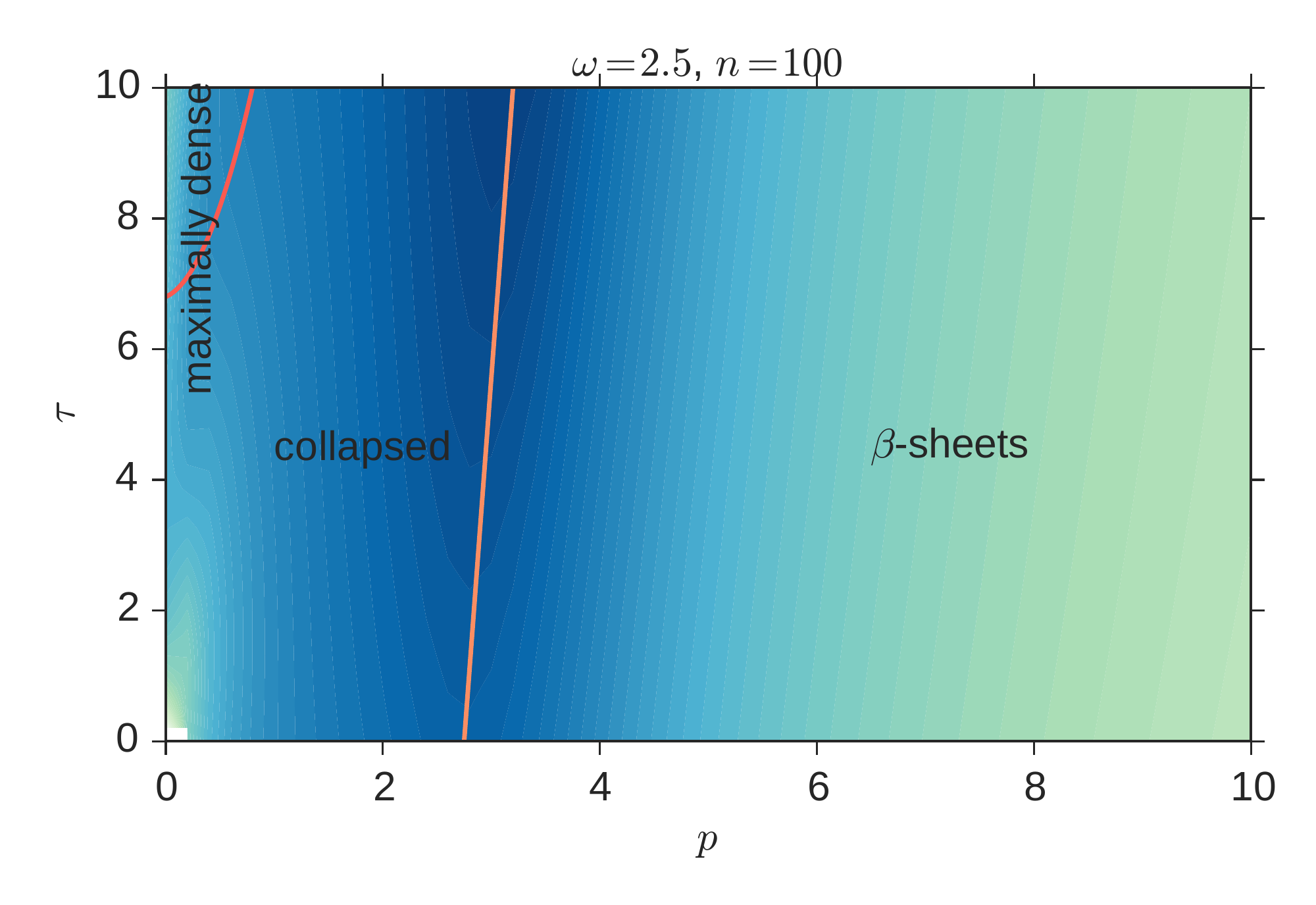}
\caption{First- and second-order transitions are shown respectively as dashed and solid lines. At $\omega=0$ the extended to  maximally dense transition is always first order except possibly at $p=0$. For other values of $\omega$ there seems to be a special value of $\omega=\omega_b$ that signals a  change from second to first order in that transition: second order for small $p$ and first order for large $p$. The extended to $\beta$ sheet transition is expected to be first order while the collapse to $\beta$ sheet transition is expected to be second order in line with previous work. The extended to collapsed phase is the canonical $\theta$-transition that is weakly second order. These plots come from an appropriate slice of the dataset 3P, except the phase diagram at $\omega = 1$ which comes from dataset SF.}
\label{fig:omega_slices}
\end{figure}

We now compare how the phase diagram changes from the one shown above for the semi-flexible VISAW model ($\omega=1$) by considering different slices of fixed $\omega$: see Figure~\ref{fig:omega_slices}.

Recalling that for $\omega=1$ we conjecture just two phases being extended and maximally dense we can now see how the other phases emerge. For $\omega<1$ there are only these two phases. It is worth noting that for $\omega=0$ the extended to maximally dense transition appears to be first order for all $p$ but as $\omega$ is increased a region of second order transition appears for small $p$. For $\omega >1$ we can first see the emergence $\omega=1.5$ of the $\beta$-sheet phase for large $p$ from the extended phase. This is in accord with the semi-flexible ISAW diagram and other $\tau$ slices discussed above. Again in accord with these $\tau$-slices we see that for $\omega=2$ and $\omega=2.5$ the extended phase has been completely replaced with the collapsed and $\beta$-sheet phases. Also we note that as $\omega$ is increased the maximally dense phase retreats to large values of $\tau$ and smaller values of $p$. This highlights the competition  of the doubly-visited site and nearest-neighbour interactions: they favour different and competing low temperature phases. We also note that for $\omega\geq 1.5$ the extended to maximally dense phase transition is completely second order for the range of $p$ and $\tau$ considered in these diagrams. 
 
\subsection{$p$ slices}
Finally we consider fixing the stiffness $p$. There are important values of the stiffness, namely $p=0, p_{BN}, 1$ that are worth considering first.

\subsubsection{$p=0$}
\begin{figure}[ht!]
\centering
\includegraphics[width=0.7\textwidth]{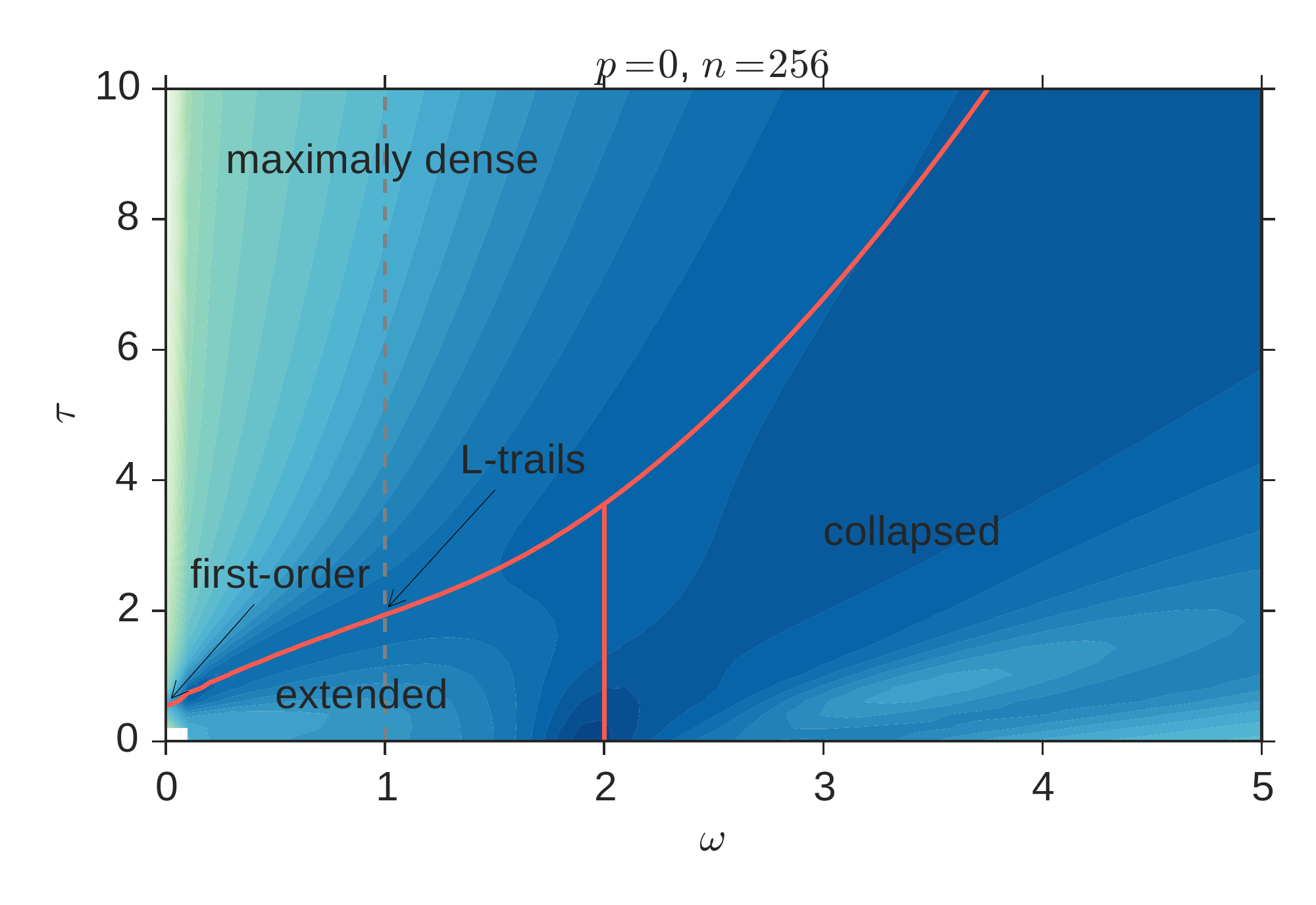}
\caption{The conjectured phase diagram for $p=0$ drawn on top of the density plot of the maximum eigenvalue of the fluctuations. For $\omega=1$ one obtains the LSAT model. There are three phases of maximally dense, extended and collapsed. The phase transition between the extended and maximally dense becomes first order as it approaches the vertical axis for small $\omega$. Otherwise the transitions are second order. The data for this plot comes from dataset VI-NN L-lattice.}
\label{fig:phase-diagram-p-eq-0}
\end{figure}

When $p=0$ all the configurations bend at each step and so effectively exist on the L-lattice. In particular, this slice contains the LSAT model for $\omega=1$. See Figure~\ref{fig:phase-diagram-p-eq-0} for the density plot of the largest eigenvalue of the fluctuations and a conjecture phase diagram. 

\begin{figure}[ht!]
\centering
\includegraphics[width=0.48\textwidth]{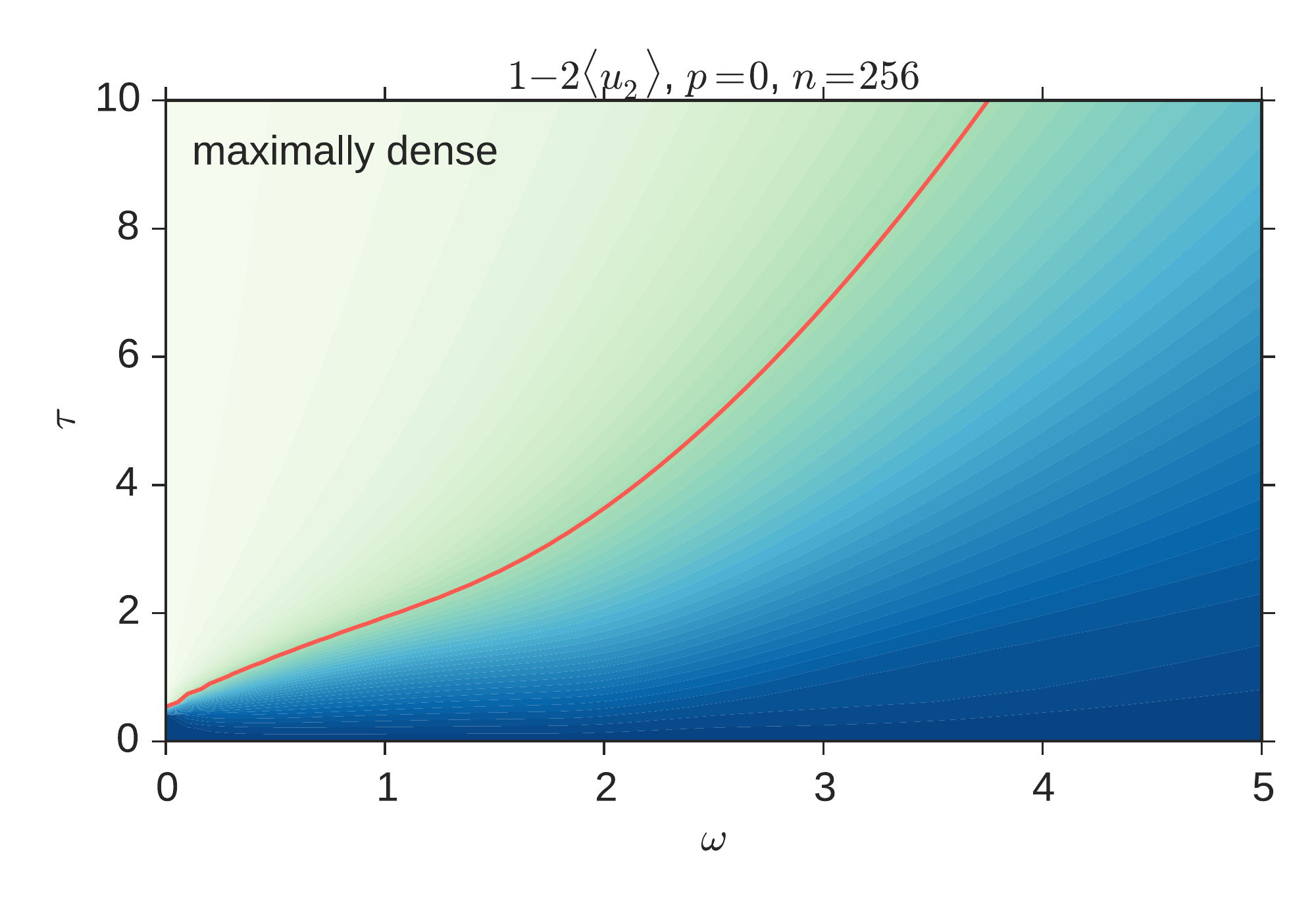}
\includegraphics[width=0.48\textwidth]{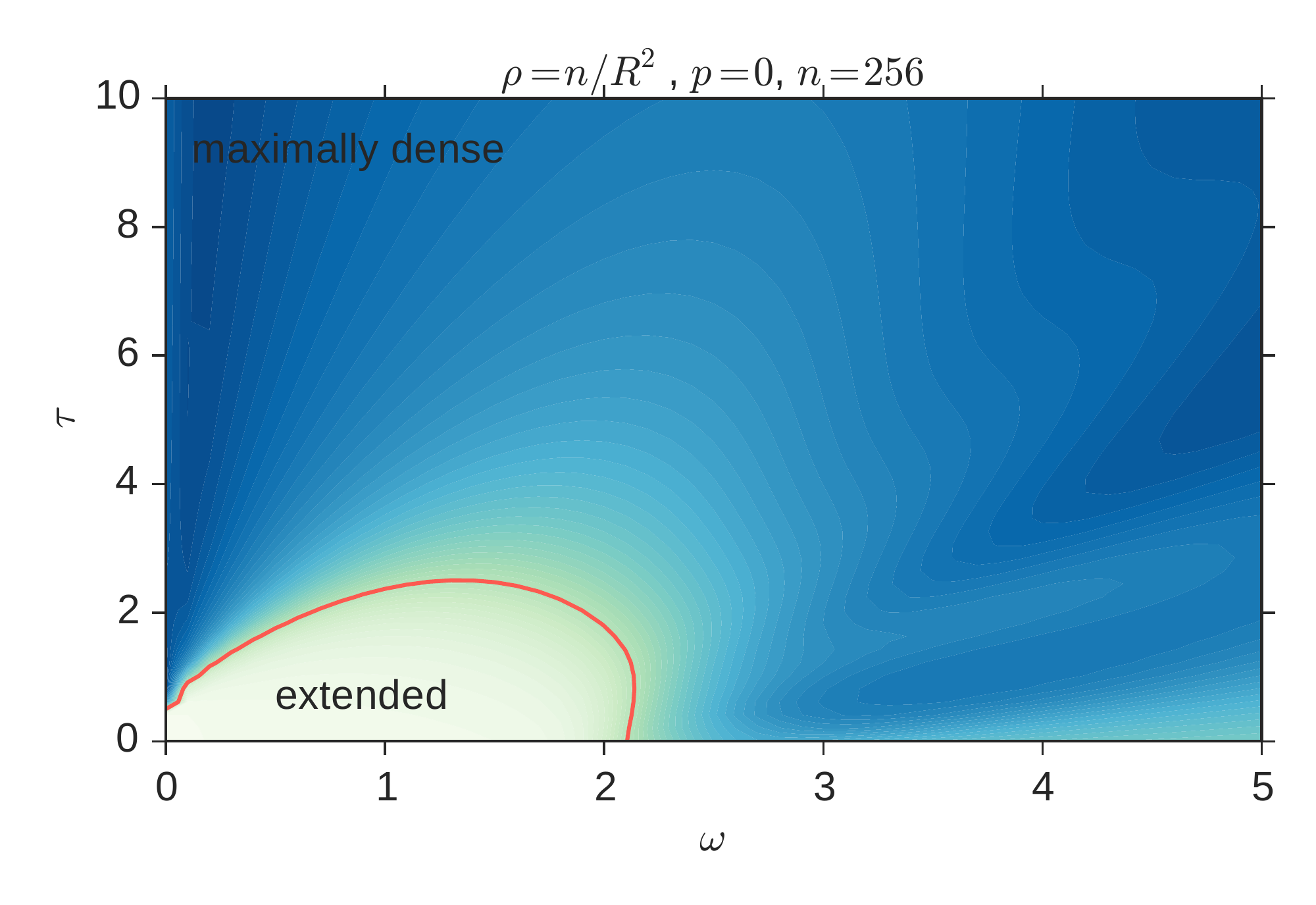}
\caption{Order parameters for $p=0$. The left diagram shows the frequency of singly-visited sites going to zero for large $\tau$ and small $\omega$, and the right diagram shows a clearly demarcated region of extended configurations where the density is small. Taken together, these diagrams indicate the presence of three distinct phases. The data for this plot comes from dataset VI-NN L-lattice.}
\label{fig:order_parameters_p-eq-0}
\end{figure}

\begin{figure}[ht!]
\centering
\includegraphics[width=0.7\textwidth]{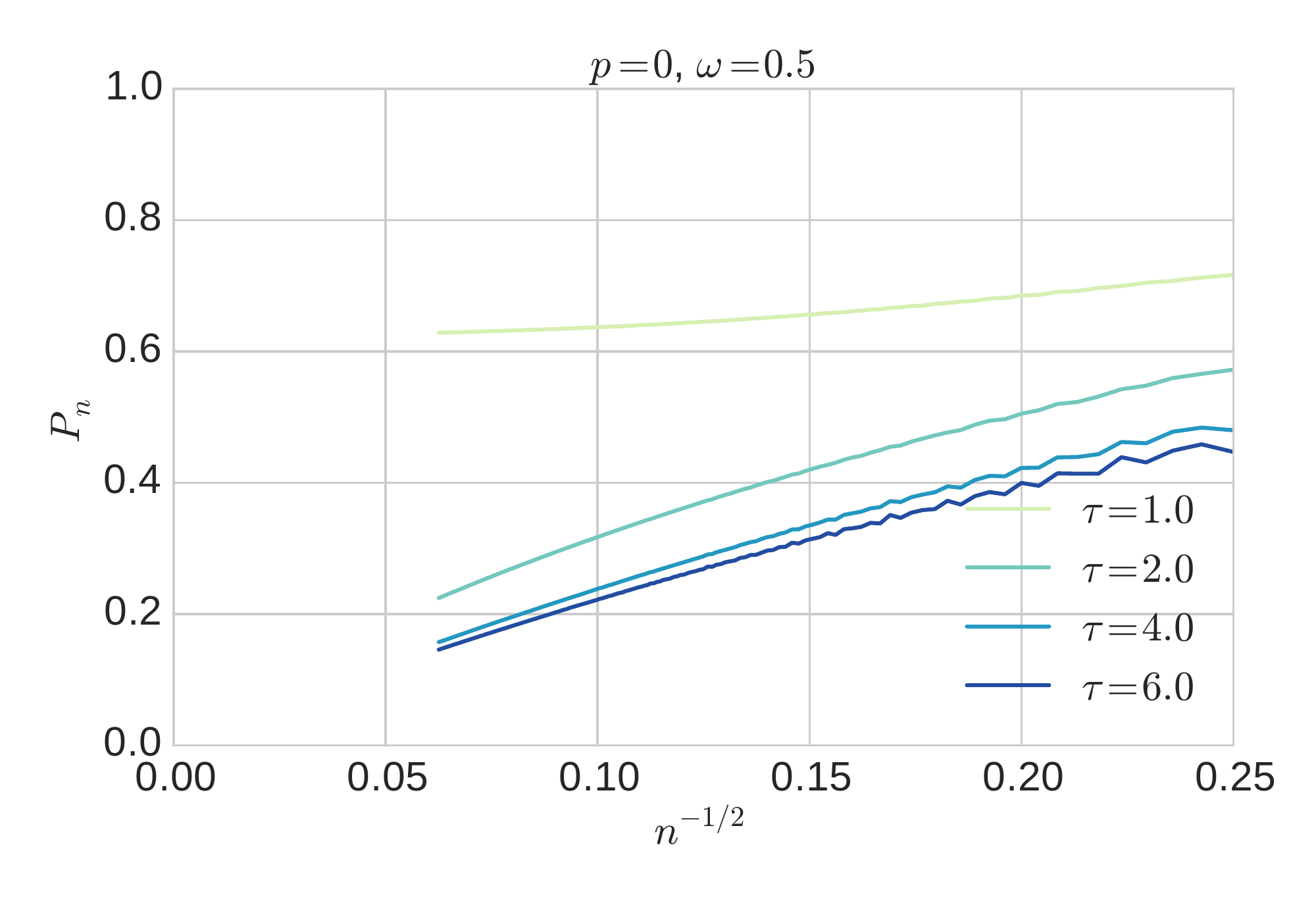}
\caption{Evidence of the maximally dense phase at $p = 0$. According to the phase diagram in Figure~\ref{fig:phase-diagram-p-eq-0}, on the vertical line $\omega = 0.5$ and for values of $\tau$ above 2, we are in the maximally dense phase. This is confirmed in the above plot which shows that the ratio of singly visited sites goes to zero. The plot on the left is obtained from dataset VI-NN L-lattice.}
\label{fig:VISAW_low_temperature_omega_eq_0_5_p_eq_0}
\end{figure}

The phase diagram is seen to contain three phases: extended, collapsed (globular) and maximally dense, see Figure~\ref{fig:order_parameters_p-eq-0}. This is as expected since the $\beta$-sheet phase exists only for sufficiently large stiffness.  The demarcation of the maximally dense phase is fairly well defined but the boundary between the extended and globular phases is very difficult to ascertain: the fluctuation data and the order parameter data given quite different phase boundaries. We have conservatively conjectured a boundary that does not change much as $\tau$ changes but we have weak evidence that this is correct.

The key observation to be made here is that the low temperature phase of the LSAT model is maximally dense! This is completely unexpected and we do not have a explanation for this. The LSAT model has a collapse transition that is expected to be in the $\theta$ universality class and not the ISAT/VISAW one and yet its low temperature collapsed phase seems to coincide with those models: recall that collapse transition has a strongly divergent specific heat in those models. We confirm that as $\omega$ is varied that the low temperature phase is indeed maximally dense: see Figure~\ref{fig:VISAW_low_temperature_omega_eq_0_5_p_eq_0}.

For non-zero $\omega$ we see that the extended to maximally dense transition is second order and only at $\omega=0$ might it turn to first order. This is in accord with the LSAT model at $\omega=1$ of course. 

\begin{figure}[h]
\centering
\includegraphics[width=0.7\textwidth]{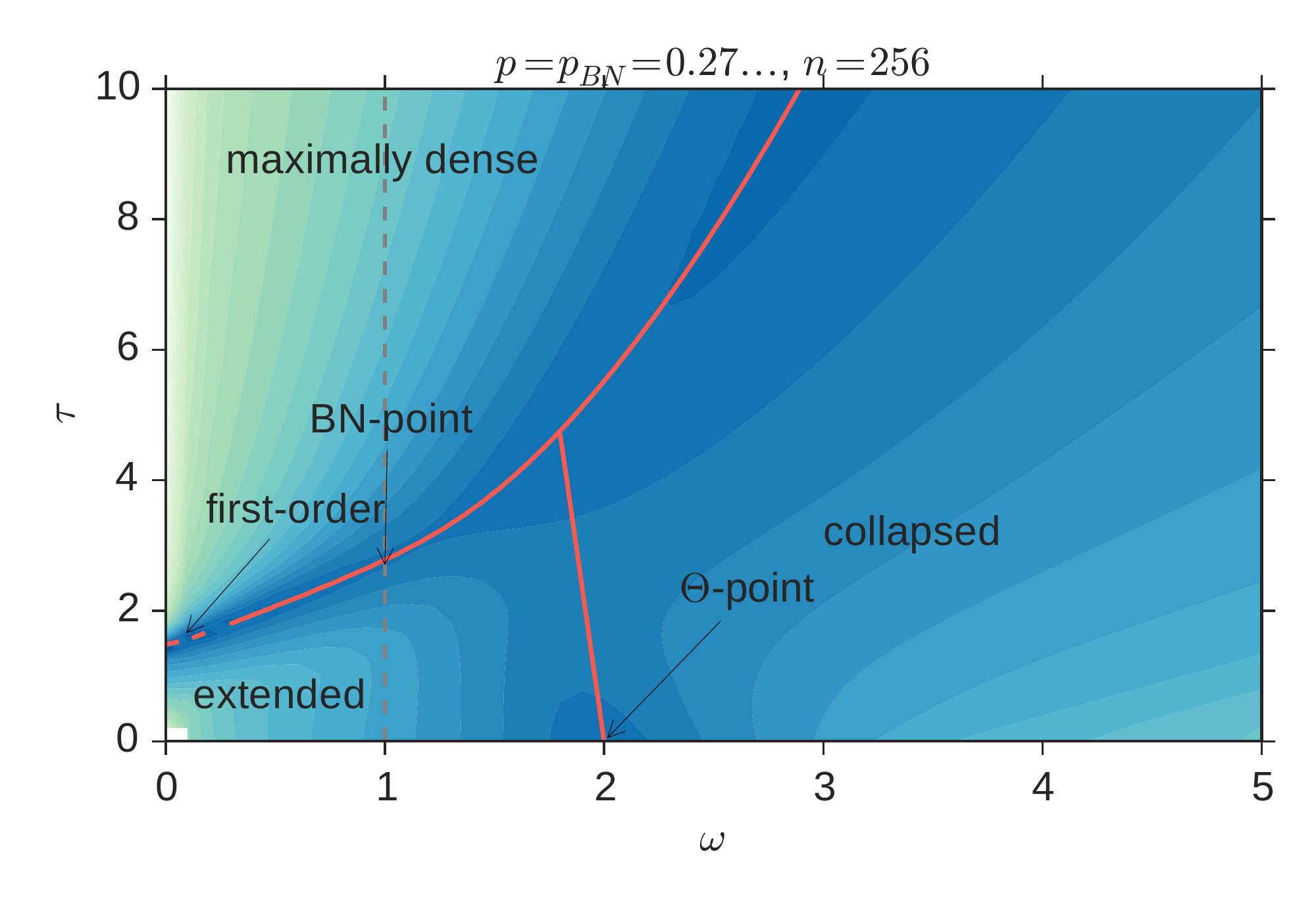}
\caption{The phase diagram for $p=p_{BN}$, as obtained from dataset VI-NN. First- and second-order transitions are shown respectively as dashed and solid lines. There are three phases, as at $p=0$, with extended, collapsed and maximally dense phases existing.}
\label{fig:phase_diagram_BN}
\end{figure}

\subsubsection{$p=p_{bn}$}
\begin{figure}[ht!]
\centering
\includegraphics[width=0.48\textwidth]{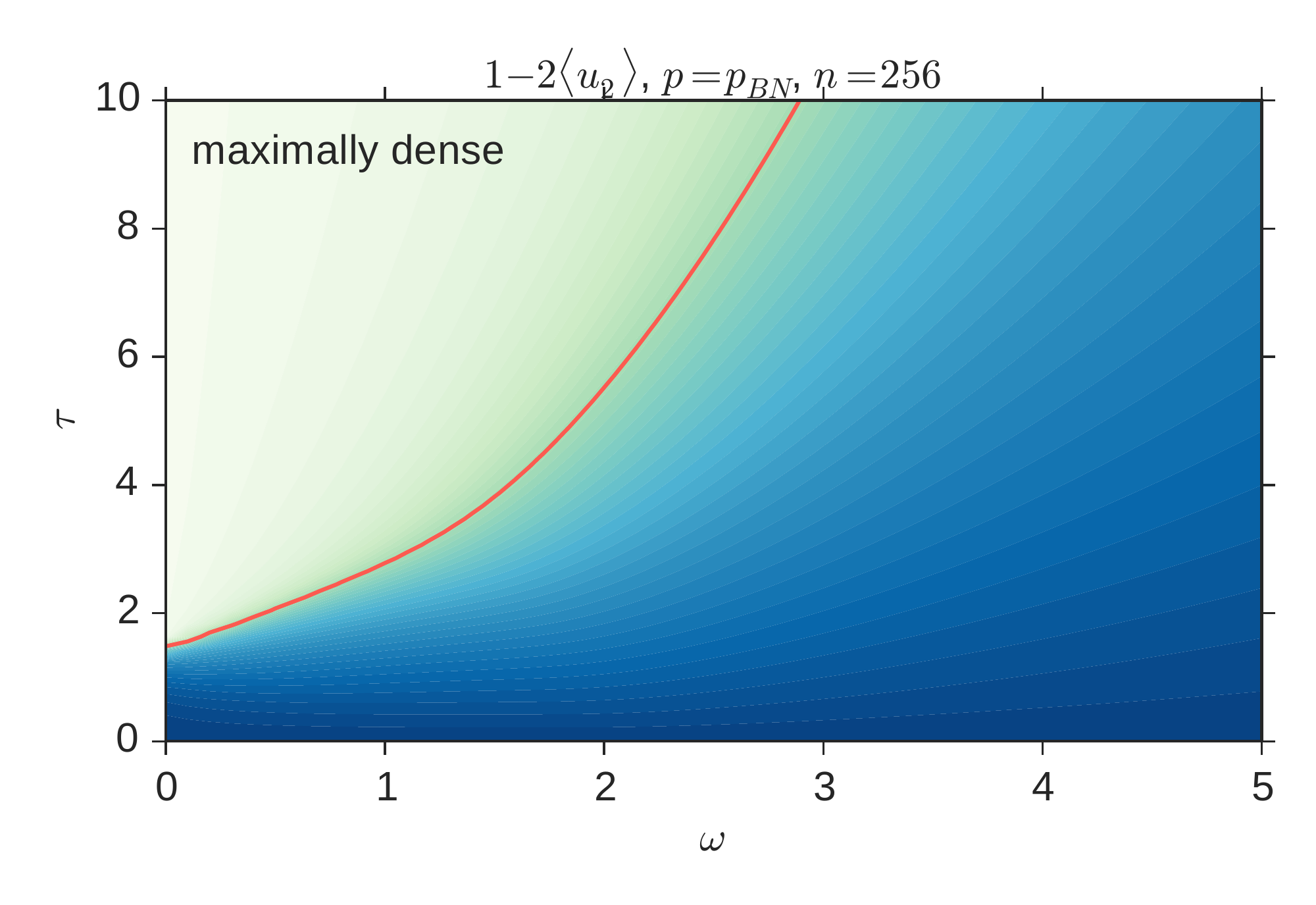}
\includegraphics[width=0.48\textwidth]{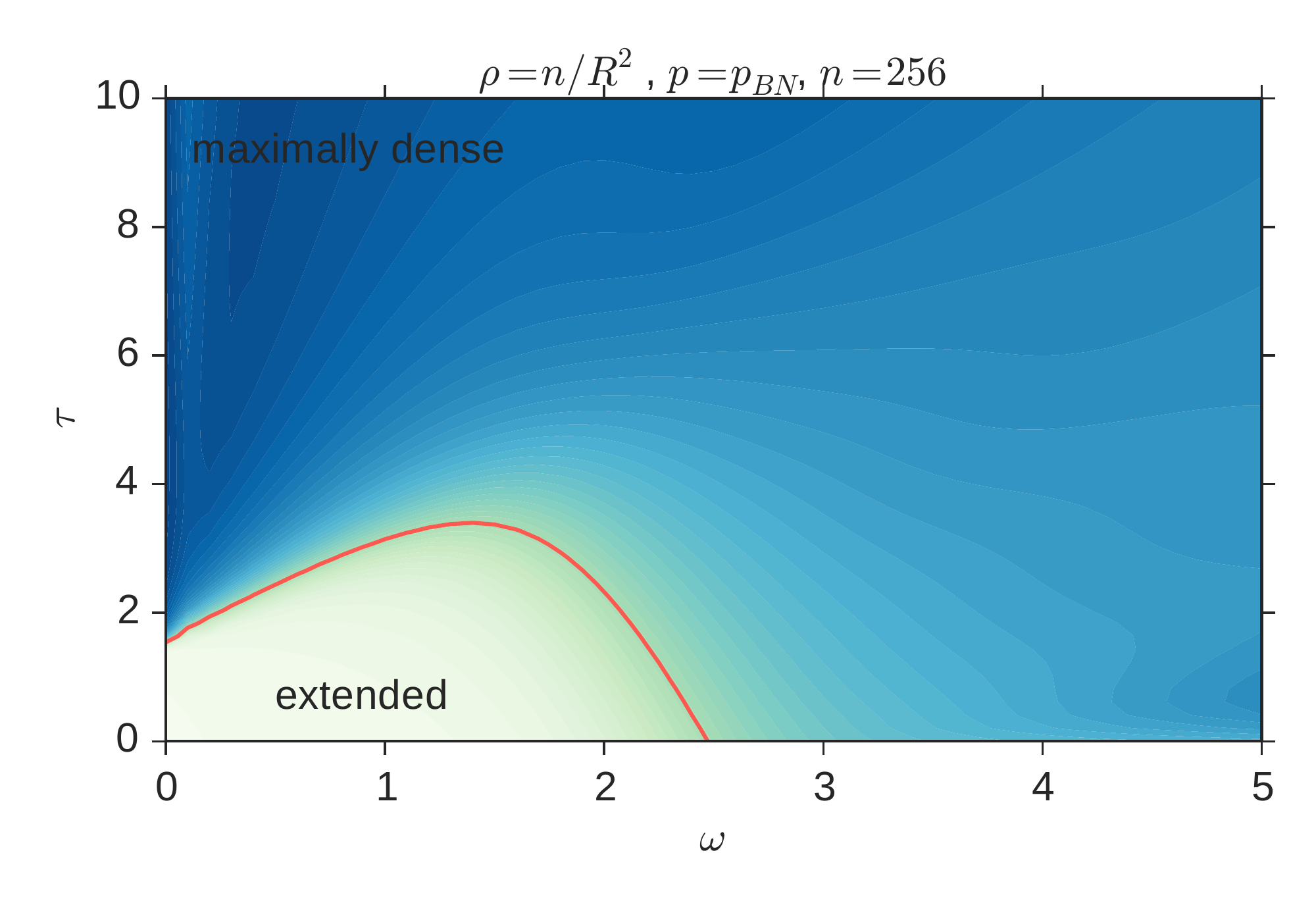}
\caption{Order parameters for $p=p_{BN}$. The left diagram shows the frequency of singly-visited sites going to zero for large $\tau$ and small $\omega$, and the right diagram shows a clearly demarcated region of extended configurations where the density is going to zero. Taken together, these diagrams indicate the presence of three distinct phases. The data for this plot comes from dataset VI-NN.}
\label{fig:order-parameters-BN}
\end{figure}

\begin{figure}
\centering
\includegraphics[width=0.7\textwidth]{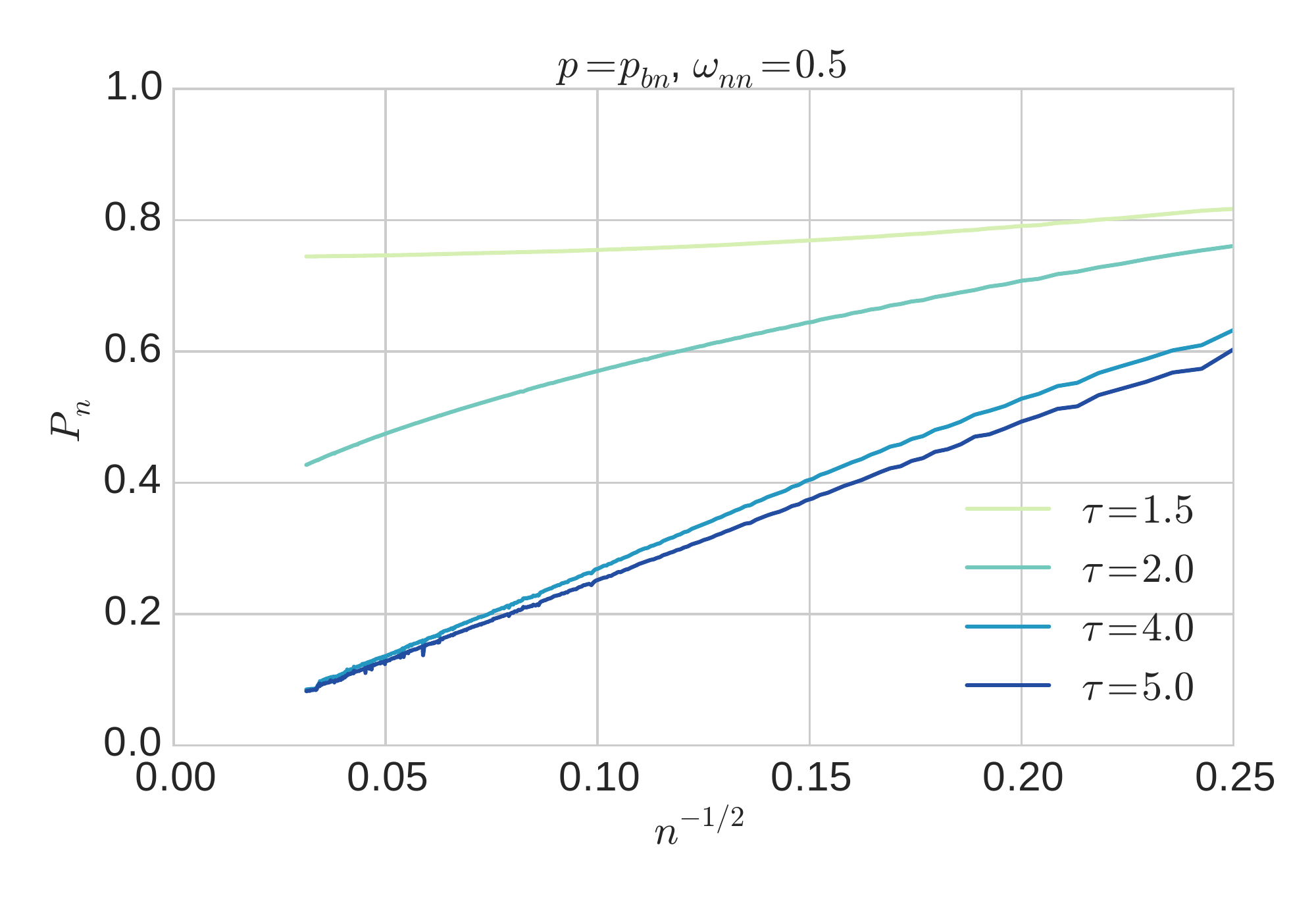}
\caption{This plot shows that the ratio of singly visited sites goes to zero for large $\tau$, suggesting that increasing $\tau$ leads to a collapsed phase that is maximally dense. This data comes from dataset VI.}
\label{fig:single-visited-BN}
\end{figure}

\begin{figure}
\centering
\includegraphics[width=\textwidth]{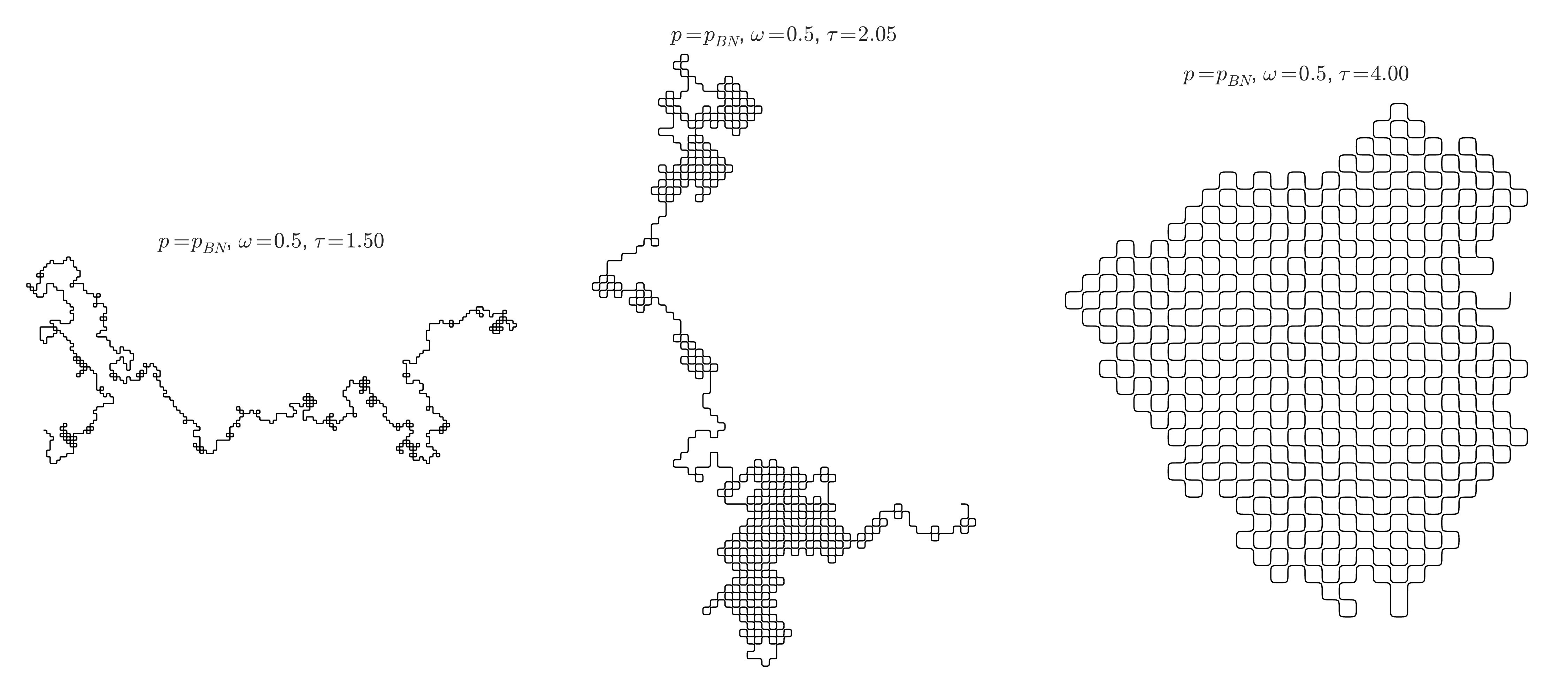}
\caption{Here are some typical configurations for $p=p_{BN}$ along the line $\omega_{nn} = 0.5$ at $\tau=1.5$, $\tau=2.05$ (at the approximate location of the transition) and $\tau=4$. The low-temperature phase is clearly compact. These configurations are obtained from dataset VI.}
\label{fig:configs_BN}
\end{figure}

We now consider fixing the stiffness to the BN-point value $p_{BN}$ so as to consider how the BN-point sits in the wider phase space varying $\omega$ and $\tau$: see Figure~\ref{fig:phase_diagram_BN}. The semi-flexible VISAW model along the line $p=p_{BN}$ is shown in  Figure~\ref{fig:phase_diagram_BN} as the line $\omega=1$.
The analysis here is done analogously to the one above for the LSAT model ($p=0$). One can see that in common with the LSAT model the low temperature phase is maximally dense, as demonstrated in Figure~\ref{fig:order-parameters-BN}, which shows diagrams for the relevant order parameters. From the scaling of the number of singly visited sites (Figure~\ref{fig:single-visited-BN}) and typical configurations (Figure~\ref{fig:configs_BN}) it follows that the low temperature phase is indeed maximally dense.   

\subsubsection{$p=1$}
\begin{figure}[ht!]
\centering
\includegraphics[width=0.7\textwidth]{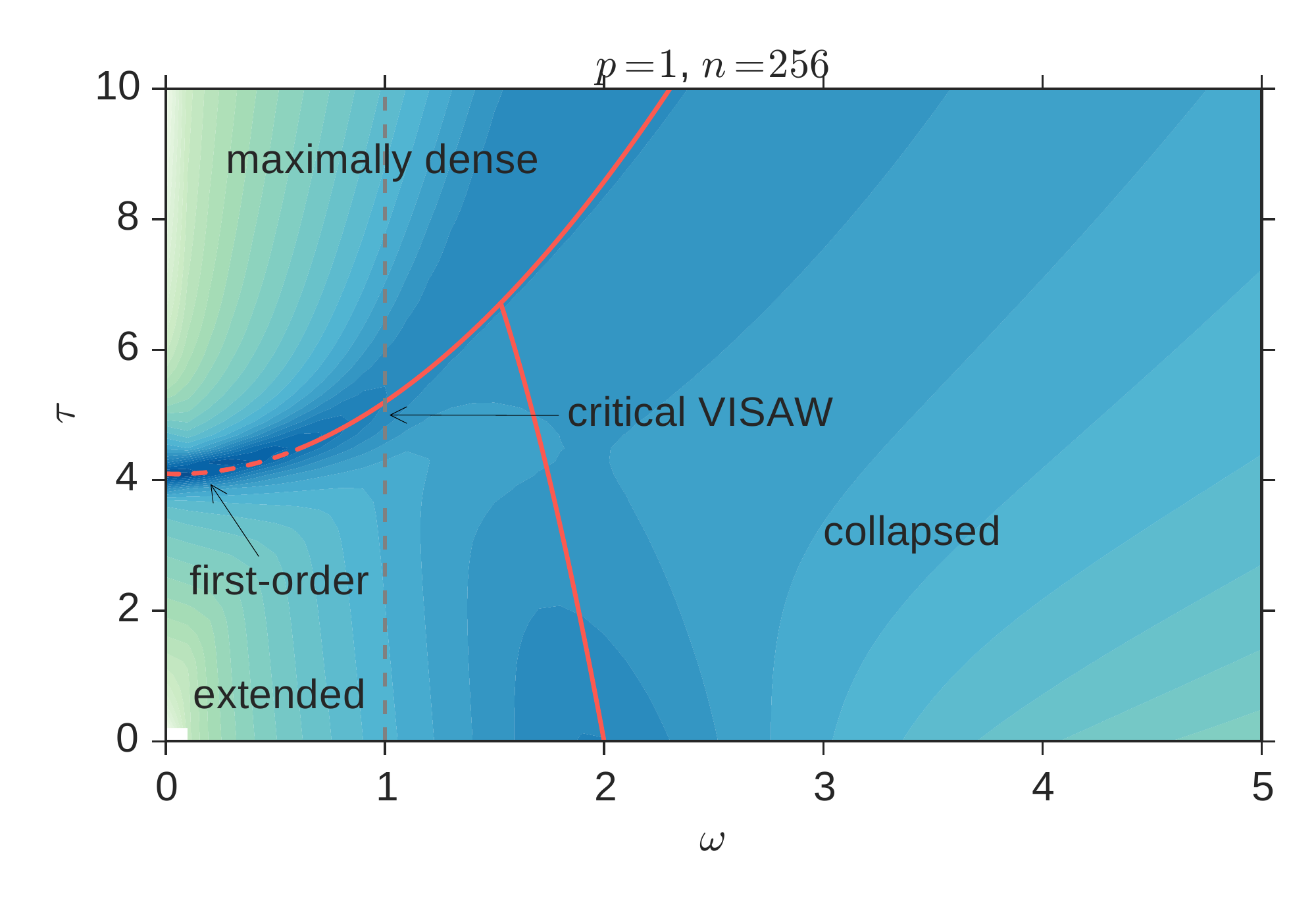}
\caption{The phase diagram for $p=1$. First- and second-order transitions are shown respectively as dashed and solid lines. This data comes from dataset VI-NN-2. The canonical ISAW model lies on the horizontal line $\tau=1$.}
\label{fig:phase-diagram_p_1}
\end{figure}

It is interesting to consider the fully-flexible case of our model ($p=1$). This model contains explicitly the fully flexible VISAW for $\omega=1$ and the standard ISAW model for $\tau=1$. See Figure~\ref{fig:phase-diagram_p_1}. Again the three phases of extended, globular and maximally dense occur. The region of the conjectured first order transition between the extended and maximally dense phase has increased significantly over $p=p_{BN}$. It may be that in the thermodynamic limit that it ends at $\omega=1$ --- a question that deserves further consideration.

\begin{figure}
\centering
\includegraphics[width=1.0\textwidth]{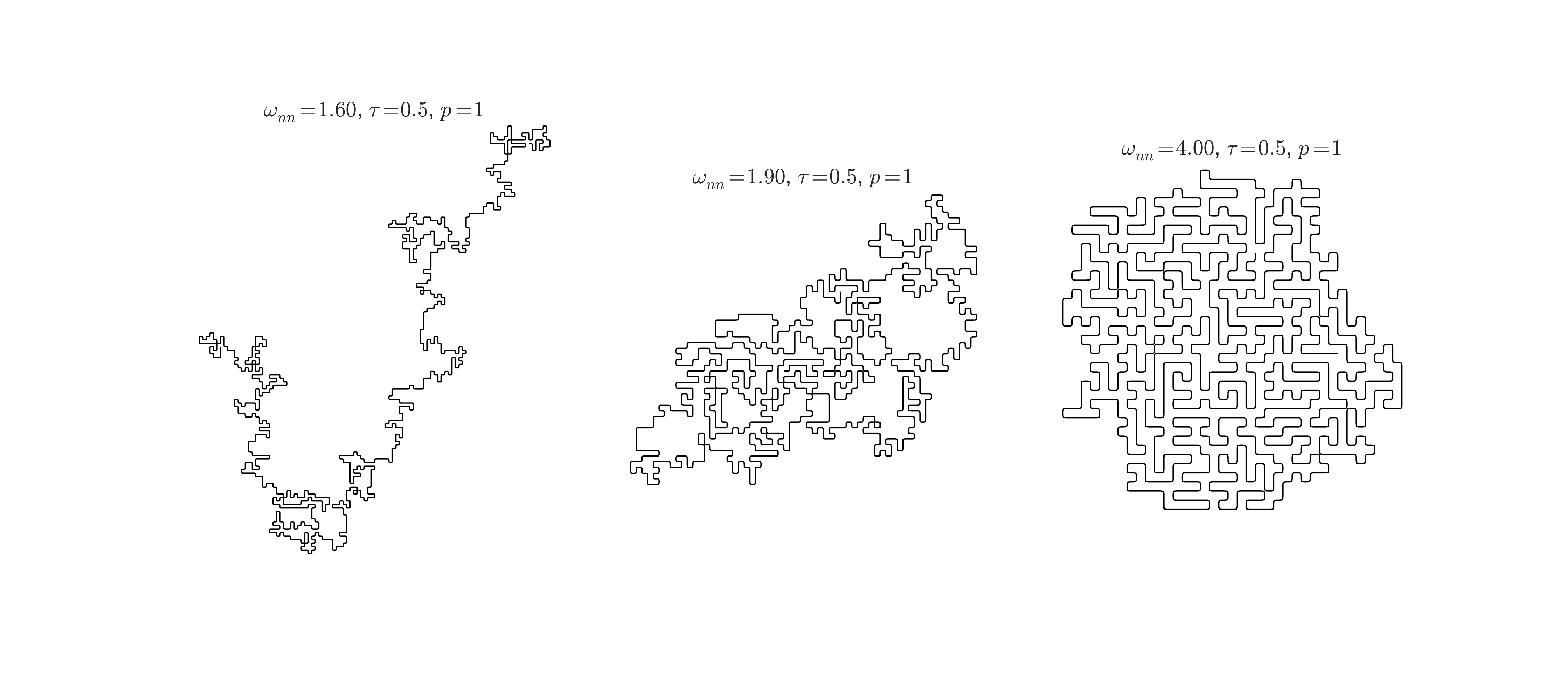}
\caption{Typical configurations for $p=1$ and $\tau= 0.5$, at $\omega=1.6$, $\omega=1.9$ and $\omega=4$. The low-temperature phase is globular: dense but with amorphous structure. These configurations come from the dataset NN-2.}
\label{fig:configs-p_1}
\end{figure}

It is worth considering the typical configurations along a line parallel in $\tau$ to the ISAW model. In Figure~\ref{fig:configs-p_1} we have illustrated some typical configurations for the line $\tau=0.5$. At low temperature (large $\omega$) we see a configuration from the globular collapsed phase which can be compared to the maximally dense and $\beta$-sheet phases. 

\subsubsection{All $p$}
\begin{figure}[ht!]
\centering
\includegraphics[width=0.45\textwidth]{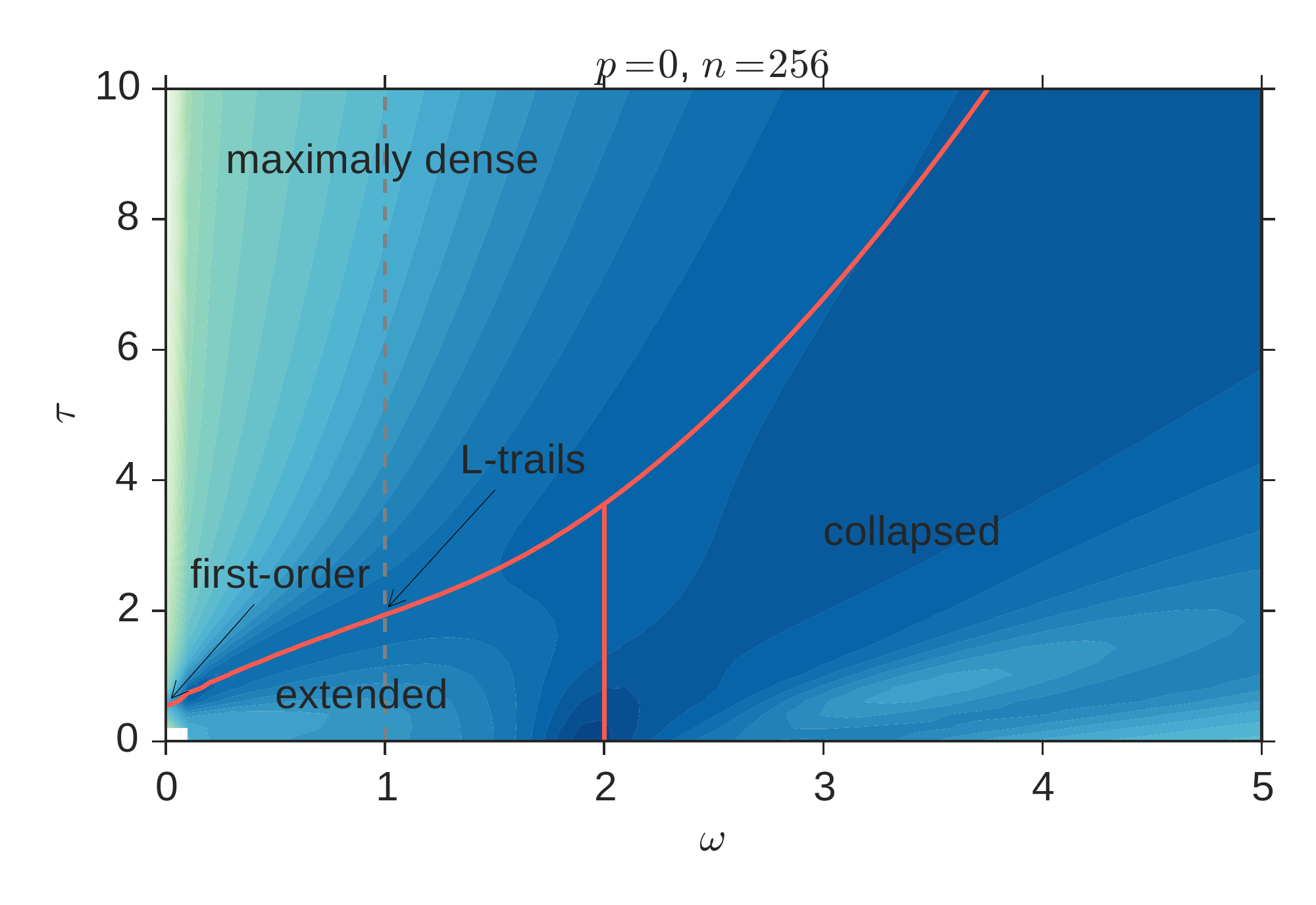}
\includegraphics[width=0.45\textwidth]{VISAW_phase_diagram_p_eq_bn.pdf}
\includegraphics[width=0.45\textwidth]{VISAW_phase_diagram_p_eq_1.pdf}
\includegraphics[width=0.45\textwidth]{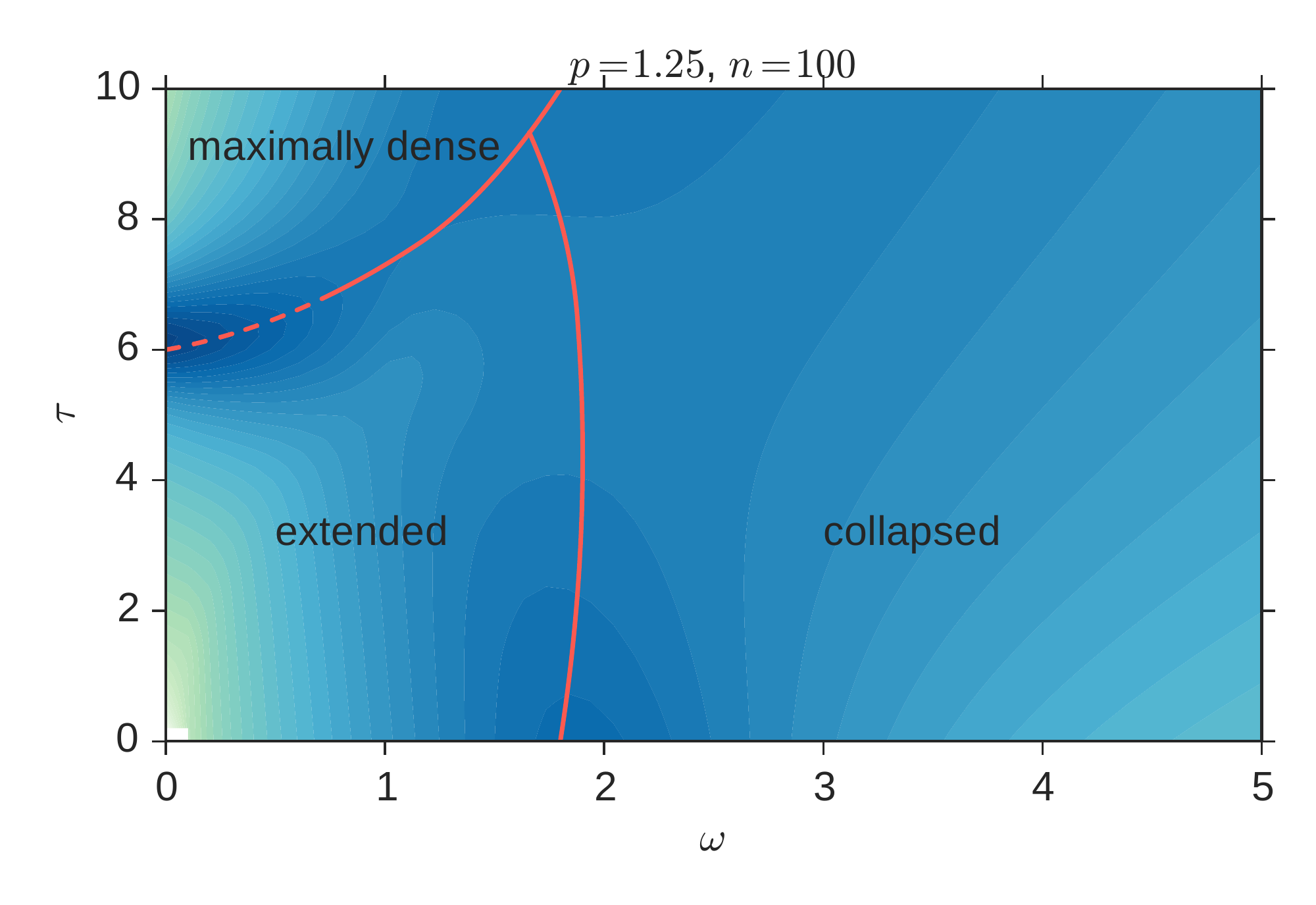}
\includegraphics[width=0.45\textwidth]{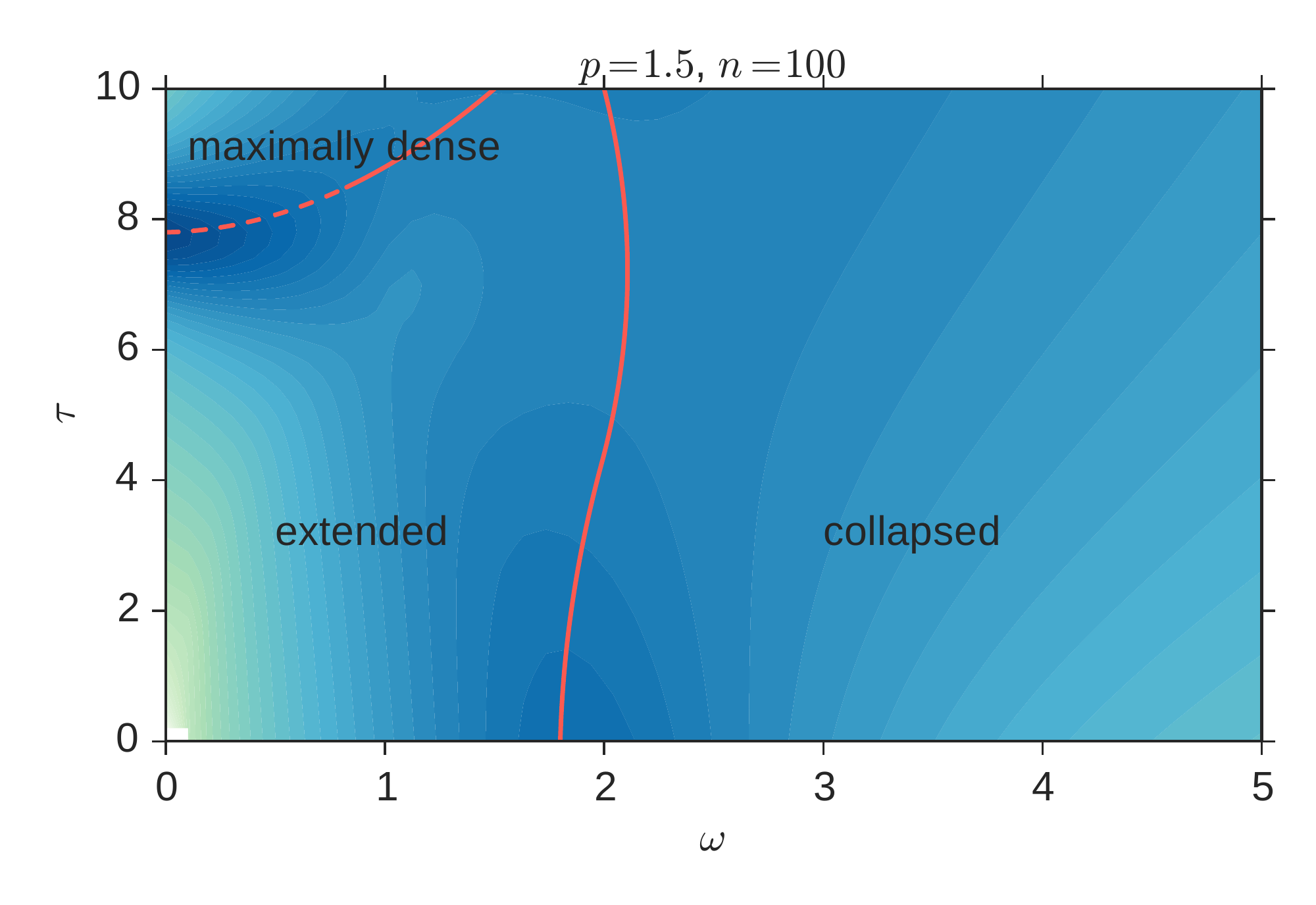}
\includegraphics[width=0.45\textwidth]{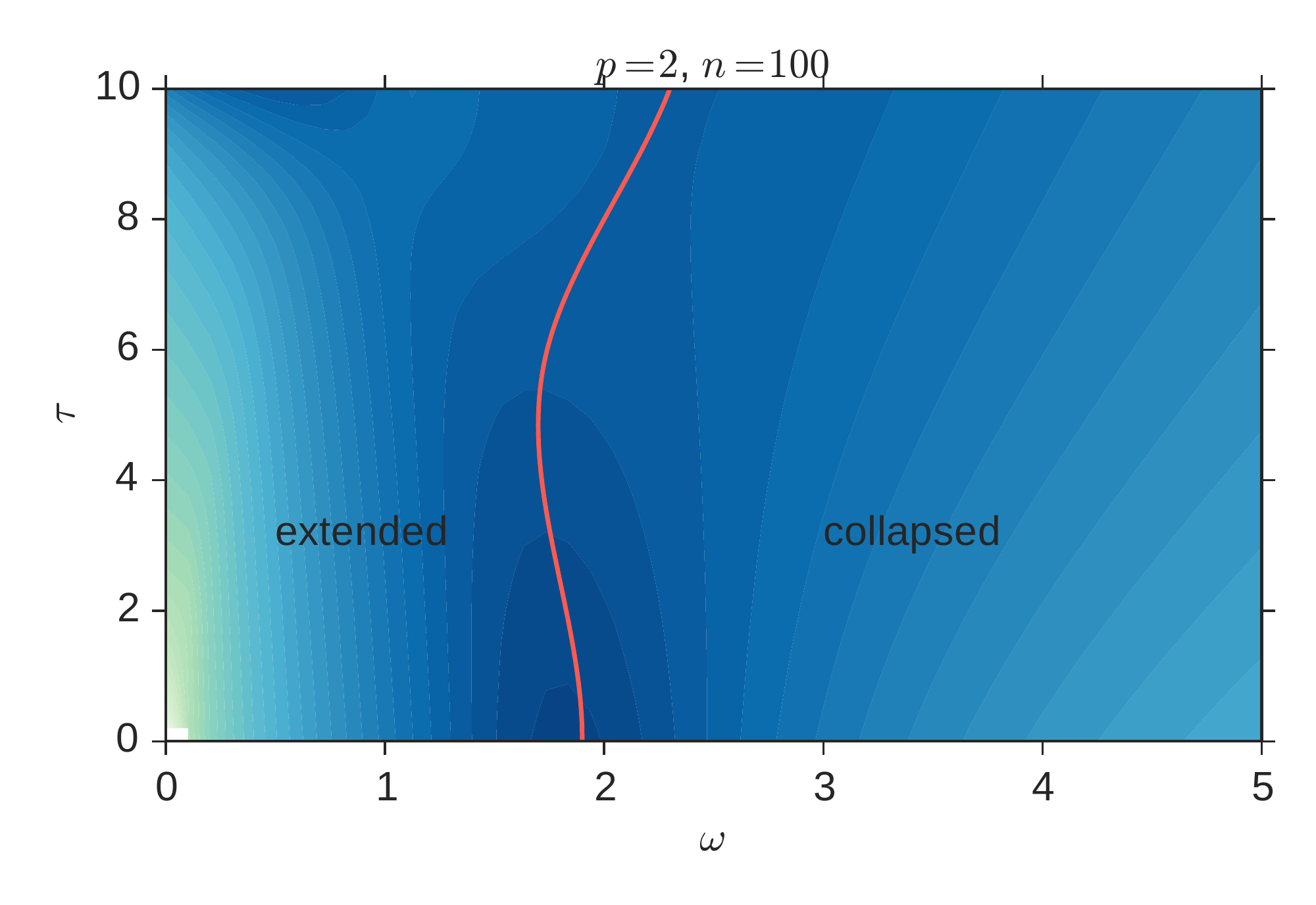}
\includegraphics[width=0.45\textwidth]{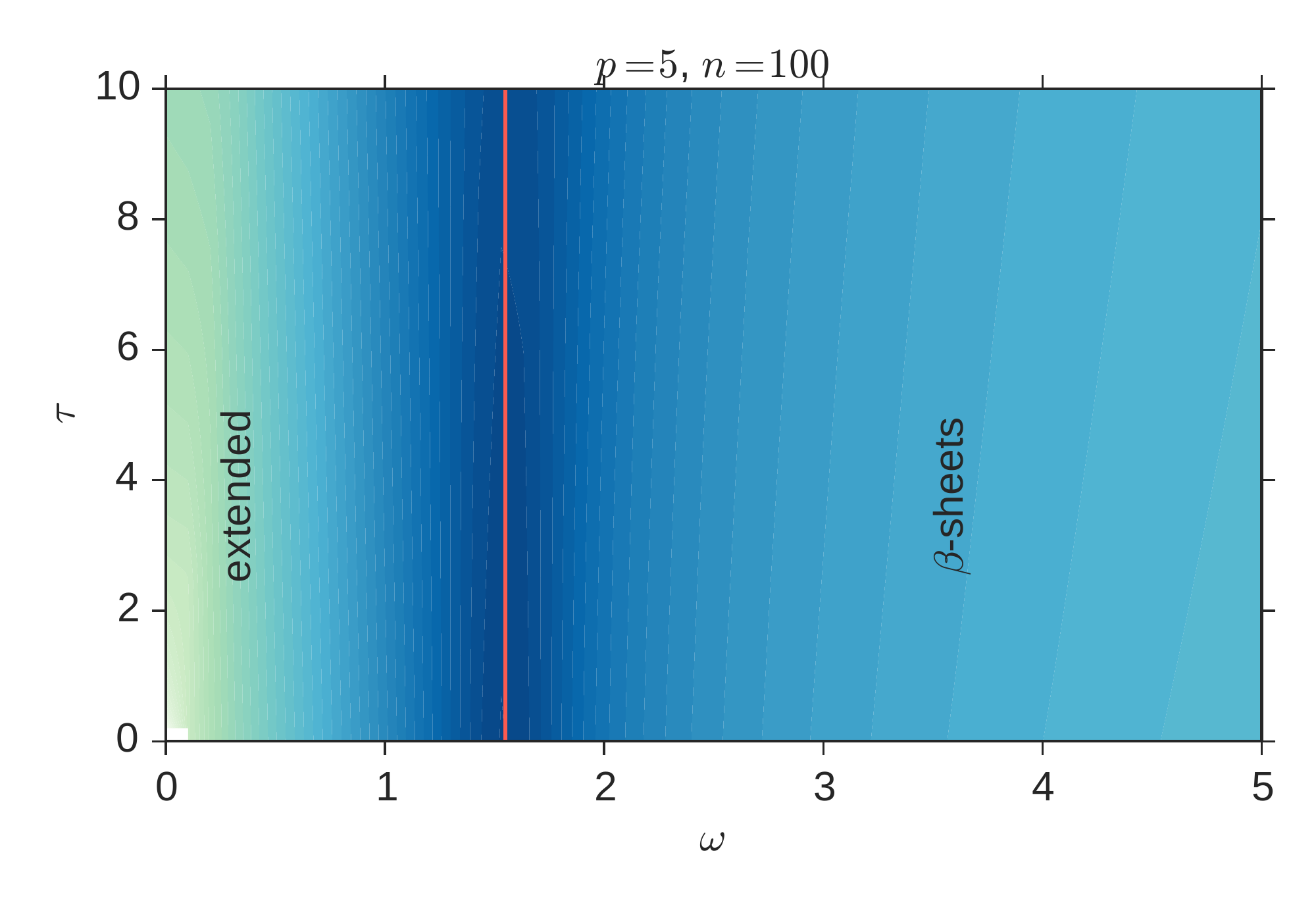}
\caption{First- and second-order transitions are shown respectively as dashed and solid lines. At $p=0$ the extended to  maximally dense transition is always second order except possibly at $\omega=0$. In line with the previously discussed $\omega$-slices there exists a  special value of $\omega=\omega_b$ that signals a  change from second to first order in that transition at other values of $p$. The first three plots come from datasets VI-NN L-lattice, VI-NN and VI-NN-2 respectively. The remaining plots come from an appropriate slice of the dataset 3P.}
\label{fig:phase-diagrams-variable-p}
\end{figure}

We now compare a fully range of $p$ values in Figure~\ref{fig:phase-diagrams-variable-p}. We see that as one increases $p$ the region of the maximally dense phase retreats to larger values of $\tau$ and smaller values of $\omega$ eventually disappearing from our diagrams' range of $\tau$ and $\omega$ once $p$ reaches $2$. The other significant feature to note is the eventual appearance of the $\beta$-sheet phase for $p=5$ at large values of $\omega$.

\section{Phase transitions}

While there are several different transitions in our model we have concentrated our investigation to the extended to maximally dense transition as the least all understood but also assessable using our data. First let us consider the other transitions. The extended to globular transition has been well studied previously, see \cite{Caracciolo2011} and reference therein, as this is the canonical polymer collapse transition. We find no evidence that it changes with parameters of our model and we conjecture that it always lies in the $\theta$-point universality class of the ISAW model. The extended to $\beta$-sheet transition has been studied previously \cite{Krawczyk2010}  and a first order transition has been found. We confirm this and see it also  is unchanged by the parameter variations in our model.
The globular to $\beta$-sheet transition is a low temperature transition that is difficult to study. Nevertheless it has been studied in \cite{Krawczyk2010} and they found $\alpha\phi\approx 0.4$: a moderately strong divergence specific heat. We are unable to provide any further precision to the previous study.
The transition between the collapsed phase and the maximally dense phase has not been studied previously and unfortunately our simulations are not long enough to elucidate this low temperature transition. The only comment we can make is that we do not see evidence that it is first order.

\clearpage

\subsection{Extended to Maximally-Dense}
\begin{figure}[ht!]
\centering
\includegraphics[width=0.7\textwidth]{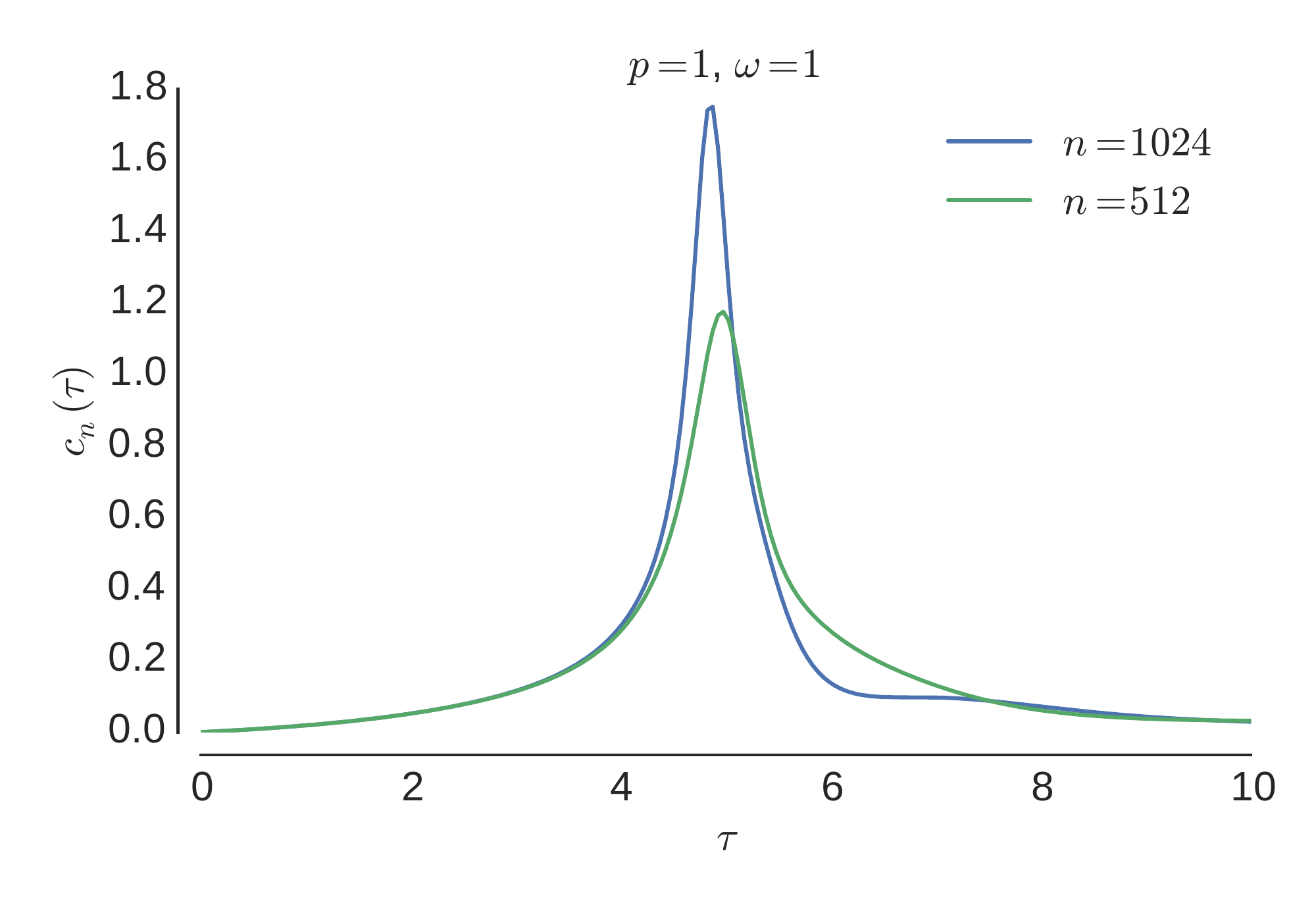}
\includegraphics[width=0.7\textwidth]{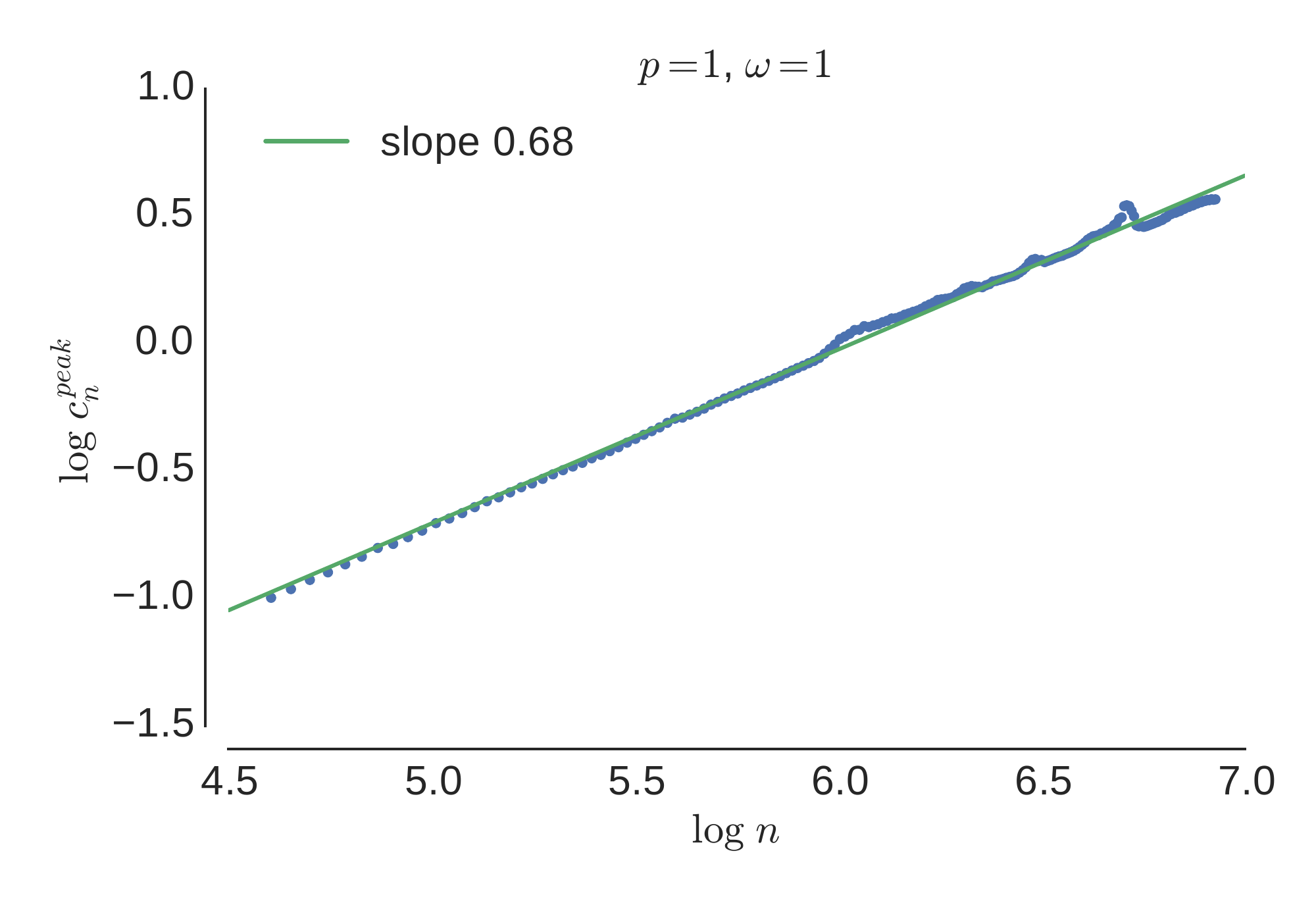}
\caption{On top  is the specific heat on the vertical line $p=1$, $\omega = 1$  as a function of $\tau$. On the bottom is the scaling of the peak heights in length, which shows an exponent close to $0.68$. These plots come from dataset VI-2.}
\label{fig:specific-heat-expon-VISAW}
\end{figure}

The transition in the VISAW model, and as we have seen it is apparently the one in the LSAT model, is the transition between the \emph{extended to maximally dense} phases. Since analysis of the VISAW model has previously found a diverging specific hear with $\alpha\phi\approx 0.69$ (we confirm this earlier estimate in Figure~\ref{fig:specific-heat-expon-VISAW}) while LSAT has a conjectured negative exponent value of $\alpha\phi=-1/7$ in common with the $\theta$-point there is no consistent value in the literature we can attribute to this transition. 

\begin{figure}[ht!]
\centering
\includegraphics[width=0.7\textwidth]{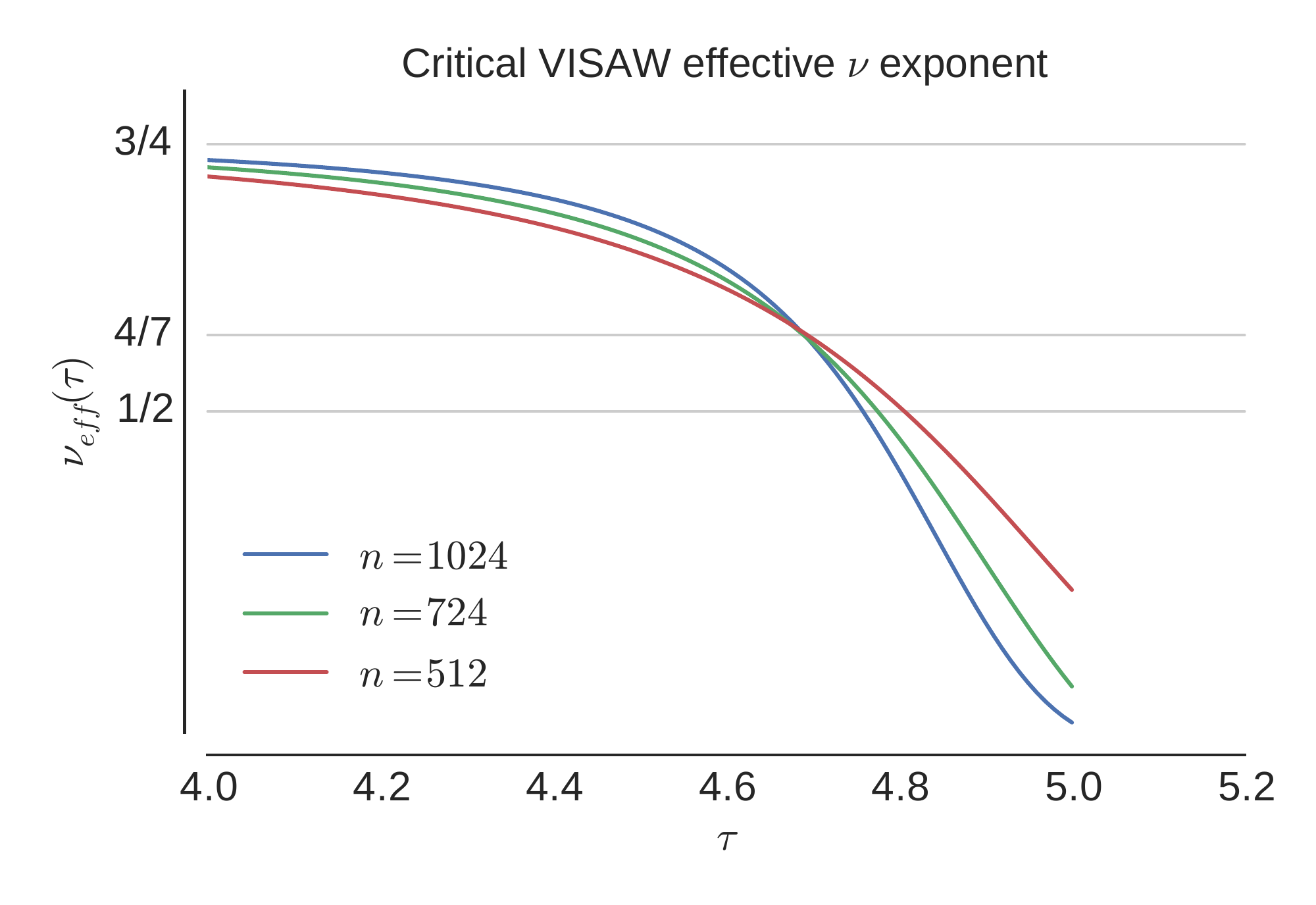}
\caption{A plot of the effective value of the exponent $\nu$ when $\omega =1$ and $p=1$, that for the VISAW model, against $\tau$ for different values of length. The different lines cross at a point close to the expected transition temperature indicating the critical value of $\nu$ to be close to the $\theta$-value of $4/7$. This plot comes from dataset VI-2.}
\label{fig:nu_VISAW}
\end{figure}

The analysis by Bl\"{o}te-Nienhuis \cite{blote1989a-a} of the semiflexible VISAW model and the subsequent numerical work by Foster and Pinettes \cite{foster2003a-a} find a variety of phase transition including first order behaviour, which indicates this transition has a richer structure that depends on model parameters. 
In fact we have found that there is a region where it is first order and a region where it is second order. Importantly  our second order transition does not have exponents that are consistent. While it is possible that the scenario is the same as in the triangular ISAT model where it has been conjectured that a first order line ends in an ISAT universality class point followed by a line of $\theta$-like transitions, our data is not precise enough to confirm this. However, that scenario would be the simplest: it begs the question of what happens at the point where the maximally dense globular and extended phases meet. It is also confused by at least one set of data, which is shown in Figure~\ref{fig:nu_VISAW}. If the fully flexible VISAW model is in the ISAT universality class, we should find a size exponent of $\nu=1/2$ with logarithmic corrections. However, the plot of the effective exponent seems to indicate that $\nu=4/7$ would be a compatible value. This confuses any possible simple conjecture.

\begin{figure}[ht!]
\centering
\includegraphics[width=0.7\textwidth]{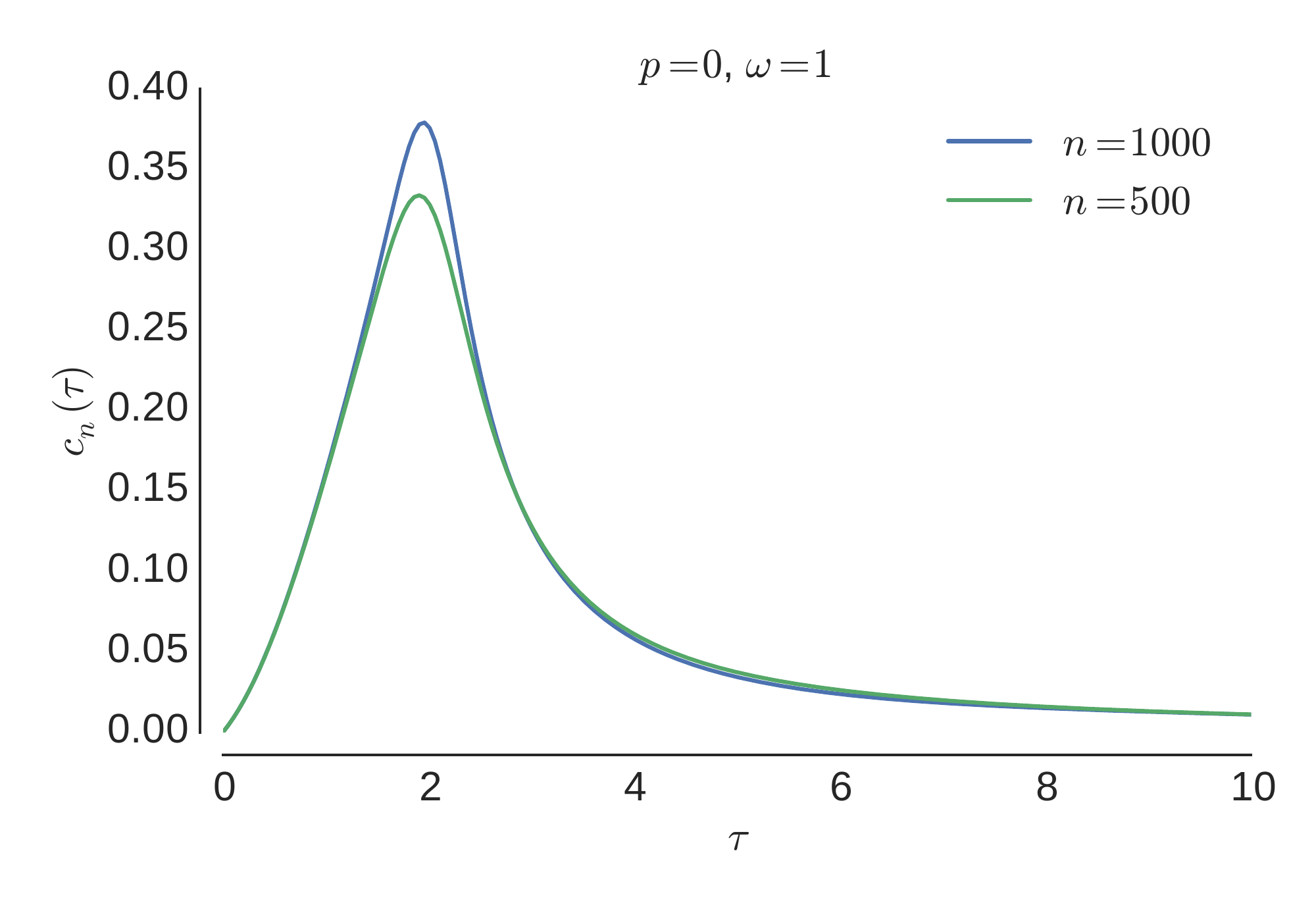}
\includegraphics[width=0.7\textwidth]{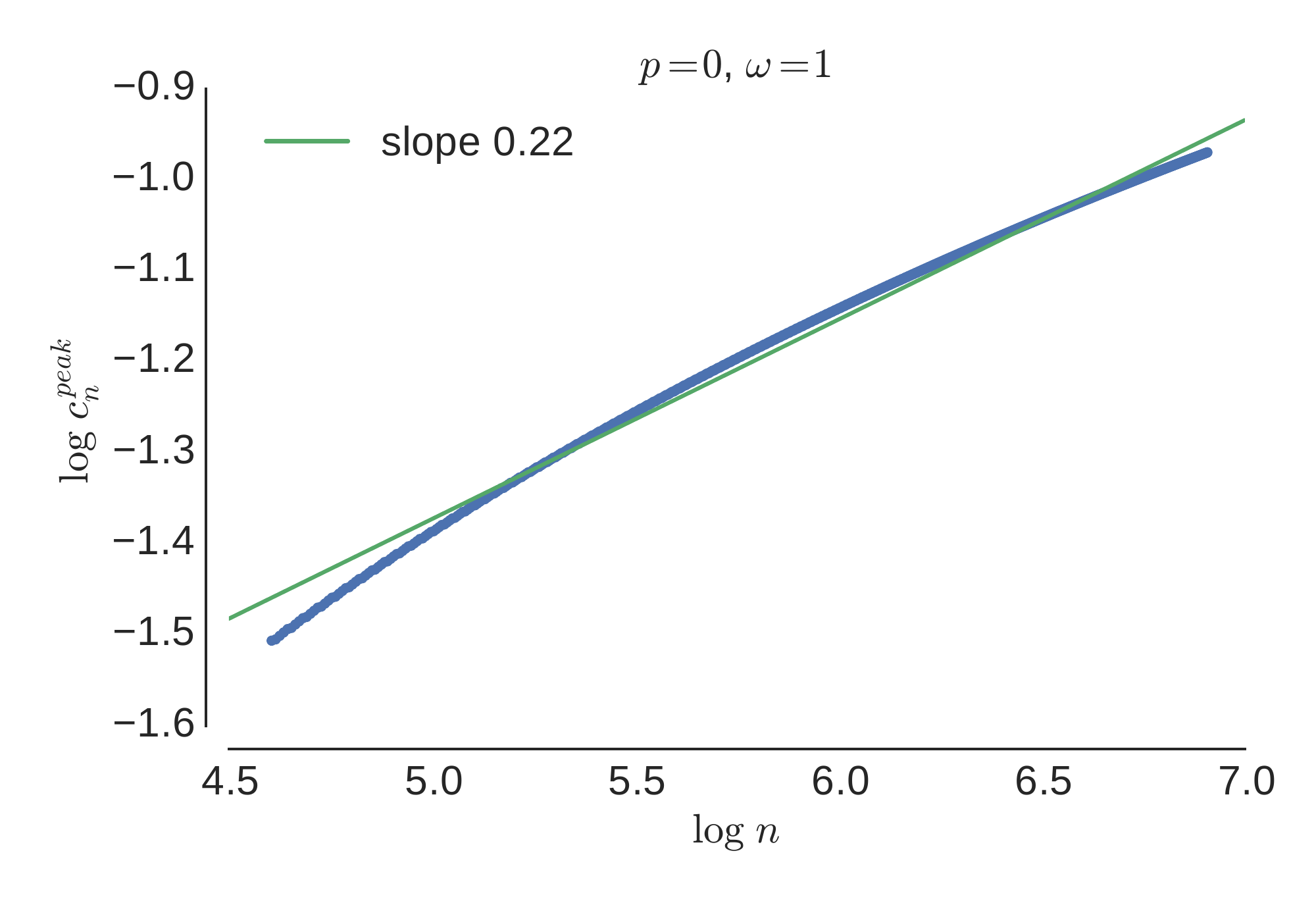}
\caption{Top: the specific heat as a function of $\tau$ on the vertical line $p=0$, $\omega = 1.0$. Bottom: the scaling of the specific heat peak with respect to the size of the walk. These plots come from dataset VI L-lattice. We believe this critical point belongs to a line of phase transitions in the $\Theta$ universality class, which is know to feature a convergent specific-heat. Therefore we interpret the low effective exponent 0.22 as the result of strong finite-size corrections to a converging specific-heat.}
\label{fig:omega_eq_1_p_eq_0}
\end{figure}

It is worthwhile noting that the estimation of the exponents for the convergent $\theta$-like transitions can be deceptive. In Figure~\ref{fig:omega_eq_1_p_eq_0} we show an effective exponent estimate for $\alpha\phi$ directly from the fluctuations which gives a positive value of around $0.22$. However, knowing that a negative exponent is likely, an estimate of the exponent using 
\begin{equation}
  (\alpha\phi)^{est} = \log_2\left(\frac{c_n^{peak} - c_{n/2}^{peak}}{c_{n/2}^{peak} - c_{n/4}^{peak}}\right)
\end{equation}
gives a value of around $-0.06$ at the lengths considered, which are up to $n=1000$. We are aware from earlier work that good estimates of $\alpha\phi$ in these problems require polymer lengths of many thousands \cite{prellberg1994a-a}.
A very similar picture is produced at $\omega=0.5$ suggesting the transition to the maximally dense phase looks $\theta$-like when $p=0$ regardless of $\omega$.

\begin{figure}[ht!]
\centering
\includegraphics[width=0.7\textwidth]{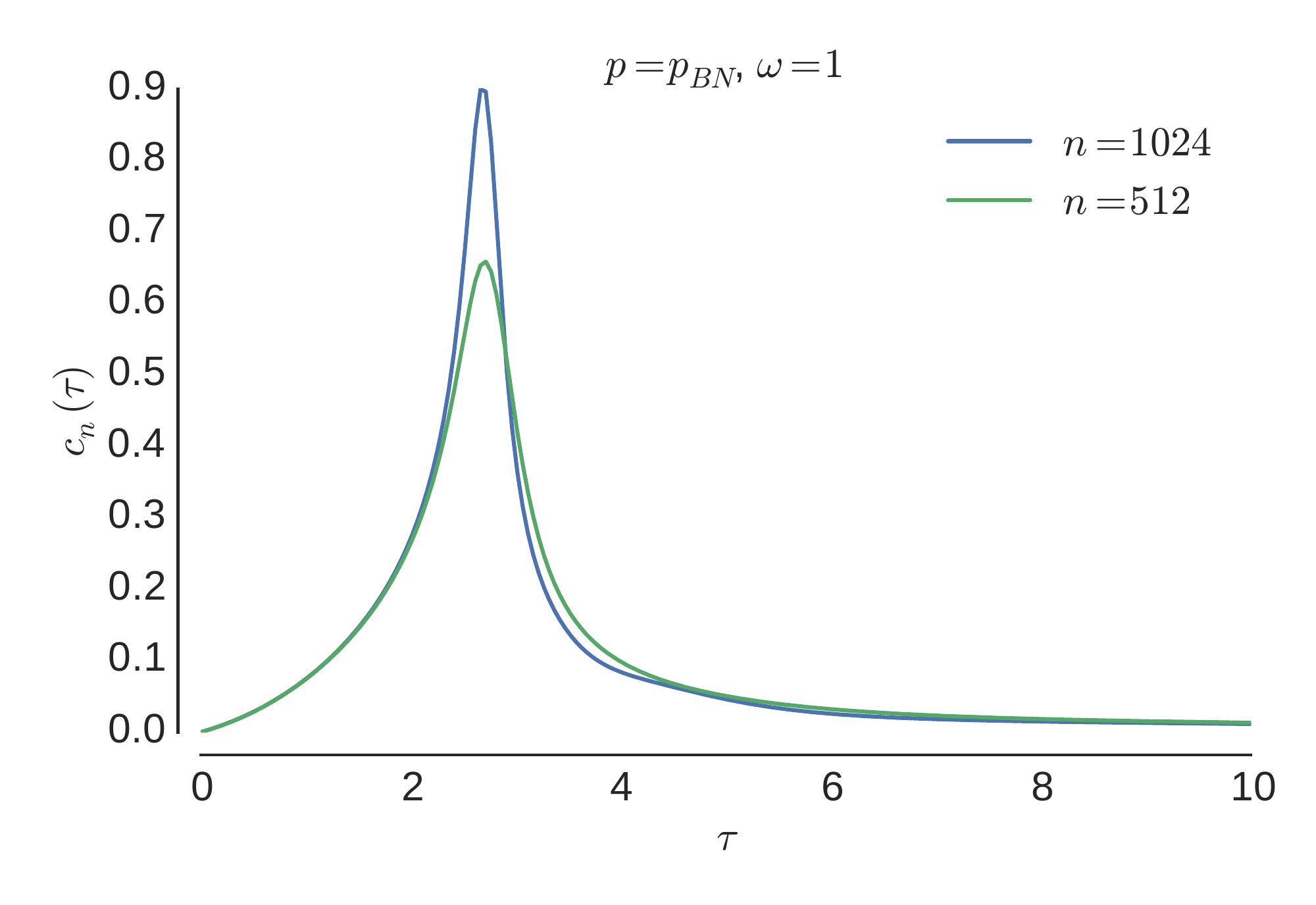}
\includegraphics[width=0.7\textwidth]{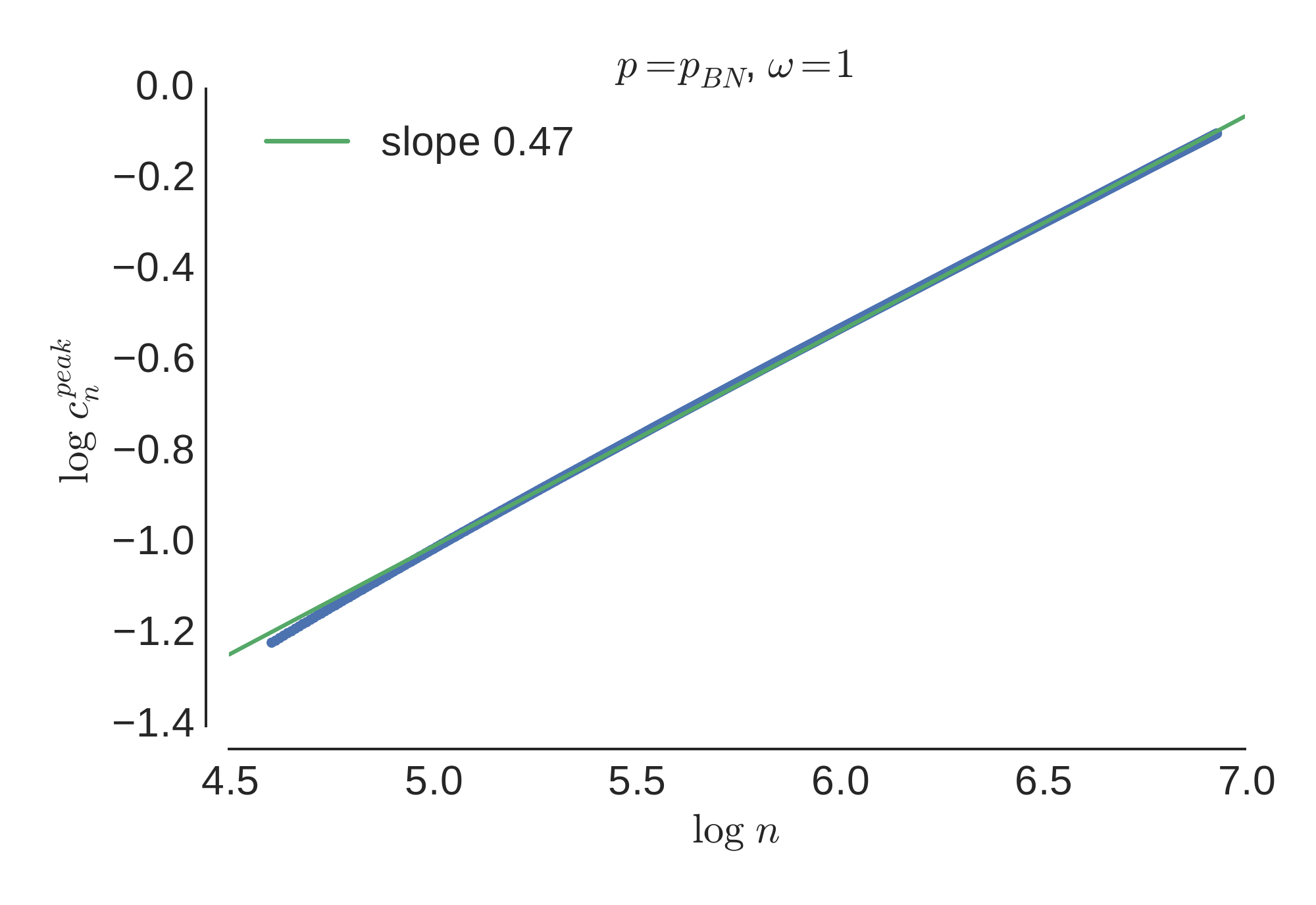}
\caption{This is the specific heat on the vertical line $p=p_{BN}$, $\omega = 1.0$. The scaling of the peak heights in length shows an exponent close to $0.47$. These plots come from dataset VI-3.}
\label{fig:specific-heat-expon-bn}
\end{figure}

Of interest has been the BN-point value of the stiffness $p_{BN}$, and in Figure~\ref{fig:specific-heat-expon-bn} we plot the specific heat at $\omega=10$ where we obtain $\alpha\phi\approx 0.47$. 

\begin{figure}[ht!]
\centering
\begin{subfigure}[b]{0.6\textwidth}
  \includegraphics[width=\textwidth]{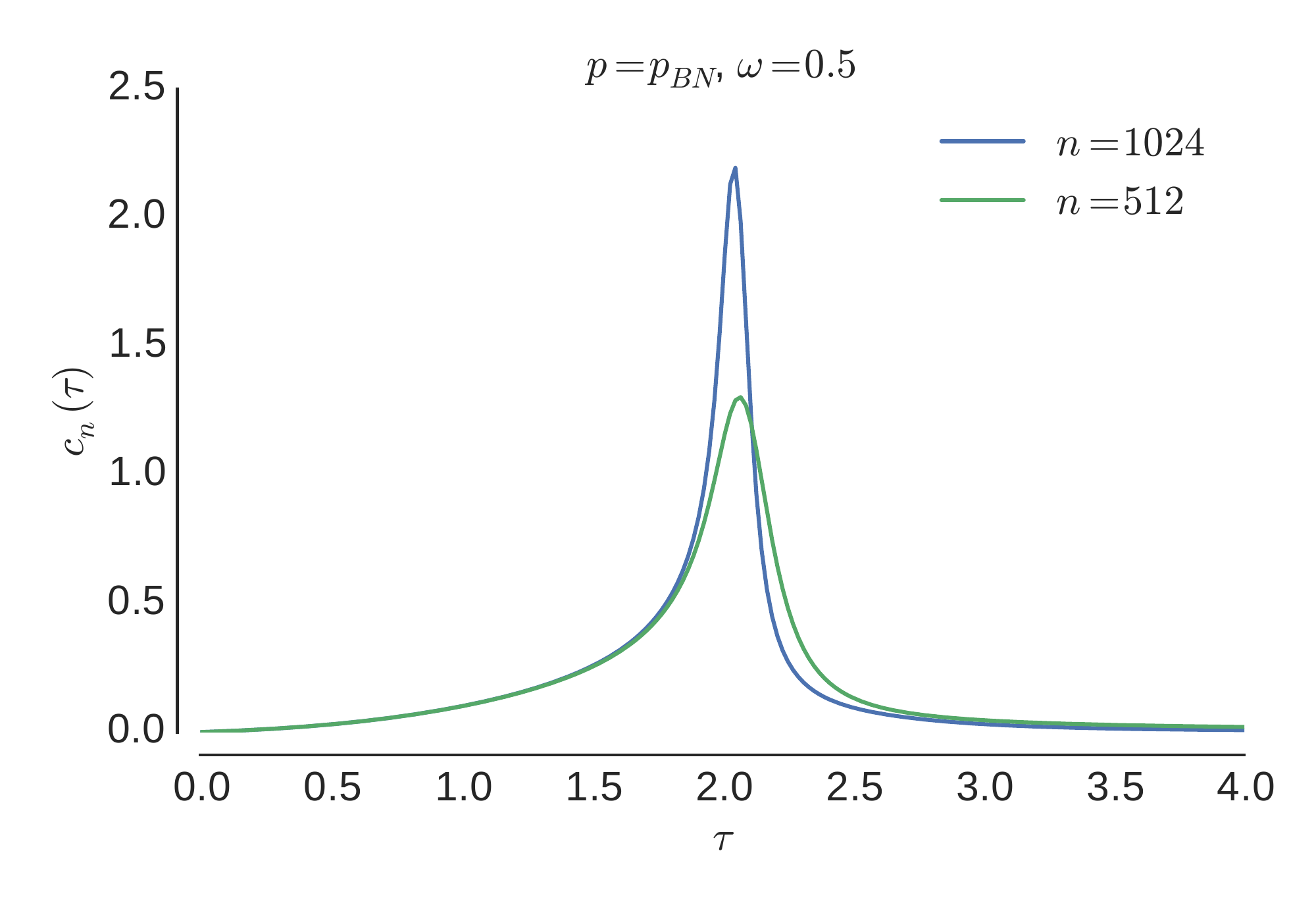}
  \caption{}\label{fig:VISAW_specific_heat_peak_omega_eq_0_5_p_eq_bn}
\end{subfigure}
\begin{subfigure}[b]{0.6\textwidth}
  \includegraphics[width=\textwidth]{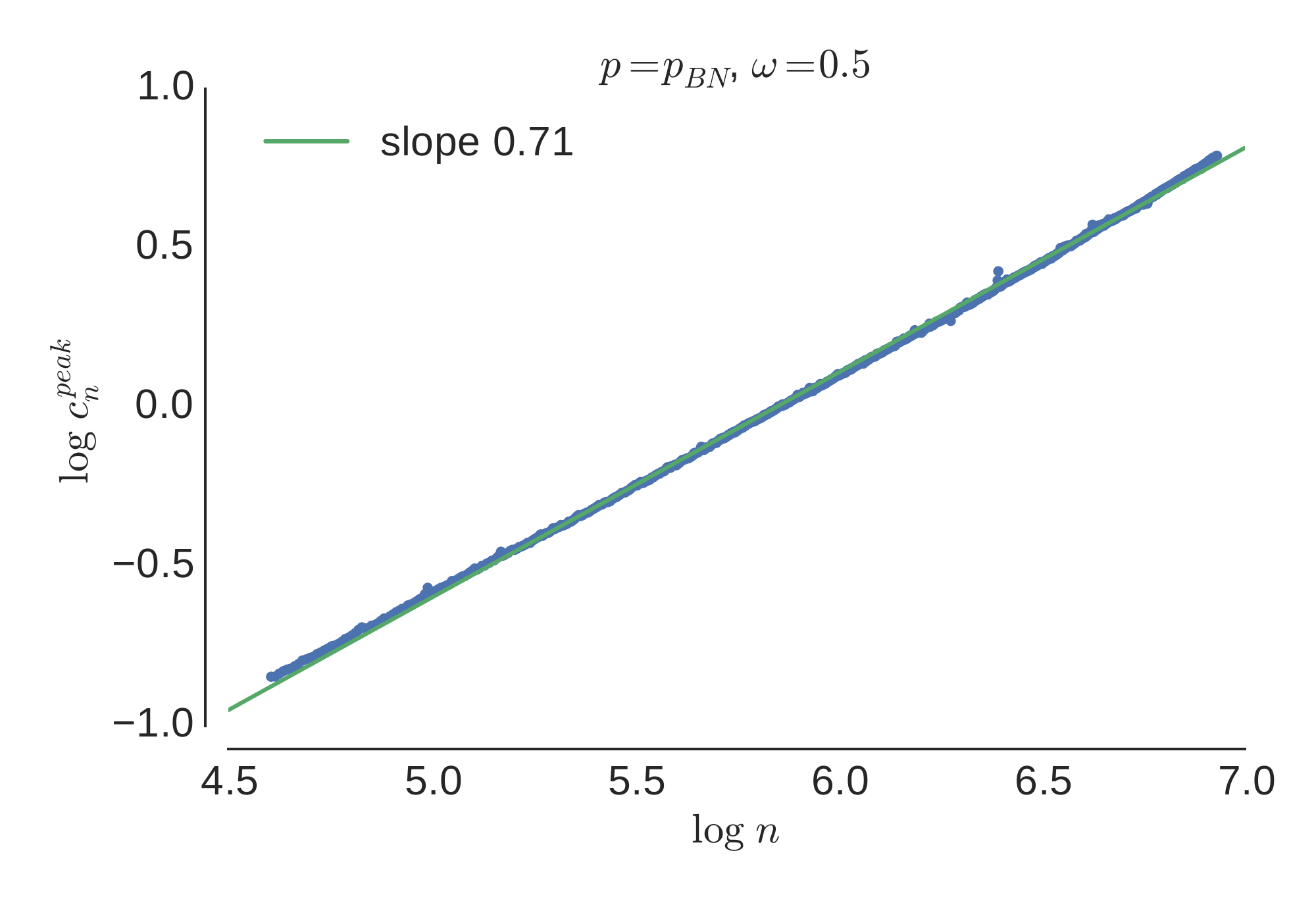}
  \caption{}\label{fig:VISAW_specific_heat_scaling_omega_eq_0_5_p_eq_bn}
\end{subfigure}
\begin{subfigure}[b]{0.6\textwidth}
  \includegraphics[width=\textwidth]{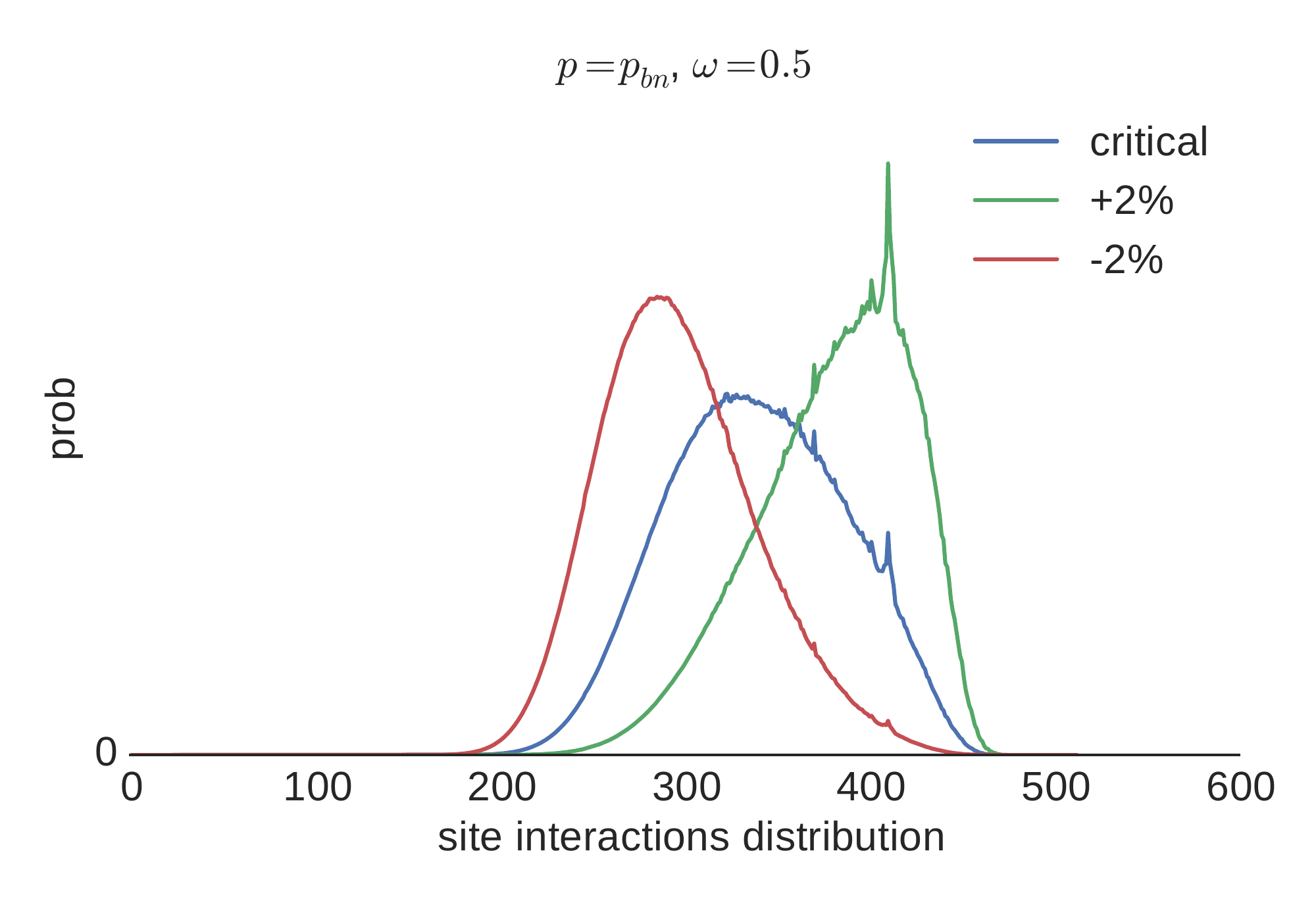}
  \caption{}\label{fig:site_interaction_distribution_omega_eq_0_5_p_eq_bn}
\end{subfigure}
\caption{(\subref{fig:VISAW_specific_heat_peak_omega_eq_0_5_p_eq_bn}) The specific heat as a function of $\tau$ on the vertical line $p=p_{BN}$, $\omega = 0.5$. (\subref{fig:VISAW_specific_heat_scaling_omega_eq_0_5_p_eq_bn}) The scaling of the specific heat peak with respect to the size of the walk. (\subref{fig:site_interaction_distribution_omega_eq_0_5_p_eq_bn}) The distribution of site interactions near the specific heat peak has a single mode. All plots come from dataset VI.}
\label{fig:omega_eq_0.5_p_eq_0}
\end{figure}

However, at $\omega=0.5$ we see in Figure~\ref{fig:omega_eq_0.5_p_eq_0} that an estimate of $0.71$ is obtained, compatible with the ISAT universality class. It could be that at $\omega=0.5$ there are just very strong finite size effects and a negative value will eventually be reached for sufficiently long lengths.

\begin{figure}[ht!]
\centering
\begin{subfigure}[b]{0.6\textwidth}
  \includegraphics[width=\textwidth]{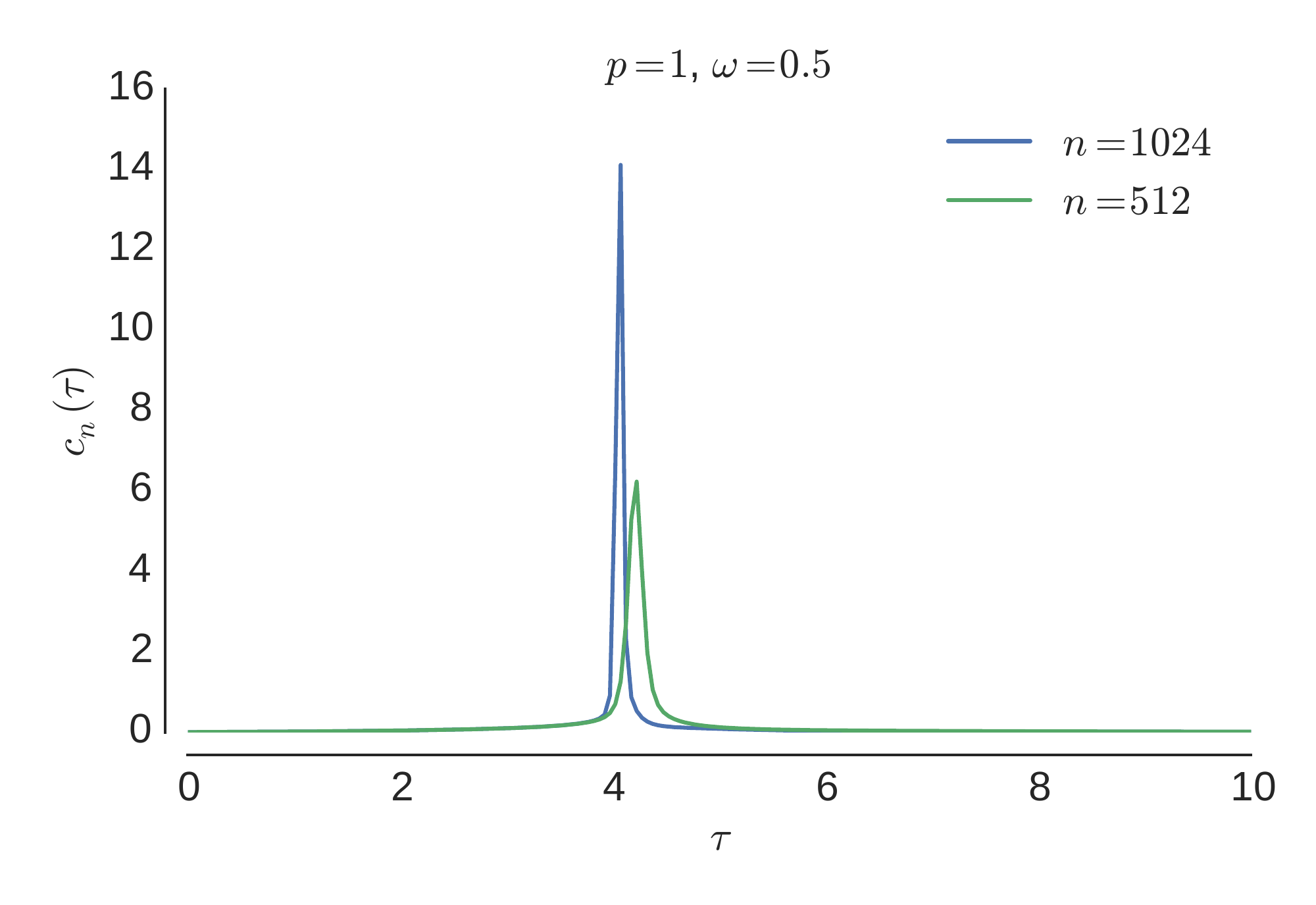}
  \caption{}\label{fig:VISAW_specific_heat_peak_omega_eq_0_5_p_eq_1}
\end{subfigure}
\begin{subfigure}[b]{0.6\textwidth}
  \includegraphics[width=\textwidth]{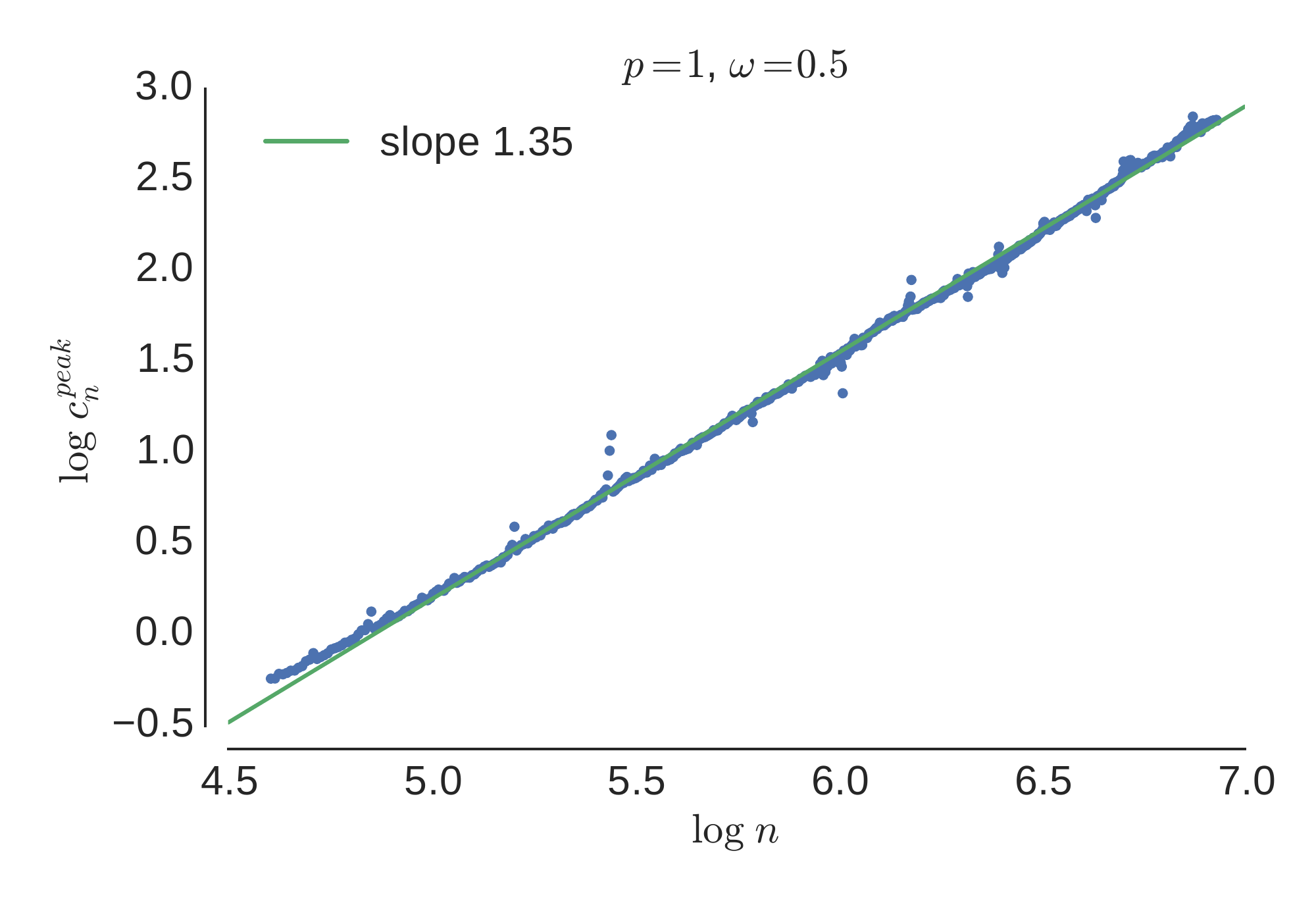}
  \caption{}\label{fig:VISAW_specific_heat_scaling_omega_eq_0_5_p_eq_1}
\end{subfigure}
\begin{subfigure}[b]{0.6\textwidth}
  \includegraphics[width=\textwidth]{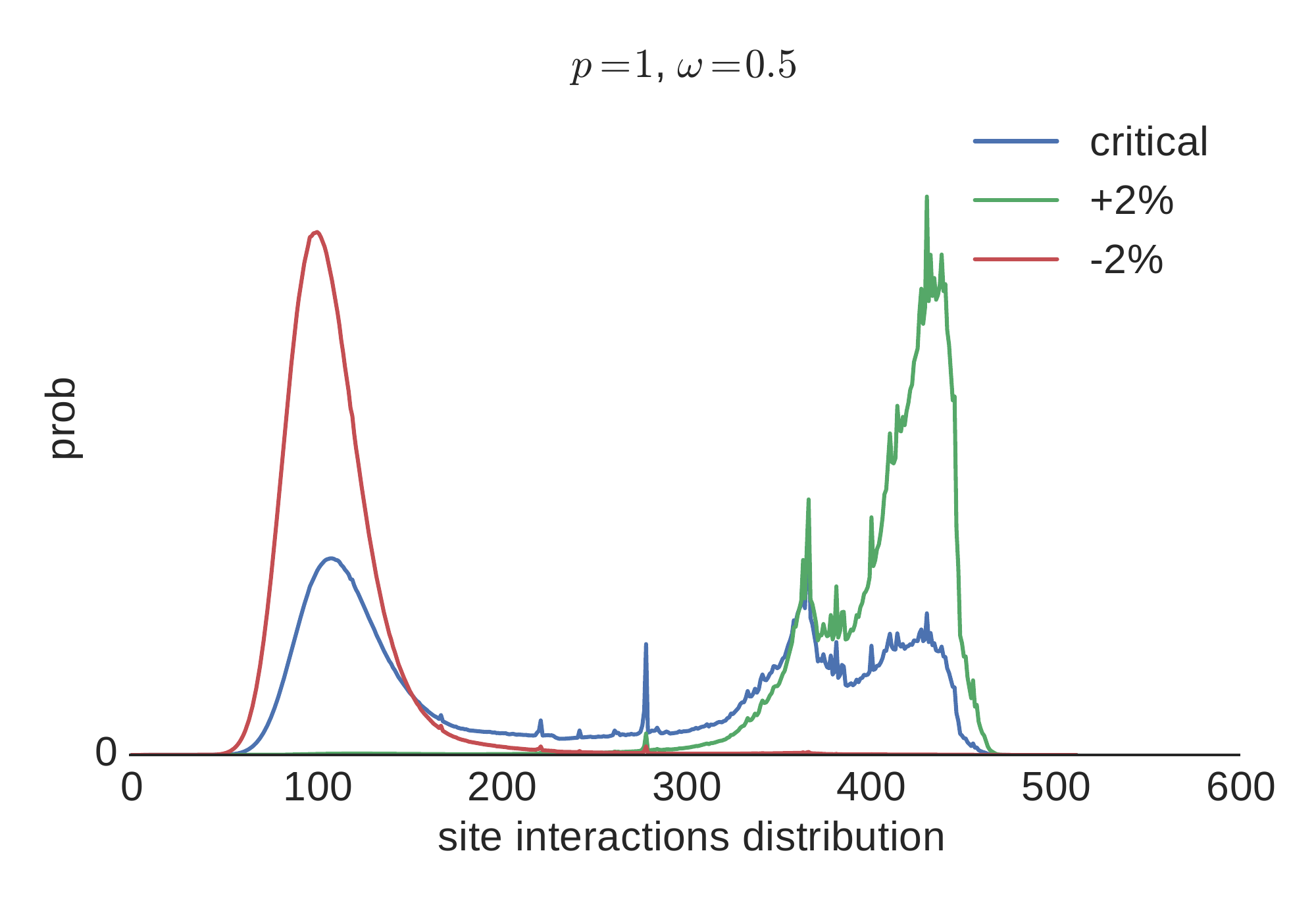}
  \caption{}\label{fig:site_interaction_distribution_omega_eq_0_5_p_eq_1}
\end{subfigure}
\caption{(\subref{fig:VISAW_specific_heat_peak_omega_eq_0_5_p_eq_1}) The specific heat as a function of $\tau$ on the vertical line $p=1$, $\omega = 0.5$. (\subref{fig:VISAW_specific_heat_scaling_omega_eq_0_5_p_eq_1}) the scaling of the specific heat peak with respect to the size of the walk. The scaling is super-linear, which is a clear indication of a first-order phase transition. (\subref{fig:site_interaction_distribution_omega_eq_0_5_p_eq_1}) The distribution of site interactions near the specific heat peak presents two clear peaks. These plots come from dataset VI-4.}
\label{fig:omega_eq_0_5_p_eq_1}
\end{figure}

\begin{figure}[ht!]
\centering
\includegraphics[width=0.5\textwidth]{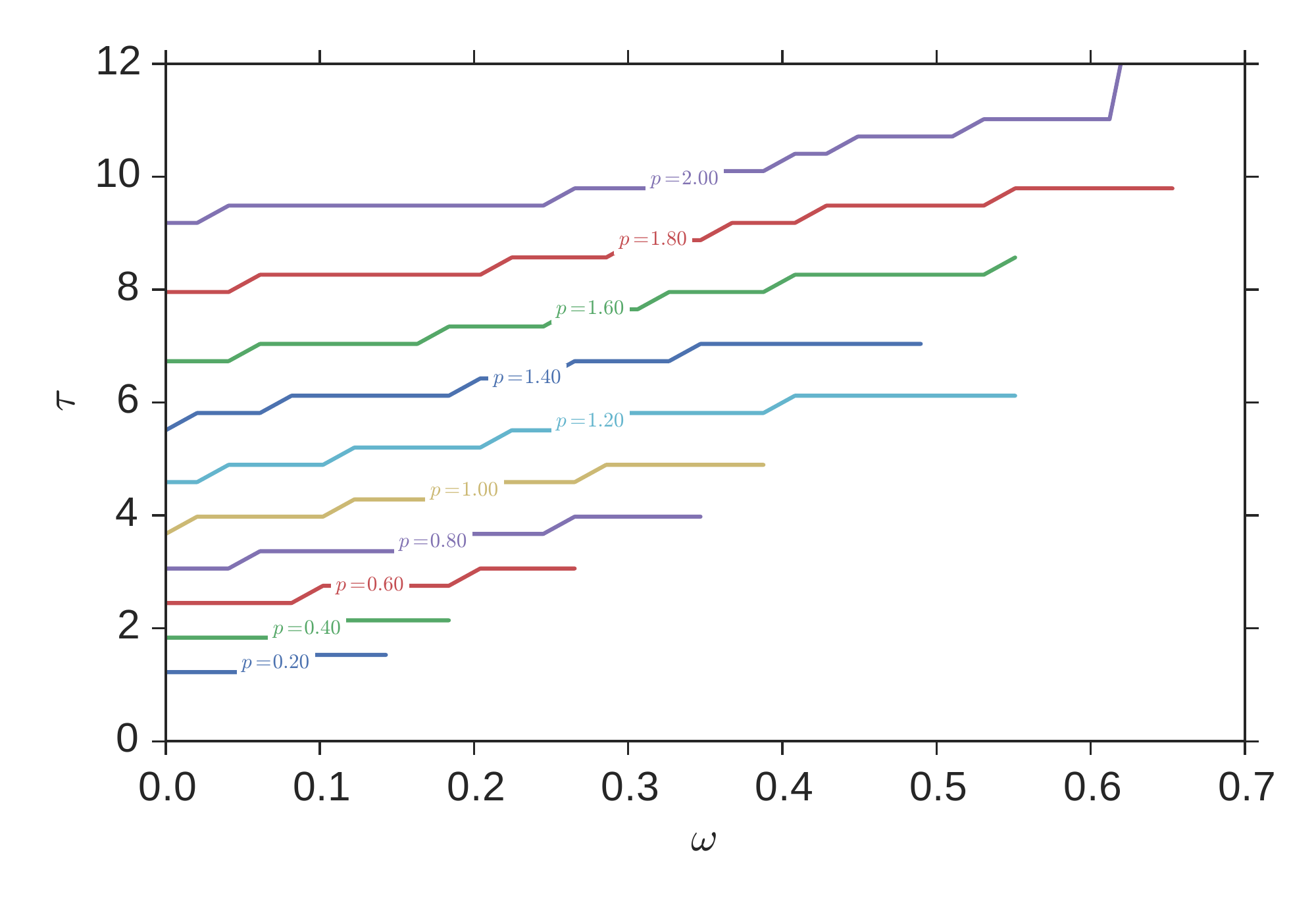}
\caption{We identified a two-dimensional region of first-order phase transitions. This plot shows, for each value of $\omega$ and $p$, the lowest value of $\tau$ for which we detected a double peak in the energy distribution. These lines have been obtained from the very short walks ($n=100$) in the dataset 3P and therefore these numbers are to be considered upper bounds (e.g. in the thermodinamic limit the lines would extend further on the right).
}
\label{fig:first-order-region}
\end{figure}

To demonstrate that the extended to maximally dense phase transition can also display first order characteristics we consider now $\omega=0.5$ with $p=1$. In Figure~\ref{fig:omega_eq_0_5_p_eq_1} we plot the specific heat and the distribution of doubly visited sites which is clearly bimodal. The specific heat looks like it is diverging super-linearly (impossible in the large length limit) which is the indication of the finite size early build up of a first-order transition. 
Finally, we estimate, using from simulations of short walks (up to length 100), the extent of the first order region for different values of $p$: this is given in Figure~\ref{fig:first-order-region}.

\clearpage

\section{Discussion}

We have considered a generalised model of polymer collapse in two dimensions that contains several well know models as subcases. We have found three low temperature phases as well as the normal high temperature extended phase that is dominated by the excluded volume effect. The three low temperature phases can be described as globular, which occurs in the canonical polymer collapse models, maximally dense, and as an anisotropic crystal phase, that has typical configurations  analogous to a three-dimensional $\beta$-sheet. The $\beta$-sheet phase and maximally dense phase do not meet. The phase transition between the extended and maximally dense phase is the most complicated and displays first order or second order characteristics depending on the parameters. More work needs to be done to fully elucidate the transitions but we now have an overall picture in a larger parameter space of the how these phases arise.

\ack

Financial support from the Australian Research Council via its support
for the Centre of Excellence for Mathematics and Statistics of Complex
Systems and through its Discovery program is gratefully acknowledged
by the authors. A L Owczarek thanks the School of Mathematical
Sciences, Queen Mary, University of London for hospitality. 

\section*{References}

\providecommand{\newblock}{}

\end{document}